\documentclass[12pt,epsf,amssymb]{article}

\usepackage{graphicx}
\usepackage{amsfonts}
\usepackage{amsmath}
\usepackage[usenames]{color}
\usepackage{amssymb}
\usepackage{amsthm}
\usepackage[mathscr]{eucal} 
\usepackage{url}

\def\N{\mathbb{N}}
\def\Z{\mathbb{Z}}
\def\Q{\mathbb{Q}}
\def\R{\mathbb{R}}
\def\C{\mathbb{C}}

\def\F{\mathbb{F}}

\begin{document}

\baselineskip 0.6cm
\newcommand{\vev}[1]{ \left\langle {#1} \right\rangle }
\newcommand{\bra}[1]{ \langle {#1} | }
\newcommand{\ket}[1]{ | {#1} \rangle }
\newcommand{\Dsl}{\mbox{\ooalign{\hfil/\hfil\crcr$D$}}}
\newcommand{\nequiv}{\mbox{\ooalign{\hfil/\hfil\crcr$\equiv$}}}
\newcommand{\nsupset}{\mbox{\ooalign{\hfil/\hfil\crcr$\supset$}}}
\newcommand{\nni}{\mbox{\ooalign{\hfil/\hfil\crcr$\ni$}}}
\newcommand{\nin}{\mbox{\ooalign{\hfil/\hfil\crcr$\in$}}}
\newcommand{\Slash}[1]{{\ooalign{\hfil/\hfil\crcr$#1$}}}
\newcommand{\EV}{ {\rm eV} }
\newcommand{\KEV}{ {\rm keV} }
\newcommand{\MEV}{ {\rm MeV} }
\newcommand{\GEV}{ {\rm GeV} }
\newcommand{\TEV}{ {\rm TeV} }

\def\diag{\mathop{\rm diag}\nolimits}
\def\tr{\mathop{\rm tr}}

\def\Spin{\mathop{\rm Spin}}
\def\SO{\mathop{\rm SO}}
\def\SU{\mathop{\rm SU}}
\def\U{\mathop{\rm U}}
\def\Sp{\mathop{\rm Sp}}
\def\SL{\mathop{\rm SL}}

\def\change#1#2{{\color{blue}#1}{\color{red} [#2]}\color{black}\hbox{}}

\theoremstyle{definition}
\newtheorem{thm}{Theorem}[subsection]
\newtheorem{defn}[thm]{Definition}
\newtheorem{notn}[thm]{Notation}
\newtheorem{exmpl}[thm]{Example}
\newtheorem{props}[thm]{Proposition}
\newtheorem{lemma}[thm]{Lemma}
\newtheorem{rmk}[thm]{Remark}
\newtheorem{anythng}[thm]{}

\begin{titlepage}
  
\begin{flushright}
IPMU18-0019
\end{flushright}
 
 \vskip 1cm
\begin{center}
 
 {\large \bf String-theory Realization of Modular Forms \\
for Elliptic Curves with Complex Multiplication}

\vskip 1.2cm
  
 Satoshi Kondo$^{1,2}$ and Taizan Watari$^2$
  
 \vskip 0.4cm
  
   {\it $^1$ Faculty of Mathematics, National Research University Higher School of Economics, 
Moscow, Russia
  \\[2mm]
 
     $^2$Kavli Institute for the Physics and Mathematics of the Universe, 
    University of Tokyo, Kashiwa-no-ha 5-1-5, 277-8583, Japan
   }
 
\vskip 1.5cm
    
\abstract{It is known that the $L$-function of an elliptic curve defined over 
$\Q$ is given by the Mellin transform of a modular form of weight 2. 
Does that modular form have anything to do with string theory? In this article, 
we address a question along this line for elliptic curves that have complex 
multiplication defined over number fields. So long as we use diagonal rational 
${\cal N}=(2,2)$ superconformal field theories for the string-theory 
realizations of the elliptic curves, the weight-2 modular form turns out 
to be the Boltzmann-weighted ($q^{L_0-c/24}$-weighted) sum of U(1) charges 
with $Fe^{\pi i F}$ insertion computed in the Ramond sector. } 
  
\end{center}
 \end{titlepage}
 

\section{Introduction}

For any elliptic curve defined over $\Q$, its Hasse--Weil 
$L$-function is given by the Mellin transform of a modular form 
of weight 2 (by Shimura--Taniyama conjecture, now a theorem). 
For many elliptic curves with complex multiplication defined over 
number fields, their Hasse--Weil $L$-functions are still given 
by multiplying the Mellin transforms of modular forms. 
In the meantime, when we define a string theory 
by using a geometry as a target space, its worldsheet formulation 
must have modular invariance; various observables computed 
in such a string theory are often modular forms. Are there any 
relations between those two kinds of modular forms? If there are, 
what are the relations precisely? We address this question\footnote{
The second named author has been led to this question through 
the application of arithmetic ideas to the cosmological constant / 
gravitino mass problem in the context of flux compactification \cite{Moore-04, 
DeWolfe-04, flux-others, flux-CM}.}
 in this article; the spirit is therefore similar to that of \cite{Schimmrigk}.

A relation between them, if there is any, cannot be as simple as 
``they are the same''.  The Hasse--Weil $L$-function is defined for  
an algebraic variety (where complex structure is specified, but metric is not), 
while there is no choice in formulating a string theory without  
specifying a metric on a target space.\footnote{a caveat: topological string theory} 
At least we need to extract some information from string theory in a way 
the results do not depend on the choice of a metric, or to find 
a way to extract for any choice of metric, to say the least. 
One will also notice that the $L$-function is defined for individual 
models defined over number fields; in arithmetic geometry, two elliptic 
curves given by $y^2 = x^3 - x$ and $y^2 = x^3 - 4 x$ are regarded different 
varieties defined over $\Q$. Those two elliptic curves are regarded the same, 
however, when we use them as a target space of string theory; coordinate 
reparametrization in $\C$ (rather than in $\Q$) washes out the difference 
between them.\footnote{In a more mathematical language, elliptic curves 
used as a target space in string theory are objects in the category of 
K\"{a}hler manifolds (if duality is ignored). On the other hand, the 
$L$-function is defined for individual objects in categories of algebraic 
varieties defined over some number fields. There is no canonical functor 
in one way or the other. 
There is just correspondence between them, where infinitely many objects 
on one side correspond to infinitely many objects on the other. }
How can a string theory with a given elliptic curve defined 
over $\C$ provide information of the $L$-functions of various models 
defined over number fields? 

In this article, we deal with a class of elliptic curves---those with 
complex multiplication with one condition (stated in 
Lemma \ref{lemma:HeckeC-LandK})---and provide an answer to the questions 
raised above. It is best to begin with a relation between character 
functions (given essentially by theta functions) of string theory 
with an elliptic curve as a target space, and the Dedekind zeta function 
$\zeta_k(s)$ of a number field $k$; the observation that the Mellin 
transform of the former is the latter [Moore \cite{Moore-98s}, \S4]
corresponds to the first line of the following diagram,  
\begin{eqnarray}
   {\rm characters} & \Longleftrightarrow & \zeta_k(s)=L(H^0_{et}(E),s),
    \nonumber \\
  ?? & \Longleftrightarrow & L(H^1_{et}(E),s).  \nonumber 
\end{eqnarray}
In section \ref{ssec:L0-combine}, we elaborate on the relation in the first line,  
because this process serves as a warming up exercise in finding out the 
appropriate string-theory objects in the second line. We will see, 
in section \ref{ssec:L1-combine}, that the objects ``??'' 
are (\ref{eq:def-f1-closed-super}, \ref{eq:def-f1-open-super}), because 
of the observation (\ref{eq:obsrv-f1}). Thms.\ \ref{thm:main-1} 
and \ref{thm:main-2} explain how the Hasse--Weil $L$-functions 
(i.e., $L(H^1_{et}(E),s)$) of arithmetic models of an elliptic curve 
with complex multiplication can be obtained from those objects defined 
in string-theory realizations. 
The objects (\ref{eq:def-f1-closed-super}, \ref{eq:def-f1-open-super})
are modular forms of weight 2 for a subgroup of ${\rm SL}(2;\Z)_{ws}$
acting on the complex structure parameter $\tau_{ws}$ of the worldsheet 
torus of string theory. The ``how can a string theory... ?'' question 
raised earlier in Introduction is also answered right after Thm. \ref{thm:main-2}.

Background materials from string theory are explained in 
sections \ref{sec:GV}, \ref{ssec:prepare-f0} and \ref{ssec:prepare-f1}, 
while sections \ref{ssec:field-def}, \ref{ssec:theta-zeta} 
and \ref{ssec:review-HW-L1} provide a quick review on materials from 
algebraic number theory we need in this article.  
Novelty may be found only in section \ref{ssec:L1-combine}.
If a reader already has introductory level knowledge on string 
theory and class field theory, he/she can proceed directly to 
section \ref{ssec:L1-combine}.

Despite the apparent style of presentation built up with 
``Theorem,'' ``Proposition'' and so forth, and despite frequent 
use of algebraic number theory, the main result reported in this 
article----what is the object ``??'' in the diagram above---is a statement 
on string theory. We have adopted a math-like style of presentation in 
this article for the purpose of making statements more precise, and 
implicit contexts explicit. 

\section{Rational CFT and Complex Multiplication}
\label{sec:GV}

Refs. \cite{Moore-98full, GV} pointed out, roughly speaking, that an elliptic 
curve has complex multiplication if and only if its string-theory realization is 
given by a rational conformal field theory (CFT). See also \cite{Wend-Chen}.
This section provides a brief review on part of the results of \cite{Moore-98full, GV} 
for readers with math background, by adopting a math-style presentation. 
We have also inserted Remark \ref{rmk:class-field-th}, which tries to give 
an idea of what elliptic curves with complex multiplication are like, for 
those who have never heard of the jargon.
\vspace{1cm}

Think of a bosonic string theory with an elliptic curve as the target space. 
This is equivalent to thinking of a $(c,\tilde{c})=(2,2)$ CFT on a worldsheet 
that has two holomorphic (left-mover) U(1) current operators and also two 
anti-holomorphic (right-mover) ones. We call such a CFT as a {\it $T^2$-target 
CFT}. $T^2$-target CFT's are parametrized by a pair $(z, \rho) \in 
{\cal H} \times {\cal H}$, where ${\cal H}$ is the upper complex half plane without 
the real axis. Here, $z$ stands for the complex structure parameter and $\rho$ 
for the complexified K\"{a}hler parameter of an elliptic curve.

\begin{notn}
For $z \in {\cal H}$, $E_z$, $[E_z]$ or $[E_z]_\C$ stands for a 
$\C$-isomorphism class of elliptic curves that can be constructed by 
$\C/(\Z \oplus z \Z)$. $\bullet$
\end{notn}

The same elliptic curve characterized by a pair 
$(z, \rho) \in {\cal H} \times {\cal H}$ can also be used as a target 
space of Type II superstring. There, we think of a $(c,\tilde{c})=(3,3)$ 
CFT with ${\cal N}= (2,2)$ supersymmetry on a worldsheet that has two holomorphic 
(left-mover) U(1) current operators in addition to the U(1) current operator 
$J_L$ in the left-mover superconformal algebra, and also two anti-holomorphic 
(right-mover) ones in addition to $J_R$ in the right-mover superconformal 
algebra. We call such a CFT as a {\it $T^2$-target
${\cal N} = (2,2)$ superconformal field theory} (SCFT).

\subsubsection{Statements from Refs. \cite{Moore-98full, GV}} 

References \cite{Moore-98full, GV} found the following, among other things. 

\begin{props}
\label{props:GV-1}
A $T^2$-target CFT is a rational CFT if and only if both of the following 
two conditions are satisfied: i) both $E_z$ and $E_\rho$ admit complex 
multiplications, and ii) $E_z$ and $E_\rho$ are isogenous. $\bullet$
\end{props}

\begin{props}
\label{props:GV-2}
A rational $T^2$-target CFT is diagonal, when either $\rho \in {\rm Aut}(E_z)$, 
or $z \in {\rm Aut}(E_\rho)$. $\bullet$
\end{props}

In the rest of this note, we only think of the cases with 
$\rho \in {\rm Aut}(E_z)$. For non-string readers, additional information 
(incl. definition) on diagonal rational CFT's are provided in section \ref{sssec:RCFT}.
For readers with physics background, it is useful to know the following 
classic results in mathematics:

\begin{rmk}
\label{rmk:class-field-th}
{\it Elliptic curves with complex multiplication by an order of an imaginary quadratic field} 
$K = \Q(\sqrt{-d_0})$ (for some positive square-free integer $d_0$) 
are classified modulo isomorphisms over $\C$ by the set 
\begin{align}
  \amalg_{f_z \in \N_{>0}} Ell({\cal O}_{f_z}),  \label{eq:classes-ell-givenK}
\end{align}
where\footnote{$D_K$ is the {\it discriminant} of an imaginary quadratic field 
$K=\Q(\sqrt{-d_0})$. When $d_0$ satisfies $d_0 \equiv 3$ mod 4, $-D_K = d_0$;
when $d_0 \equiv 1,2$ mod 4, however, $-D_K = 4d_0$.}  
\begin{align}
    Ell({\cal O}_{f_z}) = \left\{ \left. 
    \left( \begin{array}{cc} 2a & b \\ b & 2c \end{array} \right)  
    \right| \; a,b,c \in \Z, \; (a,b,c)=1, \; D_z := 4ac-b^2 = |D_K| f_z^2 \right\} 
  / {\rm SL}(2;\Z). \nonumber
\end{align}
Here, a matrix $h \in {\rm SL}(2;\Z)$ acts on a matrix $g(a,b,c)$ parametrized 
by integers $a$, $b$, $c$ as above, by $g(a,b,c) \mapsto g(a',b',c') = 
h \cdot g(a,b,c) \cdot h^T$. The corresponding $\C$-isomorphism class in 
$Ell({\cal O}_{f_z})$ is given by $E_z$ with the solution $z \in {\cal H}$ of 
\begin{align}
  a z^2 + bz + c = 0.
\end{align}
A set of integers $(a,b,c)$ that gives rise to an elliptic curve $E_z$ in this way 
is denoted by $(a_z, b_z, c_z)$.
Because the ${\rm SL}(2;\Z)$ action on a set of integers $(a_z,b_z,c_z)$ corresponds to 
the ${\rm SL}(2;\Z)$ transformation on $z \in {\cal H}$, an ${\rm SL}(2;\Z)$ orbit 
of sets of integers $(a_z,b_z,c_z)$ specifies just one $\C$-isomorphism class 
of elliptic curves. 

For any $\C$-isomorphism class $[E_z]$ of elliptic curves in $Ell({\cal O}_{f_z})$, 
the ring of endomorphism of $E_z$ is ${\rm Aut}(E_z) \cong \Z \oplus (a_z z)\Z =: 
{\cal O}_{f_z} \subset {\cal O}_K \subset K \subset \C$; here, ${\cal O}_K$ is the 
ring of algebraic integers in $K$. Such a subring ${\cal O}_{f_z}$ of ${\cal O}_K$ 
is called an {\it order}, and ${\cal O}_K = {\cal O}_{f_z=1}$ is called the 
{\it maximal order} of the imaginary quadratic field $K=\Q(\sqrt{-d_0})$.
We say that an elliptic curve $E$ whose $\C$-isomorphism class $[E]$ is in 
$Ell({\cal O}_{f_z})$ {\it has complex multiplication by ${\cal O}_{f_z}$}.  

It is also known that the ideal class group of an order\footnote{
\label{fn:ring-ray} It is a quotient of the ray ideal class group 
${\rm Cl}_K(\mathfrak{m}_f)$ with the modulus $\mathfrak{m}_f = 
(f_z)_{{\cal O}_K}$. The ray class field corresponding to the modulus 
$\mathfrak{m}_f = (f_z)_{{\cal O}_K}$ is denoted by 
$L_{\mathfrak{m}_f}=L_{(f_z)_{{\cal O}_K}}$ in this article, whereas $L_{f_z}$ 
stands for the abelian extension of $K$ satisfying 
${\rm Gal}(L_{f_z}/K)\cong {\rm Cl}_K({\cal O}_{f_z})$. 
$L_{f_z}$ is a subfield of $L_{(f_z)_{{\cal O}_K}}$. 
See \cite{Lang-EF, Moreland} for class field theory associated with 
elliptic curves.} ${\cal O}_{f_z}$, denoted by ${\rm Cl}_K({\cal O}_{f_z})$, 
is isomorphic to the set $Ell({\cal O}_{f_z})$. Table \ref{tab:cft-qif} 
shows examples of elliptic curves with complex multiplication.  
This article will not provide a review on class field theory beyond what 
is written here; instead, we will provide reference to math textbooks 
occasionally.\footnote{
A string-theorist-friendly review on global class field theory is found 
in \cite{Moore-98full, Moore-04}. Systematic expositions on global class 
field theory are also found in such textbooks as \cite{Neukirch, Lang-ANT, 
Milne-CFT}, and a little more introductory ones in \cite{Lang-EF, Moreland}. } 
$\bullet$
\end{rmk}
%
\begin{table}[tbp]
\begin{center}
\begin{tabular}{c||l|l|l|l|l}
 $|D_K|$ & $f_z = 1$ & $f_z = 2$ & $f_z = 3$ & $f_z = 4$ & $f_z = 5$ \\
 \hline
 3 & [1,1,1] & [1,0,3] & [1,1,7] & [1,0,12], [3,0,4] & [1,1,19],[3,3,7] \\
 4 & [1,0,1] & [1,0,4] & [1,0,9], [2,2,5] & [1,0,16], [4,4,5] & [1,0,25],[2,2,13] \\
 7 & [1,1,2] & [1,0,7] & $h({\cal O}_{f_z=3})=4$ & $h({\cal O}_{f_z=4})=2$ & $h({\cal O}_{f_z=5})=6$ \\
 8 & [1,0,2] & [1,0,8], [3,2,3] & $h({\cal O}_{f_z=3})=2$ & $h({\cal O}_{f_z=4})=4$ & $h({\cal O}_{f_z=5})=6$ \\
 11 & [1,1,3] & [1,0,11],[3,$\pm2$,4] & $h({\cal O}_{f_z=3})=2$ & $h({\cal O}_{f_z=4})=6$ & $h({\cal O}_{f_z=5})=4$ \\
 \hline 
 20 & [1,0,5], [2,2,3] & $h({\cal O}_{f_z=2})=4$ & $h({\cal O}_{f_z=3})=4$ & $h({\cal O}_{f_z=4})=8$ & $h({\cal O}_{f_z=5})=10$ 
\end{tabular}
\caption{\label{tab:cft-qif}Representatives $[a,b,c]$ of individual elements of 
${\rm Cl}_K({\cal O}_{f_z})$ for imaginary quadratic fields $K$ with small $|D_K|$. 
Representatives are chosen so that $0 \leq a \leq c$ and $-a < b \leq a$; when $c=a$, 
however, $0\leq b \leq a$. 
In order to save space, just the cardinality $h({\cal O}_{f_z})$ of 
${\rm Cl}_K({\cal O}_{f_z})$ is shown in part of this Table, instead of representatives 
of all the elements of ${\rm Cl}_K({\cal O}_{f_z})$.
There are two more imaginary quadratic fields, $K = \Q(\sqrt{-15})$ and 
$K=\Q(\sqrt{-19})$, that would come in between $D_K = -11$ and $D_K = -20$, 
but we omitted them, just to save space.}
\end{center}
\end{table}

Suppose that an elliptic curve $E_z$ has complex multiplication by an order 
${\rm Aut}(E_z) = {\cal O}_{f_z}$ of an imaginary quadratic field, as in 
Propositions \ref{props:GV-1} and \ref{props:GV-2}. Two string theory 
realizations with $(z, \rho)$ and $(z, \rho')$, with 
$\rho, \rho' \in {\cal O}_{f_z} 
= \Z \oplus (a_z z)\Z$ (as indicated in Proposition \ref{props:GV-2}) 
are regarded the same, when $\rho - \rho' \in \Z$ (converting $\rho$ to 
$\rho' \in \rho + \Z$ is a part of ${\rm SL}(2;\Z)$ action on the parameter $\rho$). 
This means that we can always choose $\rho$ of a rational diagonal $T^2$-target CFT 
to be an integer ($f_\rho$) multiple of $a_z z$: $\rho = f_\rho a_z z$. 
From here, therefore, follows a consequence of Propositions \ref{props:GV-1} 
and \ref{props:GV-2} phrased in a way favorable for algebraic geometers: 

\begin{props}
Think of a $\C$-isomorphism class of elliptic curves with complex 
multiplication by an order ${\cal O}_{f_z}$ of an imaginary quadratic 
field $K$, i.e., any element of (\ref{eq:classes-ell-givenK}). Then there is 
a family of bosonic string theory realizations in the form of a rational and 
diagonal $T^2$-target CFT, parametrized by $f_\rho \in \N_{>0}$. Also, 
there is a family of Type II superstring theory realizations parametrized 
by $f_\rho \in \N_{>0}$, each one of which is in the form of a rational and 
diagonal $T^2$-target ${\cal N}=(2,2)$ SCFT. $\bullet$ 
\end{props}

The parameter $f_\rho$ controls the choice of complexified K\"{a}hler form 
on the $\C$-isomorphism class of elliptic curves. Although algebraic 
geometry only deals with complex structure of a geometry and does not 
refer to the choice of a metric on it, yet there is no option in any 
string-theory realization of an object in algebraic geometry not to 
specify a metric (complexified K\"{a}hler form) on it. The Proposition 
above says that the choice of complexified K\"{a}hler form is parametrized 
by one positive integer $f_\rho \in \N_{>0}$, when we impose a condition that 
the CFT on worldsheets be rational and diagonal.\footnote{If we do not 
impose this condition, any $[\rho] \in {\cal H}/{\rm SL}(2;\Z)$ is fine.}

\subsubsection{Rudiments on Rational CFT's}
 \label{sssec:RCFT}

To describe various structures of diagonal rational $T^2$-target CFT's in 
terms of arithmetic properties of the elliptic curves with complex multiplication, 
we prepare a few important general properties of diagonal rational CFT's 
(that are not necessarily $T^2$-target).
\begin{rmk}
\label{rmk:d.CFT}
When a rational diagonal CFT is given, there is an isomorphism $\varphi_0$ 
between the algebra ${\cal A}_-$ of purely holomorphic vertex operators and 
that (${\cal A}_+$) of purely anti-holomorphic vertex operators, 
\begin{align}
 \varphi_0: {\cal A}_- \cong {\cal A}_+.
\end{align}
The Hilbert space of closed string states ${\cal H}^{\rm closed}$ in such a theory 
has a structure 
\begin{align}
  {\cal H}^{\rm closed} = \oplus_{\alpha \in iReps.} (V^-_\alpha \otimes V^+_\alpha ); 
\end{align}
here, $V^+_\alpha$ is an irreducible representation of ${\cal A}_+$, 
and there are just finite number of distinct irreducible representations 
(by definition of {\it rational} CFT's). The index $\alpha$ runs over all the 
irreducible representations. $V^-_\alpha$ is the irreducible representation 
of ${\cal A}_-$ that is regarded as $V^+_\alpha$ when the action of ${\cal A}_-$ 
is identified with that of ${\cal A}_+$ through the isomorphism $\varphi_0$ 
(by definition of the {\it diagonal}ity). $\bullet$
\end{rmk}

\begin{rmk}
In a rational $T^2$-target CFT, there exists a sublattice 
$(\Gamma_- \oplus \Gamma_+) \subset {\rm II}_{2,2}$; ${\rm II}_{2,2}$ 
is the even unimodular lattice of signature $(2,2)$, $\Gamma_-$ [resp.\ $\Gamma_+$]
is an even integral negative [resp.\ positive] definite primitive sublattice of 
${\rm II}_{2,2}$. 
When a rational $T^2$-target CFT is also diagonal, then the isomorphism 
$\varphi_0: {\cal A}_- \cong {\cal A}_+$ also induces a lattice isometry 
$\varphi_0: \Gamma_-[-1] \cong \Gamma_+$. The set of irreducible representations 
are labeled by 
\begin{align}
  iReps. \cong \Gamma_+^\vee / \Gamma_+ \cong {\rm II}_{2,2}/(\Gamma_- \oplus \Gamma_+)
  \cong \Gamma_-^\vee / \Gamma_-,
\end{align}
where $L^\vee$ stands for the dual lattice of an integral lattice $L$. $\bullet$
\end{rmk}

In the language of string theory, ${\rm II}_{2,2}$ is the lattice of charges 
under the four U(1) currents in a $T^2$-target CFT. The sublattice 
$\Gamma_-$ [resp. $\Gamma_+$] in a rational $T^2$-target CFT corresponds to 
the set of U(1) charges for which purely holomorphic [resp. purely anti-holomorphic] 
operators exist.\footnote{Due to the fact that the pair of primitive sublattices 
$\Gamma_-$ and $\Gamma_+$ fit into the even unimodular lattice ${\rm II}_{2,2}$, 
one can see that the discriminant groups and forms of $\Gamma_-[-1]$ and $\Gamma_+$ 
are identical. This condition is still weaker than the existence of a lattice 
isometry $\Gamma_-[-1] \cong \Gamma_+$. See \cite{HLOY}. 
Furthermore, the existence of a lattice isometry $\Gamma_-[-1] \cong \Gamma_+$ is 
still a weaker condition than the one for a rational $T^2$-target CFT to be diagonal. }

\begin{rmk} 
\label{rmk:open-d.RCFT}
(we keep the notation as in the previous remark) 
In a rational diagonal CFT, $\varphi_0$-{\it Cardy states} refer to 
a special class of choices of boundary conditions that can be imposed 
on worldsheets; they are in one-to-one correspondence with the set $iReps.$ 
Furthermore, when a worldsheet is in the form of a long strip and both of 
the two edges of the strip are subject to Cardy states, say, $\alpha$ and 
$\beta \in iReps.$, then the open string states on such a long strip form 
a representation of the algebra\footnote{With an abuse of notation, we also use 
$\varphi_0$ for this isomorphism.} $({\cal A}_- \times {\cal A}_+) / 
(\varphi_0{\rm ~at.bdry}) \cong {\cal A}_+$.  Therefore, the Hilbert space of 
open string states with both of the two boundary conditions being $\varphi_0$-Cardy 
states consists of 
a direct sum of $V^o_\alpha$, the irreducible representation of the algebra 
that is isomorphic to $V^+_\alpha$ when seen as a representation of ${\cal A}_+$.
$\bullet$
\end{rmk}

\subsubsection{More Statements from Ref. \cite{GV}}

With the preparation in remark \ref{rmk:class-field-th} and 
section \ref{sssec:RCFT}, we are now ready to write down more 
statements in \cite{GV}: 
\begin{props}
Think of a diagonal rational $T^2$-target CFT, with $\rho = f_\rho a_z z$. 
Then the characters of the irreducible representations of the algebra ${\cal A}_+$
and ${\cal A}_-$ are given by 
\begin{align}
 {\rm ch}_{V^+_\alpha}(\bar{q}  = e^{-2\pi i \bar{\tau}} ) & \;
   := {\rm Tr}_{V^+_\alpha}[ e^{-2\pi i \bar{\tau} (\tilde{L}_0-\tilde{c}/24)} ] 
   = \chi_{\alpha}(\bar{q}), \\
 {\rm ch}_{V^-_\alpha}(q  = e^{2\pi i \tau} ) & \; := {\rm Tr}_{V^-_\alpha}[ e^{2\pi i \tau (L_0-c/24)}] 
   = \chi_{\alpha}(q), 
\end{align}
where 
\begin{align}
   \chi_\alpha(e^{2\pi X})  = \frac{f_0(X;\alpha)}{(\eta(e^{2\pi X}))^2}, \qquad 
  f_0\left( X ; \alpha \right) := \sum_{p \in \alpha} e^{2\pi i X \frac{(p,p)}{2}};
\end{align}
$\eta(q) = q^{\frac{1}{24}} \prod_{n=1}^\infty (1-q^n)$ is the Dedekind $\eta$-function, 
and $L_0$ [resp. $\tilde{L}_0$] is one of the generators of the Virasoro algebra 
in the left-moving [resp. right-moving] sector.  
$\alpha \in iReps. \cong \Gamma_+^\vee / \Gamma_+$ labels irreducible 
representations on the left-hand sides, but is regarded as the subset of $\Gamma_+^\vee$ on 
the right-hand side; the positive definite intersection form $(-, -)$ of 
the lattice $\Gamma_+^\vee$ is used in the exponent. 

The intersection form of $\Gamma_+$ is given by a matrix 
\begin{align}
  f_\rho \left[ \begin{array}{cc} 2a_z & b_z \\ b_z & 2c_z \end{array} \right]
  = 2 a_z f_\rho
    \left[ \begin{array}{cc} 1 & -z_1 \\ - z_1 & |z|^2 \end{array} \right]
\end{align}
when a set of generators of $\Gamma_+$ is chosen appropriately; 
here, ${\rm Re}(z)=:z_1$ and ${\rm Im}(z)=: z_2 > 0$. 
It follows that 
\begin{align}
{\rm discr}(\Gamma_+) = f_\rho^2 D_z = f_\rho^2 f_z^2 |D_K|.  \qquad \qquad \bullet
\end{align}
\end{props}

\begin{props}
\label{props:char-open}
In a diagonal rational $T^2$-target CFT with $\rho = f_\rho a_z z$, 
the open string states subject to $\varphi_0$-Cardy states\footnote{
Cardy states are certain class of boundary conditions to be imposed on 
string theories on a worldsheet with a boundary. It may sound strange to 
refer to a boundary condition as a ``state,'' but it is legitimate, because 
choice of such a boundary condition can be regarded as a choice of 
a state in a vector space on which linear operators act, in a certain perspective 
in string theory. Because we do not need to exploit this perspective in this note, 
Cardy states can be understood as a certain class of choices of boundary 
conditions imposed on a string theory on a worldsheet with a boundary. \\
In order to define Cardy states, one needs to specify the linear combination of operators from ${\cal A}_-$ and ${\cal A}_+$ whose kernel the boundary states are in. 
The prefix ``$\varphi_0$-'' specifies the linear combination. } 
are in the representation of the algebra 
$\varphi_0: ({\cal A}_- \times {\cal A}_+/(\varphi_0{\rm ~at.bdry}) 
\cong {\cal A}_+$, and the irreducible representations are labeled by 
(see below for notation)
\begin{align}
  \alpha \in \Lambda_{\rm Cardy}/\Lambda_{\rm winding} \cong iReps. = \Gamma_+^\vee/\Gamma_+.
\end{align}
The character of the irreducible representation $V^0_\alpha$ of 
such open string states is given by 
\begin{align}
 {\rm ch}_{V^o_\alpha}(e^{-2\pi t}) := {\rm Tr}_{V^o_\alpha}[ e^{-2\pi t (L_0-c/24)} ] 
  = \chi_{\alpha}(e^{-2\pi t}).  
\end{align}
$L_0$ is one of the Virasoro generators for the open string sector. $\bullet$
\end{props}

Just for the purpose of keeping track of the set of all the irreducible 
representations of an algebra ${\cal A}_+$ and anything isomorphic to it, 
it is not necessary to introduce yet another set $\Lambda_{\rm Cardy}/\Lambda_{\rm winding}$,
which is isomorphic to the set $\Gamma_+^\vee/\Gamma_+$. In string theory, 
however, the set $\Lambda_{\rm Cardy}/\Lambda_{\rm winding}$ labeling the irreducible 
representations of open string states should be regarded as the character group 
of $\Gamma_+^\vee/\Gamma_+$ labeling the irreducible representations of 
close string states; they are the same finite set (set theoretically), though. 

In the language of string theory, the lattice $\Gamma_+$ [resp.\ $\Gamma_+^\vee$] 
can be regarded also as all the possible values of the right-mover 
momentum $\sqrt{(\alpha'/2)} \; k_+$ of all the purely anti-holomorphic operators 
[resp.\ all the operators]. $[\sqrt{(\alpha'/2)} \; k_+^\C]: \Gamma_+ \rightarrow \C$ 
is an embedding where the absolute-value-square $|\; [\sqrt{(\alpha'/2)} \; k_+^\C](p)
 \; |_\C^2$ in $\C$ for $p \in \Gamma_+$ reproduces the self intersection 
$(p,p)$ in $\Gamma_+$.

$\Lambda_{\rm Cardy}$ and $\Lambda_{\rm winding}$ are also rank-2 lattices, and 
$\Lambda_{\rm Cardy} \cong \Gamma_+^\vee$, $\Lambda_{\rm winding} \cong \Gamma_+$ 
as lattices. In string theory, the geometry of elliptic curve (target space) 
is given by 
\begin{align}
 [\C/\Lambda_{\rm winding}]_\C \cong [E_z]_\C ;
\end{align}
$\Lambda_{\rm Cardy} = \Lambda_{\rm winding}^\vee$ is a lattice of torsion points of 
the elliptic curve $\C/\Lambda_{\rm winding}$. When we refer to open string states 
on a strip-shape worldsheet with two boundaries subject 
to $\varphi_0$-Cardy states, we think of states of an open string with one 
boundary fixed at the origin of $\Lambda_{\rm Cardy}/\Lambda_{\rm winding}$ and 
the other boundary at another point in $\Lambda_{\rm Cardy}/\Lambda_{\rm winding}$. 
The number of torsion points of $\Lambda_{\rm Cardy}/\Lambda_{\rm winding}$ is 
\begin{align}
  [ \Lambda_{\rm Cardy} : \Lambda_{\rm winding} ] = [\Gamma_+^\vee : \Gamma_+] = 
     f_\rho^2 D_z = \# [iReps.].
\end{align}

The combination of the isometry $\Lambda_{\rm Cardy} \cong \Gamma_+^\vee$ 
and an embedding $[\sqrt{(\alpha'/2)} \; k_+^\C]: \Gamma_+^\vee \rightarrow \C$ 
gives an embedding of $\Lambda_{\rm Cardy} \rightarrow \C$. It is natural for string 
theorists to use $[\sqrt{2/\alpha'} \Delta X^\C/(2\pi)]: 
\Lambda_{\rm Cardy} \rightarrow \C$ as the notation of this embedding, but we use 
a simpler notation $\Omega': \Lambda_{\rm Cardy} \rightarrow \C$ instead in the 
following. The following information is written down here, as we use it later on:
\begin{align}
 \Omega'(\Lambda_{\rm winding}) & \; = \sqrt{2a_z f_\rho} (\Z \oplus z \Z) 
    = [\sqrt{(\alpha'/2)} \; k_R^\C](\Gamma_+)  \subset \C \\
  \Omega'(\Lambda_{\rm Cardy}) & \; = \frac{\sqrt{2a_z f_\rho}}{f_\rho}
    \left( \frac{2 a_z z + b_z}{D_z} \Z \oplus
        \frac{b_z + 2c_z z}{D_z} \Z \right) \subset \C.
\end{align}
%

\section{$\zeta_k(s)$: the $L$-function for $H^0_{et}(E)$}
\label{sec:forH0}

\subsection{Field of Definition of Arithmetic Models}
\label{ssec:field-def}

\begin{defn}
Let $X$ be an algebraic variety defined over\footnote{In a more colloquial language, 
$X$ is a subvariety of a projective space over a number field $k$, given by 
a set of defining equations whose coefficients are in $k$.} a number field $k$. 
When we wish to emphasize the choice of the field of definition, we also write $X/k$.
When $k'/k$ is a field extension, and $[X']_{k'}$ is an $k'$-isomorphism class of 
algebraic varieties defined over $k'$, $X/k$ is said to be {\it a model of} $[X']_{k'}$, 
if the base change\footnote{In a more colloquial language, 
this is an operation to include all the points where the affine coordinates take 
value in $k'$ (although defining equations have coefficients in $k$).}
of $X$, $X \times_{{\rm Spec}(k)} {\rm Spec}(k')$, belongs to $[X']_{k'}$. 
\end{defn}

Although we wish to deal with a $\C$-isomorphism class of a variety  
$[X]_\C$ as a target space in string theory, arithmetic geometry deals with its 
models $X/k$ defined over some number field $k$. 
From the perspective of arithmetic geometry, one might be interested in 
$X/k$ in the category of algebraic varieties defined over a number field $k$, but its 
{\it string-theory realizations} depend only on the $\C$-isomorphism class $[X]_{\C}$
of which $X/k$ is a model. 

Let $X$ be an $n$-dimensional algebraic variety defined over a number field $k$. 
One can then define\footnote{The definition of the $L$-function is written down 
in [Shimura \cite{Shimura-AV}, 19.1--6] for the case $X$ is an abelian variety, 
and $i=1$; for more general cases, we need to refer to etale cohomology groups
to define $L(H^i_{et}(X),s)$. When $X$ is an elliptic curve, however, the $L$-function 
for $i=1$ has a more intuitive characterization in terms of the number of 
$\F_{\mathfrak{p}}$(residue field)-points of the reduction of $X$ at a non-zero prime ideal 
$\mathfrak{p}$ of $k$. } $2n$ $L$-{\it functions for} $X/k$, $L(H^i_{et}(X),s)$ for 
$i=0,1,\cdots, 2n$, each one of which is associated with the $i$-the cohomology 
(denoted symbolically) $H^i(X)$ of $X$. In the case $X/k$ is a curve, $n=1$, 
the $L$-function for $H^0(X)$ is the Dedekind zeta function $\zeta_k(s)$ of the 
field of definition, $k$. The $L$-function for $H^2(X)$ is $\zeta_k(s-1)$. 
When we refer to Hasse--Weil $L$-function, or simply the $L$-function, that 
is meant to be the $L$-function associated with $H^1(X)$. The notation 
$L(X/k,s)$ is also used for $L(H^1_{et}(X),s)$. 

The Dedekind zeta function $\zeta_k(s)$ of a number field $k$ is so well-understood 
an object for a given number field $k$ that few people will find it interesting 
to think of an elliptic curve $E$ defined over $k$, think of string-theory 
realizations of $[E/k]_\C$, and then write down $\zeta_k(s)$ in terms of 
characters of states in the string-theory realizations of $[E/k]_\C$. We still do so
in this section as a warming-up exercise for section \ref{sec:forH1}, where 
we study how the $L$-functions associated with $H^1(E)$ are related to 
string-theory realizations of $E$. 

The following facts are well-known. 
\begin{thm} [e.g., Shimura \cite{Shimura-AA}, Thm. 5.7]
Let $[E]_\C \in Ell({\cal O}_{f_z})$ be a $\C$-isomorphism class of elliptic curves 
with complex multiplication by an order ${\cal O}_{f_z}$ of an imaginary quadratic 
field $K$. Then it has a model defined over the ring class field 
\begin{align}
  L_{f_z} = K(j([E]))
\end{align}
of $K$, where $j([E])$ is the $j$-invariant of $[E]$; 
this number field $L_{f_z}$ is determined uniquely from $K$ and 
$f_z \in \N_{>0}$, and is independent of choice of $[E]$ in $Ell({\cal O}_{f_z})$. 
An example of models over $L_{f_z}$ is a Weierstrass model given 
by (e.g., Silverman \cite{Silverman-Ell}, Prop. III.1.4(c))
\begin{align}
  y^2 + xy = x^3 - \frac{36}{j([E])-1728}x - \frac{1}{j([E])-1728},
   \label{eq:Weierstrass}
\end{align}
although this is not the only possible model of $E$ defined over $L_{f_z}$. 

Obviously, there exists a model of $E$ defined over $L$ for any extension field $L$ 
of $L_{f_z}$; we can use the base change of $E/L_{f_z}$ to obtain a model over $L$.
 $\bullet$ 
\end{thm}

\begin{thm}
Let $[E]_\C \in Ell({\cal O}_{f_z})$ be a $\C$-isomorphism class 
of elliptic curves with complex multiplication by ${\cal O}_{f_z}$. Then it also 
has a model defined over a number field 
\begin{align}
 F_{f_z}^{[E]} := \Q(j([E]));
\end{align}
the Weierstrass equation given by (\ref{eq:Weierstrass}) is an example of 
such models. The field of definition $F_{f_z}^{[E]}$ is a degree-$h({\cal O}_{f_z})$ 
extension over $\Q$, and $L_{f_z}$ is a degree-2 extension of $F_{f_z}^{[E]}$. 
The field $F_{f_z}^{[E]}$ is not necessarily stable under conjugation in 
${\rm Aut}(\overline{\Q}/\Q)$; the subfield $F_{f_z}^{[E]} \subset \overline{\Q}$ 
depends on which one of $[E] \in Ell({\cal O}_{f_z})$ is used (although 
$L_{f_z} \subset \overline{\Q}$ does not). For any $[E] \in Ell({\cal O}_{f_z})$, 
however, there exists a unique element $\rho_{[E]} \in {\rm Gal}(L_{f_z}/\Q)$ 
that generates ${\rm Gal}(L_{f_z}/F_{f_z}^{[E]}) \cong \Z/2\Z$. $\bullet$
\end{thm}

\begin{notn}
The extension field $L_{f_z}$ for $f_z = 1$ of an imaginary quadratic field $K$ 
may also be denoted by $H_K$ or simply by $H$, and called the {\it Hilbert class 
field of} $K$. The subfield $F_{f_z}^{[E]} \subset H_K$ may also be denoted by 
$F_K^{[E]}$, $F_K$, $F^{[E]}$ or $F$, when $K$ and/or $[E]$ is fixed.  $\bullet$
\end{notn}

In this section, we will focus on the cases where $k/K$ is an abelian extension 
containing $L_{f_z}/K$ (including $k=L_{f_z}$ cases), and write down theh relation 
between characters of string-theory realizations of $[E/k]_\C$ and $\zeta_k(s)$. 

\subsection{Preparation in String Theory}
\label{ssec:prepare-f0}

The embedding $\Omega': \Lambda_{\rm winding} \otimes \Q \rightarrow \C$ was determined 
so that the absolute-value-square norm $| \quad |_\C^2$ reproduces the 
intersection form of $\Gamma_+ \cong \Lambda_{\rm winding}$. It is more convenient 
to modify the embedding by rescaling it, so that the image of the lattices 
$\Gamma_+ \cong \Lambda_{\rm winding}$ and $\Gamma_+^\vee \cong \Lambda_{\rm Cardy}$ 
fit within $K = \Q(\sqrt{-d_0}) = \Q(z) \subset \overline{\Q} \subset \C$. 
Let us take $\Omega := [(C^{-1} \sqrt{2a_z f_\rho}z_2 i)] \cdot \Omega'$, 
where $C \in \Q$ is some constant. Then 
\begin{align}
 \Omega (\Lambda_{\rm Cardy}) & \; = C^{-1} ( \Z \oplus z \Z) \subset \overline{\Q} \subset \C, \\
 \mathfrak{b}_z := 
 \Omega (\Lambda_{\rm winding}) & \; = C^{-1} f_\rho 
     ((2a_z z + b_z)\Z \oplus (b_z z + 2c_z) \Z )
     \subset \overline{\Q} \subset \C. 
\end{align}
As long as $C \in \Q$, $\mathfrak{b}_z$ and $\Omega(\Lambda_{\rm Cardy})$ are now 
regarded as rank-2 lattices within $K \subset \C$. 
We fix $C \in \Q$ so that ${\cal O}_K$ is contained in $\Omega(\Lambda_{\rm Cardy})$; 
we need to take 
\begin{align}
 C \in  \frac{f_z}{{\rm GCD}\left(a_z,\frac{f_z - b_z}{2}\right)} \Z \quad 
 {\rm if~}D_K{\rm ~is~odd}, \qquad 
 C \in \frac{f_z}{{\rm GCD}(a_z, b_z/2)} \Z \quad 
  {\rm if~}D_K{\rm ~is~even}; 
\end{align}
in the rest of this article, we use the minimum positive value of $C$. 
One can also see that $C^{-1} {\cal O}_{f_z} \subset \Omega(\Lambda_{\rm Cardy})$.

Here are a couple of bilinear forms introduced on an imaginary quadratic field $K$.
\begin{align}
 (-,-)_{K/\Q}: K \times K \ni (x,y) \mapsto & \; (x,y)_{K/\Q} := {\rm Tr}_{K/\Q}(xy) \in \Q, \\
 \vev{-,-}_{K/\Q}: K \times K \ni (x,y) \mapsto & \; \vev{x,y}_{K/\Q} := 
   (x\bar{y}+\bar{x}y) \in \Q \subset \C, \\
 & \; \vev{x,x}_{K/\Q} = 2|x|_\C^2, \\
 (-,-)_{\Gamma_+ \otimes \Q}: K \times K \ni (x,y) \mapsto & \; (x,y)_{\Gamma_+ \otimes \Q} := 
    C^2 \frac{a_z}{f_\rho D_z} \vev{x,y}_{K/\Q}, \\
 & \; (x,x)_{\Gamma_+ \otimes \Q} = C^2 \frac{a_z}{f_\rho D_z} \vev{x,x}_{K/\Q}.
\end{align}
The ratio between the last two norms, $\vev{-,-}_{K/\Q}$ and $(-,-)_{\Gamma_+ \otimes \Q}$, 
is due to the rescaling factor $C^{-1} \sqrt{2a_z f_\rho} z_2 i$ between $\Omega$ and 
$\Omega'$; $(2a_z f_\rho)z_2^2 = f_\rho D_z/(2a_z)$. Using these relations, 
$\Omega(\Lambda_{\rm Cardy})$ can be characterized in terms of $\mathfrak{b}_z$
and vice versa as 
\begin{align}
 \Omega(\Lambda_{\rm Cardy}) =  \frac{f_\rho D_z}{C^2 a_z} \mathfrak{b}_z^*, \qquad 
 \mathfrak{b}_z = \frac{f_\rho D_z}{C^2 a_z}  \Omega(\Lambda_{\rm Cardy})^*. 
\end{align}
Here, $\mathfrak{a}^*$ for a full-rank $\Z$-lattice $\mathfrak{a} \subset 
K \subset \C$ is the dual lattice with respect to the intersection form 
$\vev{-,-}_{K/\Q}$.  The notation $\mathfrak{a}^\vee$ is reserved for 
the dual lattice with respect to the intersection form $(-,-)_{K/\Q}$.

Modular invariance of CFT on worldsheet is an important principle in string theory. 
In a diagonal rational CFT, characters $\{\chi_\alpha \}_{\alpha \in iReps.}$ can be regarded
as a vector-valued modular form; by introducing a vector space 
$\C[iReps.] = {\rm Span}_\C \{ e_\alpha \; | \alpha \in iReps. \}$ with a basis that 
consists of formal elements $e_\alpha$'s in one-to-one correspondence with the set of 
irreducible representations $iReps.$, 
\begin{align}
 \sum_\alpha e_\alpha \; \chi_\alpha(q) \in M_{wt=0}({\rm SL}(2;\Z), \C[iReps.]).
\end{align}
$\chi_\alpha$'s can be regarded as characters of the left-mover chiral algebra 
(where the argument is $q=e^{2\pi i \tau}$), those of the right movers, 
and those of open string states subject to $\varphi_0$-Cardy states 
(where the argument is $q=e^{-2\pi t}$). The modular invariance of the closed string partition function is 
ensured by the cancellation between the left-mover $\C[iReps.]$ representation matrix 
of the worldsheet-${\rm SL}(2;\Z)$ group and that of the right-mover. 

The power series expansion of $\chi_\alpha(q)$'s with respect to $q$, however, 
begins with a fractional power term, $q^{-\frac{c}{24}}$; the central charge is $c=2$ 
in $T^2$-target CFT's. An object mathematically nicer is 
$\sum_\alpha e_\alpha [\chi_\alpha \eta^2]$, when the Fourier expansion 
begins with the $1=q^0$ term. Now 
\begin{align}
 \sum_{\alpha \in iReps.} e_\alpha [\chi_\alpha \eta^2] = \sum_{\alpha \in iReps.} f_0(\tau; \alpha) 
   \in M_{wt=1}({\rm SL}(2;\Z), \C[iReps.]).
\end{align}
It may look artificial to multiply $\eta^2$ for the purpose of getting 
the integer-power leading term in the Fourier expansion, but this nice object 
shows up naturally in superstring version of the diagonal rational $T^2$-target CFT's. 
In the closed string Ramond sector, 
\begin{align}
  f_0(\tau_{ws};\alpha) = (-i) 
   {\rm Tr}_{V_\alpha^-;{\rm R}} \left[ F_L e^{\pi i F_L} q^{L_0-c/24} \right], 
  \qquad q = e^{2\pi i \tau_{ws}}, 
\end{align}
and in the open string Ramond sector, 
\begin{align}
 f_0(i t_{ws}; \alpha) = (-i)
    {\rm Tr}_{V_\alpha^o; {\rm R}}[Fe^{\pi i F} q^{L_0-c/24}], 
    \qquad q = e^{-2\pi t_{ws}};
\end{align}
here, $\tau_{ws} \in {\cal H}$ is the complex structure parameter of worldsheet 
torus, and $t_{ws} \in \R_{>0}$ the parameter of the shape of a cylinder; 
$F_L$ and $F$ are the fermion number operators on the closed string left-moving 
sector and open string sector, respectively.\footnote{Here, we think of 
a worldsheet made of a strip with width $\pi$ and length $2\pi t_{ws}$; it is made 
periodic to be a cylinder in the ``$2\pi t_{ws}$'' direction.}
It is evident in the closed string language that 
$\sum_\alpha e_\alpha f_0(\tau_{ws};\alpha)$ is a vector-valued modular form of 
weight 1; on a worldsheet torus $\Sigma$ with 
the complex structure $\tau_{ws}$ with the odd spin structure, think of the partition function with the action modified from $S$ to $\widetilde{S}$ by a parameter $u \in \C$:
\begin{align}
 Z(\tau_{ws}, u) & := \int_{{\rm Map}(\Sigma, E(z,\rho))} e^{i\widetilde{S}(u,\bar{u})}, \\
 i \widetilde{S}(u,\bar{u}) & \; = i S + 
  2\pi i \int_{\Sigma} \frac{d^2\sigma}{{\rm Im}(\tau_{ws})}
    \left(u J_L(\sigma) - \bar{u}J_R(\sigma)\right). 
  \label{eq:modify-WSaction-1}
\end{align}
This ``partition function'' transforms as \cite{KYY, Kachru}
\begin{align}
 Z\left(\frac{a\tau_{ws} + b}{c\tau_{ws}+d}, \frac{u}{c\tau_{ws}+d} \right) = 
 \mathbb{E}\left[ \frac{(1/2)cu^2}{c\tau_{ws}+d} - \frac{(1/2)c\bar{u}^2}{c\bar{\tau}_{ws}+d}
           \right] Z(\tau_{ws},u),  
\label{eq:pathintegral-modular-1}
\end{align}
where $\mathbb{E}[X] := e^{2\pi i X}$, under the worldsheet ${\rm SL}(2;\Z)_{ws}$ 
transformation acting on $\tau_{ws}$ through 
$\tau'_{ws} = \frac{a\tau_{ws}+b}{c\tau_{ws}+d}$. Because 
\begin{align}
 \frac{-1}{(2\pi)^2} \left[\frac{\partial}{\partial u}\frac{\partial }{\partial \bar{u}} 
   Z(\tau_{ws},u,\bar{u})\right]_{u=\bar{u}=0} =
    \sum_\alpha f_0(\tau_{ws};\alpha) f_0(-\bar{\tau}_{ws}; \alpha), 
\end{align}
we can use (\ref{eq:pathintegral-modular-1}) to see, 
\begin{align}
  \sum_{\alpha} f_0(\tau'_{ws} ;\alpha)f_0(-\bar{\tau}'_{ws}; \alpha) = 
   (c\tau_{ws}+d)(c\bar{\tau}_{ws}+d) 
  \left[ \sum_\alpha f_0(\tau_{ws}; \alpha) f_0(-\bar{\tau}_{ws}; \alpha) \right],
\end{align}
a property satisfied by a vector-valued modular form of weight 1. 
 
As we are working on the $T^2$-target diagonal rational CFT's, we know a lot more 
about the functions $f_0(\alpha)$, $\alpha \in iReps.$.
The vector-valued modular form $\sum_\alpha e_\alpha f_0(\tau_{ws};\alpha)$ is given by 
congruent theta functions associated with the rank-2 even positive definite lattice 
$\Gamma_+ \cong \Lambda_{\rm winding}$
\begin{align}
 f_0(\tau_{ws};\alpha) = \sum_{\xi \in \alpha} e^{2\pi i \tau_{ws} \frac{C^2 a_z}{f_\rho D_z} 
       \frac{\vev{\xi,\xi}_{K/\Q}}{2} }, 
 \qquad  \alpha \in \Omega(\Lambda_{\rm Cardy}) / \mathfrak{b}_z \cong iReps.
        \cong \Gamma_+^\vee/\Gamma_+,
\end{align}
using the rescaled embedding $\Omega: \Lambda_{\rm winding} \otimes \Q 
\rightarrow \mathfrak{b}_z \otimes \Q \subset K \subset \C$. They are none other than 
a vector-valued modular form in the Weil representation $\rho_{\Gamma_+}$ 
of ${\rm SL}(2;\Z)_{ws}$ associated with the lattice $\Gamma_+ \cong \Lambda_{\rm winding}$.

\subsection{Theta Functions and Zeta Functions}
\label{ssec:theta-zeta}

The Mellin transform of the Riemann theta function yields the Riemann zeta 
function (e.g., [Koblitz \cite{Koblitz}, II \S4], [Lang \cite{Lang-ANT}, XIII \S1]).  
This classical result has been generalized for any number field $k$, and is  
found in textbooks (e.g., Lang \cite{Lang-ANT}, Chap. XIII). 
We wish to find a relation, however, between $\zeta_k(s)$ and such functions as 
$f_0(\tau_{ws};\alpha)$ obtained from string-theory realizations 
of an arithmetic variety $E/k$. Here, we provide a brief review of 
known mathematical facts that serve for this purpose. 

Let $k$ be an abelian extension of $K$ that contains the ring class field $L_{f_z}/K$ 
(where $f_z \in \N_{>0}$). The Dedekind zeta function $\zeta_k(s)$ can be 
reconstructed from functions that can be defined 
in terms of the imaginary quadratic field $K$, by (e.g., 
[Lang \cite{Lang-ANT}, Thm. XII.1], [Serre \cite{Serre-CourseA}, Prop. VI.13])
\begin{align}
  \zeta_{L_{f_z}}(s) & \; = \prod_{\chi \in {\rm Char}[{\rm Gal}(L_{f_z}/K)]} \left(
     \sum_{\mathfrak{K} \in {\rm Cl}_K((f_z))} \vev{\chi, \mathfrak{K}} \zeta_K(s,\mathfrak{K})
         \right), \label{eq:zeta-ringL-zetaK4class} \\
  \zeta_{k}(s) & \; = \prod_{\chi \in {\rm Char}[{\rm Gal}(k/K)]} \left(
     \sum_{\mathfrak{K} \in{\rm Cl}_K(\mathfrak{m}_f)} \vev{\chi, \mathfrak{K}} \zeta_K(s,\mathfrak{K})
         \right), \label{eq:zeta-genk-zetaK4class}
\end{align}
apart from the Euler factors for prime ideals $\mathfrak{P} \in {\rm Spec}({\cal O}_{L_{f_z}})$ 
[resp. $\in {\rm Spec}({\cal O}_k)$] in the fiber of the support of the modulus 
$(f_z)_{{\cal O}_K}$ [resp. $\mathfrak{m}_f$]; 
here, $\mathfrak{m}_f$ is an integral ${\cal O}_K$ ideal\footnote{The subscript ${}_f$
here is a reminder that we are referring to the finite part of a modulus. Almost 
all the subscripts ${}_f$ in this article, such as those in $\mathfrak{c}_f$, 
$\mathfrak{C}_f$, $\phi_f$, and $\chi_f$, are reminders for the finite part. 
When the same letter ``f'' is used in the form of $f_z \in \N$, and $L_{f_z}$, however, 
it originates from a German word meaning conductor.} such that the ray class field 
corresponding to $\mathfrak{m}_f$ contains $k$. Here, $\chi$ runs over the characters 
of the abelian group ${\rm Gal}(L_{f_z}/K)$ or ${\rm Gal}(k/K)$, which are quotient 
of the abelian group ${\rm Cl}_K((f_z)_{{\cal O}_K})$ or ${\rm Cl}_K(\mathfrak{m}_f)$.  
For a congruent ideal class $\mathfrak{K} \in {\rm Cl}_K(\mathfrak{m}_f)$ for 
an integral ${\cal O}_K$ ideal $\mathfrak{m}_f$ (which includes the case of 
$\mathfrak{m}_f=(f_z)_{{\cal O}_K}$), 
\begin{align}
  \zeta_K(s,\mathfrak{K}) := \sum_{I \in \mathfrak{K}} \frac{1}{(\N I)^s}.
\end{align}
By writing $I \in \mathfrak{K}$, we mean that the sum is over integral 
${\cal O}_K$ ideals $I$ that belong to the class $\mathfrak{K}$. 
$\N I$ is the norm of the ideal $I$. 

There is a particularly simple formula for $\zeta_K(s, \mathfrak{K})$'s, 
when $f_z=1$ and $k = H_K$. Let us take one integral 
${\cal O}_K$ ideal $\mathfrak{a}(\mathfrak{K})$ from each ideal class 
$-\mathfrak{K} \in {\rm Cl}_K$, and fix a set of these representatives 
$\{ \mathfrak{a}(\mathfrak{K}) \; | \; \mathfrak{K} \in {\rm Cl}_K \}$ once 
and for all. Then\footnote{
Note also that there are only finitely many units in ${\cal O}_K$ 
when $K$ is an imaginary quadratic field. That is, $\#[{\cal O}_K^\times] < \infty$.} 
 (e.g., [Lang \cite{Lang-ANT}, Chap. XIII])
\begin{align}
  \zeta_K(s,\mathfrak{K}) = \frac{(\N\mathfrak{a}(\mathfrak{K}))^s}{\#[{\cal O}_K^\times]}
     \sum_{\xi \in \mathfrak{a}(\mathfrak{K})} \frac{1}{|\xi|_\C^{2s}}, 
  \label{eq:zeta-K4classes-easy-1}
\end{align}
where one embedding $\sigma: K \hookrightarrow \C$ is fixed implicitly 
(as we have done already in section \ref{ssec:prepare-f0});
an idea here is that integral ideals in the class $\mathfrak{K}$ can be 
listed up in the form of $(\xi)_{{\cal O}_K}\mathfrak{a}(\mathfrak{K})^{-1}$. 

The functions $\zeta_K(s,\mathfrak{K})$ for 
$\mathfrak{K} \in {\rm Cl}_K(\mathfrak{m}_f)$ are obtained as Mellin transforms 
of congruent theta functions associated with the number field $K$. To see this, 
let us start off by introducing congruent theta functions associated with 
a number field $L$. The ring of algebraic integers ${\cal O}_L$ in $L$ 
can be regarded as a lattice by using 
\begin{align}
\vev{-,-}_{L/\Q}: {\cal O}_L \times {\cal O}_L \ni (x,y)
 \mapsto \sum_{a=1}^{r_1}\rho_a(xy) + \sum_{b=1}^{r_2} \left( 
    \sigma_b(x)\bar{\sigma}_b(y) + \bar{\sigma}_b(x)\sigma_b(y) \right)  
\end{align}
as the intersection form; here $\rho_a: L \hookrightarrow \R$ with 
$a=1,\cdots, r_1$ are real embeddings of $L$, and $\sigma_b$ and $\bar{\sigma}_b$ 
with $b=1,\cdots, r_2$ are imaginary embeddings $L\hookrightarrow \C$ forming 
$r_2$ pairs under complex conjugation in $\C$; $r_1+2r_2=[L:\Q]$. 
This lattice $({\cal O}_L, \vev{-,-}_{L/\Q})$ is integral, and in particular, 
it is even when $L$ is a totally imaginary field. For a sublattice $\Lambda$ of 
this lattice $({\cal O}_L, \vev{-,-}_{L/\Q})$ and $x \in \R^{[L:\Q]}/\Lambda$, 
we set 
\begin{align}
  \vartheta_L(\Lambda, x) = \vartheta_L(\tau; \Lambda, x) := 
    \sum_{w \in x} e^{2\pi i \tau \frac{\vev{w,w}_{L/\Q}}{2}} = \sum_{w \in x} q^{\frac{\vev{w,w}_{L/\Q}}{2}}.
   \label{eq:def-theta-cong-P=0}
\end{align}
The definition of the functions $\vartheta_L(\Lambda, x)$ is possible, in fact, 
without referring to a number field $L$ or the lattice $({\cal O}_L, \vev{-,-}_{L/\Q})$;
we just need an integral lattice $\Lambda$ as a part of data for the definition. 
So, one may drop the subscript ${}_L$.
We will be interested in those theta functions for an imaginary quadratic field $L=K$ 
along with $x$ placed at torsion points of $L \otimes_\Q \R/\Lambda$, often within 
$\Lambda^*/\Lambda \subset L \otimes_\Q \R/\Lambda$.

Now, let us get started for the case of $k=H_K$; $f_z = 1$ and 
$\mathfrak{m}_f = {\cal O}_K$. The Mellin transformation of the congruent theta 
functions associated with $K$ (Lang \cite{Lang-ANT}, Chap. XIII),
\begin{align}
 \frac{1}{\#[{\cal O}_K^\times]}
 \int_0^{+\infty} \left[ \vartheta_K(it;\mathfrak{a}(\mathfrak{K}),0) - 1 \right] 
     t^s \frac{dt}{t} =   \frac{\Gamma(s)}{(2\pi)^s}
   \frac{1}{\#[{\cal O}_K^\times]}\sum_{\xi \in \mathfrak{a}(\mathfrak{K})} \frac{1}{|\xi|_\C^{2s}}
      = \frac{\Gamma(s)}{(2\pi)^s} \frac{\zeta_K(s,\mathfrak{K})}{\N \mathfrak{a}(\mathfrak{K})^s},
  \label{eq:Mellin-zetaDdk-theta}
\end{align}
provide all the components $\zeta_K(s,\mathfrak{K})$ (for $\mathfrak{K} \in {\rm Cl}_K$)
necessary in reconstructing $\zeta_{H_K}(s)$ 
through (\ref{eq:zeta-K4classes-easy-1}, \ref{eq:zeta-ringL-zetaK4class}).

Here, we record a property of the functions $\vartheta_K(\Lambda,0)$ as we use it 
in section \ref{sssec:L0-Hilbert}:
\begin{props} [Iwaniec \cite{Iwaniec}, Thm. 10.8; Miyake \cite{Miyake}, Cor. 4.9.5(2)]
\label{props:automrph-theta0} 
Let $L$ be a totally imaginary field. 
Let $\Lambda$ be a rank-$2r_0$ $\Z$-lattice in $({\cal O}_L, \vev{-,-}_{L/\Q})$, and 
$N_\Lambda \in \N_{>0}$ be the level\footnote{
The {\it level} $N_\Lambda$ of an integral lattice $\Lambda$ is 
$N_\Lambda := {\rm Min}\{ N \in \N_{>0} \; | \; \Lambda^*[N] {\rm ~is~integral~} \}$.
Here, $\Lambda^*$ stands for the dual lattice of $\Lambda$.}
of $\Lambda$. Then $\vartheta_L(\Lambda,0)$ is a modular form of weight $r_0$ for 
$\Gamma_0(2N_\Lambda)$ with a multiplier system (homomorphism) 
$\underline{\chi}: \Gamma_0(2N_\Lambda) \rightarrow \{ \pm 1\}$. 
The subgroup $\Gamma_0(2N_\Lambda) \subset {\rm SL}(2;\Z)$ acts on the argument $\tau$ 
as used in the definition (\ref{eq:def-theta-cong-P=0}) of theta functions through 
${\rm SL}(2;\Z)$. $\bullet$ 
\end{props}
This means that $\zeta_K(s,\mathfrak{K})$ can be regarded as the Mellin transform 
of a modular form of weight 1 for $\Gamma_0(M)$ with some appropriately chosen 
integer $M$ (apart from subtraction of $1$).

Let us now move on to more general cases, where $k=L_{f_z}$ with $f_z>1$, 
or $[k:L_{f_z}] > 1$. In order to reconstruct $\zeta_k(s)$ from the 
information of how many ideals ${\cal O}_K$ has, 
the idea behind (\ref{eq:zeta-K4classes-easy-1}) needs to be generalized. 
First, we still choose a set of representatives $\{\mathfrak{a}(\mathfrak{K}) \; 
| \; \mathfrak{K} \in {\rm Cl}_K(\mathfrak{m}_f)\}$ satisfying 
$\mathfrak{a}(\mathfrak{K}) \in -\mathfrak{K}$; then $\mathfrak{a}(\mathfrak{K})$ is 
prime to $\mathfrak{m}_f$ by definition. Let us assume that all the representatives 
are integral ${\cal O}_K$ ideals (not just fractional ideals) although this assumption 
is only for the sake of simplifying the presentation here. Second, for integral 
${\cal O}_K$ ideals $\mathfrak{a}$ and $\mathfrak{m}_f$ that are relatively prime, let 
us also introduce the following notations:
\begin{align}
  \mathfrak{a}_{1}(\mathfrak{m}_f) & \; := \left\{ \xi \in \mathfrak{a} \; | \; 
        {}^\exists \epsilon \in {\cal O}_K^\times \; {\rm s.t.~} \epsilon \xi \equiv 1 
    {\rm ~mod~}\mathfrak{m}_f \right\},   \label{eq:def-subset-1-mod}\\
  \mathfrak{a}_{G_0}(\mathfrak{m}_f) & \; := \left\{ \xi \in \mathfrak{a} \; | \; 
       {}^\exists \epsilon \in {\cal O}_K^\times, \; {}^\exists a \in G_0 \; {\rm s.t.~} 
       \epsilon \xi \equiv a {\rm ~mod~} \mathfrak{m}_f \right\}, 
  \label{eq:def-subset-subgroup-mod}
\end{align}
where $G_0$ is a subgroup of $[{\cal O}_K/\mathfrak{m}_f]^\times$ in the 
multiplication law. We will also use a notation $\mathfrak{a}_\Z(\mathfrak{m}_f)$ 
for $\mathfrak{m}_f = (f_z)_{{\cal O}_K}$ for some $f_z \in \N_{>0}$ in this article, 
which is meant to be $\mathfrak{a}_{G_0}((f_z)_{{\cal O}_K})$ introduced above, 
with $G_0 \subset [{\cal O}_K/(f_z)_{{\cal O}_K}]^\times$ chosen to be the image of 
all integers in $\Z$ prime to $f_z$.
Now, with these preparations, a general version of the idea 
behind (\ref{eq:zeta-K4classes-easy-1}) is stated as follows: the list of 
integral ideals $I$ that are prime to $\mathfrak{m}_f$ and belong to 
$\mathfrak{K} \in {\rm Cl}_K(\mathfrak{m}_f)$ are in one-to-one correspondence\footnote{
$[\mathfrak{a}(\mathfrak{K})]_1(\mathfrak{m}_f)$ is the $\mathfrak{a}=
\mathfrak{a}(\mathfrak{K})$ case of (\ref{eq:def-subset-1-mod}).} 
with $[\xi] \in [\mathfrak{a}(\mathfrak{K})]_1(\mathfrak{m}_f)/{\cal O}_K^\times$; the 
correspondence is given by $I \cdot \mathfrak{a}(\mathfrak{K}) = (\xi)_{{\cal O}_K}$ 
in the ideal group of $K$ (e.g., [Lang \cite{Lang-ANT}, Chap. XIII]).
So, the expression (\ref{eq:zeta-K4classes-easy-1}) for $\zeta_K(s,\mathfrak{K})$ with 
$\mathfrak{K} \in {\rm Cl}_K$ is replaced by 
\begin{align}
  \zeta_K(s,\mathfrak{K}) = \frac{(\N\mathfrak{a}(\mathfrak{K}))^s}{\#[{\cal O}_K^\times]}
     \sum_{\xi \in [\mathfrak{a}(\mathfrak{K})]_1(\mathfrak{m}_f)} \frac{1}{|\xi|_\C^{2s}}, 
  \qquad \mathfrak{K} \in {\rm Cl}_K(\mathfrak{m}_f).
  \label{eq:zeta-K4classes-easy-2}
\end{align}

These zeta functions $\zeta_K(s,\mathfrak{K})$ for 
$\mathfrak{K} \in {\rm Cl}_K(\mathfrak{m}_f)$ are obtained as the Mellin 
transform as in (\ref{eq:Mellin-zetaDdk-theta}), but now in the form of 
\begin{align}
  \frac{1}{\#[{\cal O}_K^\times]} \int_0^\infty \frac{dt}{t} \; t^s 
    \sum_{y \in [[\mathfrak{a}(\mathfrak{K})]_1(\mathfrak{m}_f)]_{\mathfrak{a}(\mathfrak{K})\mathfrak{m}_f}}
         \vartheta_K(it; \mathfrak{a}(\mathfrak{K})\mathfrak{m}_f, y)
  = \frac{\Gamma(s)}{(2\pi)^s}
    \frac{\zeta_K(s,\mathfrak{K})}{(\N\mathfrak{a}(\mathfrak{K}))^s}.
  \label{eq:Mellin-zetaDdk-theta-gen}
\end{align}
Here, $[[\mathfrak{a}(\mathfrak{K})]_1(\mathfrak{m}_f)]_{\mathfrak{a}(\mathfrak{K})\mathfrak{m}_f}$
stands for the image\footnote{Note that 
$[\mathfrak{a}(\mathfrak{K})]_1(\mathfrak{m}_f) \subset \mathfrak{a}(\mathfrak{K})$
has a periodicity $\mathfrak{a}(\mathfrak{K})\mathfrak{m}_f$ and hence $\mathfrak{a}(\mathfrak{K})\mathfrak{m}_f$ or any abelian subgroup $\mathfrak{m}$ of it acts on 
$[\mathfrak{a}(\mathfrak{K})]_1(\mathfrak{m}_f)$ by the addition law. 
} 
 of $[\mathfrak{a}(\mathfrak{K})]_1(\mathfrak{m}_f)$ in the quotient map 
$\mathfrak{a}(\mathfrak{K}) \rightarrow \mathfrak{a}(\mathfrak{K})/
\mathfrak{a}(\mathfrak{K})\mathfrak{m}_f$.

Here, we record a property of the functions $\vartheta_K(\Lambda, y)$ for a torsion 
point $y \in \R^{[L:\Q]}/\Lambda$, as we use it in section \ref{sssec:L0-nonmaximal}. 
\begin{props}[Iwaniec \cite{Iwaniec}, Cor. 10.7; Miyake \cite{Miyake}, Cor. 4.9.4] 
\label{props:automrph-theta0-gen}
Let $\Lambda$ be a rank-$2r_0$ $\Z$-lattice in $({\cal O}_L, \vev{-,-}_{L/\Q})$ of 
a totally imaginary field $L/\Q$ with $[L:\Q] = 2r_0$, and $N_\Lambda$ its level. 
Let $\underline{N} \in \N_{>0}$ be divisible by $N_\Lambda$. Then, 
%
%
for any $x \in \underline{N}^{-1} \Lambda$, $\vartheta_L(\Lambda, x)$ is a modular form 
of weight $r_0$ for $\Gamma(4\underline{N})$ without a non-trivial multiplier system. 
The group $\Gamma(4\underline{N}) \subset {\rm SL}(2;\Z)$ acts on the argument $\tau$ 
as we used in (\ref{eq:def-theta-cong-P=0}) through ${\rm SL}(2;\Z)$. 

When $x \in \Lambda^* /\Lambda$, in particular, we can choose 
$\underline{N}=N_\Lambda$. 

The same statement holds true, when $\Gamma(4\underline{N})$ is replaced by 
$\Gamma_1(4\underline{N})$ and a non-trivial multiplier system (homomorphism)
$\underline{\chi}: \Gamma_1(4\underline{N}) \rightarrow S^1$ is allowed.  
When $x \in {\cal O}_L/\Lambda$, we can take $\underline{N}=N_\Lambda$ (as above), 
and moreover, $(x,x)/2 \in \Z$, from which one can conclude that the multiplier 
system $\underline{\chi}$ is trivial (cf \cite{Iwaniec, Miyake}). $\bullet$
\end{props}

This is not to say, however, that $\zeta_K(s,\mathfrak{K})$ for $\mathfrak{K} \in 
{\rm Cl}_K(\mathfrak{m}_f)$ with a non-trivial modulus $\mathfrak{m}_f$ cannot be 
the Mellin transform of a modular form for $\Gamma_0(M_*)$ for any choice of 
$M_* \in \N_{>0}$.
We do not pursue this question in detail in this article, but at least 
in Example \ref{exmpl:zeta-nonmaximal}, we see that the set 
$[\mathfrak{a}(\mathfrak{K})]_1(\mathfrak{m}_f)$ can be decomposed into a sum of 
$\Z$-sublattices of ${\cal O}_K$ (not necessarily ${\cal O}_K$-ideals), so the 
Proposition \ref{props:automrph-theta0} can be used, instead of 
Prop. \ref{props:automrph-theta0-gen}. 

\subsection{Combining Them Together}
\label{ssec:L0-combine}

Having done preparations in sections \ref{ssec:prepare-f0} and \ref{ssec:theta-zeta}, 
let us now combine them together to find out how the Dedekind zeta function 
$\zeta_k(s)$ of the field of definition of an elliptic curve $E/k$ is 
reconstructed from functions obtained in the string-theory realizations of $E/k$. 
In section \ref{sssec:L0-Hilbert}, we begin with the cases of elliptic curves 
with complex multiplication by the maximal order ${\cal O}_K$ of a quadratic 
imaginary field $K$, and the field of definition is $k=L_{f_z=1} = H_K$. 
More general cases are treated in section \ref{sssec:L0-nonmaximal}.

\subsubsection{Cases with $k=H_K$}
\label{sssec:L0-Hilbert}

Note first that the lattices $\Lambda_{\rm winding}$ and $\Lambda_{\rm Cardy}$ have been 
embedded into $K$, which is also identified (by a fixed imaginary embedding 
$\sigma: K \hookrightarrow \C$) with a subset of $\C$. The character functions 
$f_0(\tau_{ws}; \alpha)$ in a string-theory realization of $E/k$ is closely related 
to the congruent theta functions in (\ref{eq:def-theta-cong-P=0}). To be concrete, 
choose one elliptic curve $E/H_K$ which has complex multiplication by ${\cal O}_K$. 
We focus on its string realizations whose $f_\rho$ satisfies 
\begin{align}
  {\rm LCM}(\mathfrak{a}(\mathfrak{K})_{\mathfrak{K} \in {\rm Cl}_K} ) \; 
     | \; \mathfrak{b}_z .
  \label{eq:cond-frho-L0-Hilb}
\end{align}
Then [Moore \cite{Moore-98s}, \S4]
\begin{align}
   \sum_{\pi(\alpha) = 0} f_0\left(\tau_{ws}; \alpha \right)
  = \vartheta_K(\tau; \mathfrak{a}(\mathfrak{K}),0), \qquad 
   \tau = \frac{C^2a_z}{f_\rho D_z} \tau_{ws}, 
  \label{eq:sum-f0-L0-Hilb}
\end{align}
where $\pi: \Lambda_{\rm Cardy}/\Lambda_{\rm winding} \rightarrow 
\Omega(\Lambda_{\rm Cardy})/\mathfrak{a}(\mathfrak{K})$ is the quotient map, which is 
well-defined when the condition (\ref{eq:cond-frho-L0-Hilb}) is satisfied. 
Consequently, we see that $\zeta_{H_K}(s)$ is obtained by summing up the character 
functions $f_0(\tau_{ws};\alpha)$ of irreducible representations $\alpha$ of the chiarl 
algebra ${\cal A}_- \cong {\cal A}_L$ in those string realizations in a way specified 
by (\ref{eq:sum-f0-L0-Hilb}) first;
the sum of $f_0(\alpha)$'s is now a single component modular form of 
$\Gamma_0(2N_{\mathfrak{a}(\mathfrak{K})})$ (see Prop. \ref{props:automrph-theta0}), 
although $f_0(\alpha)$'s as a whole forms a vector-valued modular form of 
the entire ${\rm SL}(2;\Z)_{ws}$. Remaining steps are to take a Mellin transformation 
of the sum with respect to the imaginary part of the complex structure 
parameter $\tau_{ws}$ of a worldsheet torus, and then to sum them and multiply 
them as in (\ref{eq:zeta-ringL-zetaK4class}). 

\begin{table}[tbp]
\begin{center}
\begin{tabular}{c||cccccc}
$-D_K$ & $z = w_K$ & $[a,b,c]$ & $C$ & $\Omega(\Lambda_{\rm Cardy})$ & 
   $\mathfrak{b}_z/f_\rho$ & $iReps.$(if $f_\rho=1$) \\
\hline 
3 & $\frac{-1+\sqrt{3}i}{2}$ & [1,1,1] & 1 & $\Z + w_K \Z$ & 
   $(2w_K +1)\Z + (2+w_K)\Z$ & $\Z/3\Z$ \\ 
4 & $i$ & [1,0,1] & 1 & $\Z + w_K \Z$ & 
   $2\Z + 2w_K \Z$ & $\Z/(2\Z) \times \Z/(2\Z)$ \\
8 & $\sqrt{2}i$ & [1,0,2] & 1 & $\Z + w_K \Z$ & 
   $4\Z + 2w_K \Z$ & $\Z/(4\Z) \times \Z/(2\Z)$ \\
\end{tabular}
\caption{\label{tab:123}Embedding of the lattices $\Lambda_{\rm Cardy}$ and $\Lambda_{\rm winding}$ 
into $K \subset \C$, for the unique (trivial) element of $Ell({\cal O}_K)$ with 
$K = \Q(\sqrt{-d_0})$, $d_0=1,2,3$. For elliptic curves $[\C/\mathfrak{b}_z]_\C$ corresponding 
to the trivial element of $Ell({\cal O}_K)$, we can use the value of $z$ for one ($w_K$) of the 
two generators of ${\cal O}_K$ over $\Z$, ${\cal O}_K = \Z + w_K\Z$. In all the three examples
here, the relation $\Omega(\Lambda_{\rm Cardy}) = {\cal O}_K$ holds, although this is not always the 
case (see Table \ref{tab:d=5}).}
\end{center}
\end{table}
\begin{exmpl}
For any one of the imaginary quadratic fields $K=\Q(\sqrt{-1})$, $\Q(\sqrt{-2})$ 
and $\Q(\sqrt{-3})$, there is just one $\C$-isomorphism class of elliptic curves 
with complex multiplication by ${\cal O}_K$ ($f_z = 1$). Put differently, 
${\rm Cl}_K = \{ [0]\}$. We choose an ideal $\mathfrak{a}([0])$ to be ${\cal O}_K$.
In all these examples, one can confirm explicitly\footnote{In fact, this is a special 
case of a more general statement. For the $\C$-isomorphism class of elliptic curves 
that correspond to the trivial element in $Ell({\cal O}_K) \cong {\rm Cl}_K$ (that is 
not necessarily trivial), one can verify that $\mathfrak{b}_z/f_\rho = \mathfrak{d}_{K/\Q}$, 
the different of the extension $K/\Q$. The {\it different} $\mathfrak{d}_{E/F}$ associated 
with extension of number fields $E/F$ is always an integral ideal of ${\cal O}_E$.
So, $(\mathfrak{a}([0])= {\cal O}_K)|\mathfrak{d}_{K/\Q} = \mathfrak{b}_z/f_\rho$.}
 that ${\cal O}_K \subset \mathfrak{b}_z/f_\rho$. 
So, any string realization with $f_\rho \in \N_{>0}$ satisfies the 
condition (\ref{eq:cond-frho-L0-Hilb}). More detailed data are found in 
Table \ref{tab:123}. 

When we choose the minimal $f_\rho=1$ string realization,  
\begin{align}
 \vartheta_K(\tau; {\cal O}_K,0) & \; = \sum_{\alpha \in iReps.} f_0(\tau_{ws}; \alpha), 
   \qquad \tau = \tau_{ws}/|D_K|
\end{align}
for all these three examples, because $C=1$, $a_z = 1$ and 
$\Omega(\Lambda_{\rm Cardy}) = {\cal O}_K$. $\bullet$ 
\end{exmpl}

\begin{figure}[tbp]
 \begin{center}
 \begin{tabular}{ccc}
 \includegraphics[width=.35\linewidth]{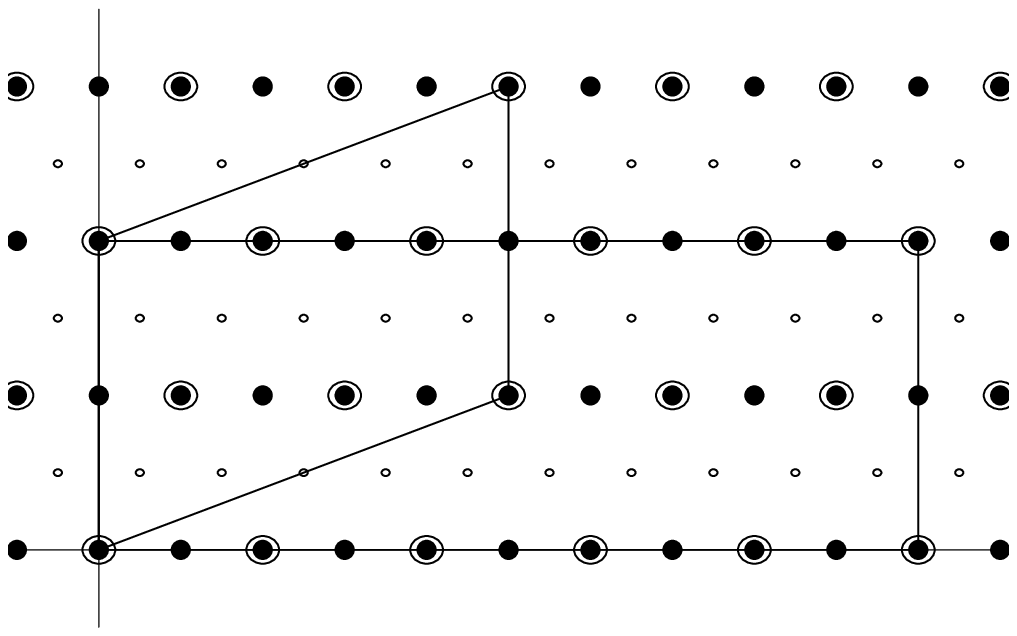} & $\qquad$ & 
 \includegraphics[width=.25\linewidth]{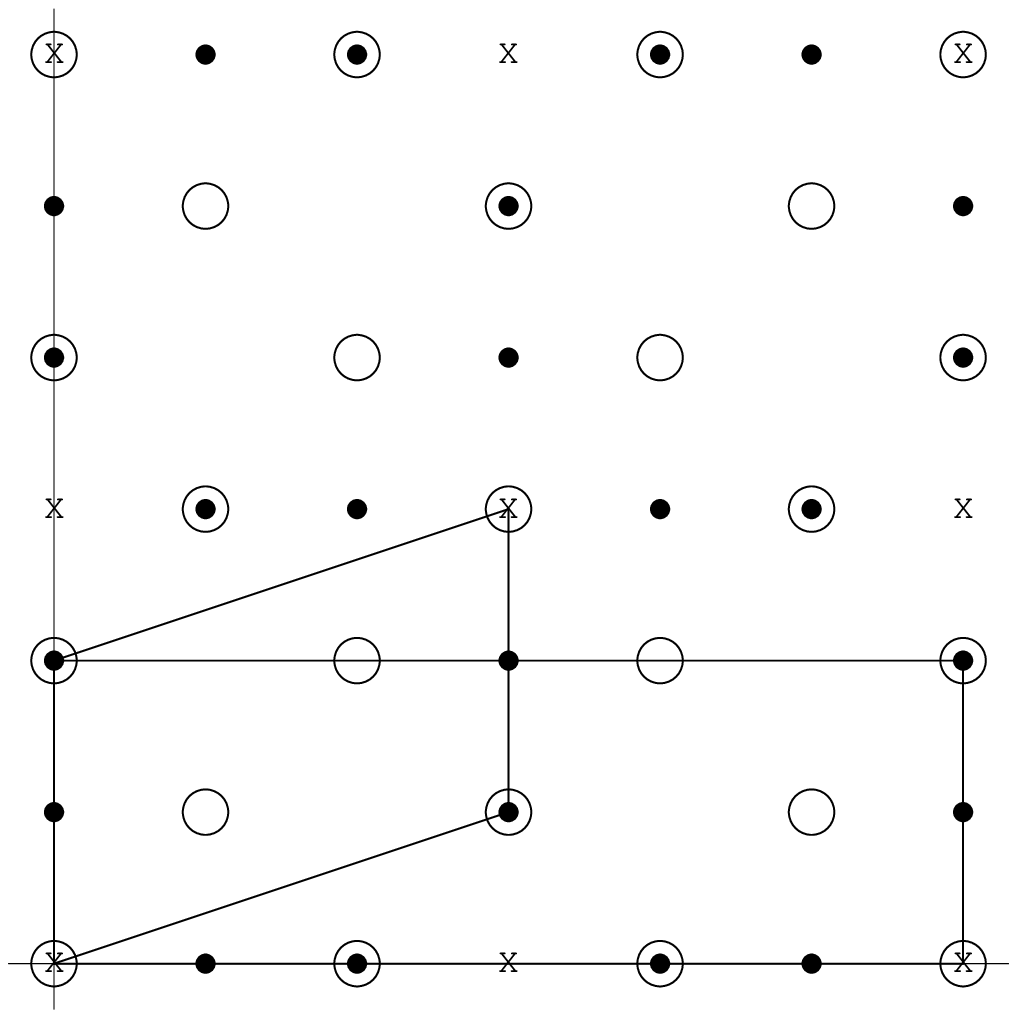} \\
 (a) & & (b) 
 \end{tabular}
\caption{\label{fig:L0-X}
(a) is an illustration for Example \ref{exmpl:d=5-zeta} and 
(b) for \ref{exmpl:zeta-nonmaximal}. In (a), ${\cal O}_K=\mathfrak{a}([0])$ 
and $\mathfrak{a}([1])$ in $K=\Q(\sqrt{-5})\subset \C$ are shown by large 
black dots and large open circles, respectively. $\Omega(\Lambda_{\rm Cardy})=
{\cal O}_K$ for $z = \sqrt{5}i$, whereas $\Omega(\Lambda_{\rm Cardy})$
for $z = (-1+\sqrt{5}i)/2$ consists both of ${\cal O}_K$ and the tiny open circles 
in (a). The unit cell $\Omega(\Lambda_{\rm winding})/f_\rho$ is also 
indicated for the two inequivalent $[E_z]_\C$'s in $Ell({\cal O}_K)$ in (a). 
In (b), where $K=\Q(\sqrt{-1})$ and $f_z = 3$, ${\cal O}_K$ ideals $(f_z)_{{\cal O}_K}$  
and $\mathfrak{a}([1])$ for $[1] \in {\rm Cl}_K({\cal O}_{f_z=3})$ are indicated by 
x and $\circ$, respectively. The set $[{\cal O}_K]_\Z((f_z))$ consists of dots. 
Also shown are the unit cell $\Omega(\Lambda_{\rm winding})/f_\rho$ for the two inequivalent 
$[E_z]_\C$'s in $Ell({\cal O}_{f_z=3})$.   By enlarging those unit cells by 
$f_\rho =3$, all the points of $\Omega(\Lambda_{\rm winding})$ are to be contained 
in $f_z \mathfrak{a}([1])=(3+3i)_{{\cal O}_K}$, the $\circ$-and-x points in (b).}
 \end{center}
\end{figure}

\begin{exmpl}
\label{exmpl:d=5-zeta}
For the case $K=\Q(\sqrt{-5})$, there are two inequivalent $\C$-isomorphism 
classes of elliptic curves with complex multiplication by ${\cal O}_K$. 
The lattices $\Omega(\Lambda_{\rm Cardy})$ and $\mathfrak{b}_z$ in our embedding 
$\Omega$ are shown in Table~\ref{tab:d=5}. 

For the two ideal classes ${\rm Cl}_K \cong \Z/(2\Z) =: \{ [0], [1] \}$ in this 
case, we can choose $\mathfrak{a}([0]) = {\cal O}_K$, and 
$\mathfrak{a}([1]) = (2\Z + (1+\sqrt{5}i)\Z)$, which is not a principal ideal. 
For this choice of representatives, $\{ \mathfrak{a}([0]), \mathfrak{a}([1])\}$, 
the condition (\ref{eq:cond-frho-L0-Hilb}) is satisfied for any string realizations 
with $f_\rho \in \N_{>0}$ (see Figure~\ref{fig:L0-X}~(a)). So, let us use the string-theory 
realization with the minimum choice $f_\rho = 1$. Note that we still think of 
two string realizations, one for each $[E] \in Ell({\cal O}_K)$. \\
In order to obtain $\vartheta(\mathfrak{a}([0]),0)$ [resp. $\vartheta(\mathfrak{a}([1]),0)$], 
$f_0(\tau_{ws};\alpha)$'s need to be summed up over $\#2$ [resp. $\#3$] irreducible 
representations $\alpha \in iReps.$ in those string-theory realizations; 
see Table \ref{tab:d=5}.
Although $\vartheta(\mathfrak{a}(\mathfrak{K}),0)$'s are expressed differently 
in terms of $f_0(\tau_{ws};\alpha)$'s for the two string-theory realizations, 
the functions $f_0(\tau_{ws};\alpha)$'s for the two realizations are also different. 
In the end, $\vartheta(\mathfrak{a}(\mathfrak{K}),0)$'s should not depend on which 
string-theory realization (which one of $[E] \in Ell({\cal O}_K)$) is used to reconstruct them. 
After all, one and the same zeta function $\zeta_k(s)$ is being re-constructed for 
the common Hilbert class field $k = K(j([E])) = \Q(\sqrt{-5},\sqrt{-1})$. $\bullet$
\end{exmpl}
\begin{table}[tbp]
\begin{center}
\begin{tabular}{cc||ccc|ccc}
$z$ & $[a,b,c]$ & $C$ & $\Omega(\Lambda_{\rm Cardy})$ & $\mathfrak{b}_z$ &
$\#1$ & $\# 2$ & $\# 3$ \\
\hline
$\sqrt{5}i = w_K$ & [1,0,5] & 1 & $\Z+w_K\Z$ & $10\Z+2w_K\Z$ & 1 & 20 & 10 \\
$\frac{-1+\sqrt{5}i}{2} = \frac{-1+w_K}{2}$ & [2,2,3] & 1 & $\Z+\frac{1+w_K}{2}\Z$ & 
  $2w_K\Z+(5+w_K)\Z$ & 2 & 10 & 5 
\end{tabular}
\caption{\label{tab:d=5}
Two $\C$-isomorphism classes of elliptic curves with complex multiplication by 
the maximal order ${\cal O}_K$, $K=\Q(\sqrt{-5})$. Embedded lattices $\Lambda_{\rm Cardy}$ 
and $\Lambda_{\rm winding}$ in $K \subset \C$ are shown here, relatively to 
${\cal O}_K = \Z + w_K\Z$, where $w_K = \sqrt{5}i$; 
$f_\rho = 1$ is used for $\mathfrak{b}_z$.
The last three columns are     
$\#1 = [\Omega(\Lambda_{\rm Cardy}):{\cal O}_K]$, 
$\#2 = [{\cal O}_K: \mathfrak{b}_z] = [\mathfrak{a}([0]) : \mathfrak{b}_z]$ 
and $\#3 = [\mathfrak{a}([1]) : \mathfrak{b}_z]$. }
\end{center}
\end{table}

We have seen that the Dedekind zeta function $\zeta_{k=H_K}(s)$ can be written 
down by using the combinations (\ref{eq:sum-f0-L0-Hilb}) of a string-theory realization 
of an arithmetic model $E/H_K$ (when $[E] \in Ell({\cal O}_K)$). 
Prop. \ref{props:automrph-theta0} says that those combinations of $f_0$'s are
weight-1 modular forms of $\Gamma_0(2N_{\mathfrak{a}(\mathfrak{K})})$, 
where $N_{\mathfrak{a}(\mathfrak{K})}$ is the level\footnote{For 
$\mathfrak{a}([0]) = {\cal O}_K$, $N_{\mathfrak{a}(\mathfrak{K})} = |D_{K/\Q}|$.} 
of the even lattice 
$\mathfrak{a}(\mathfrak{K})$, which is a sublattice of the even lattice 
$({\cal O}_K, \vev{-,-}_{K/\Q})$. When we see those combinations as functions of 
$\tau_{ws}$, as natural in string-theory perspective, however, only the common 
subset of $\Gamma_0(2N_{\mathfrak{a}(\mathfrak{K})})$ and the worldsheet 
${\rm SL}(2;\Z)_{ws}$ transformation,  
\begin{align}
  {\rm SL}(2;\Z)_{\rm ws} \cap
   \left( \begin{array}{cc} (f_\rho D_z/C^2a_z) & \\ & 1 \end{array} \right)
  \Gamma_0(2N_{\mathfrak{a}(\mathfrak{K})})   
   \left( \begin{array}{cc} (C^2a_z/f_\rho D_z) & \\ & 1 \end{array} \right),
  \label{eq:common-modular-grp}
\end{align}
is evident. 
So, in particular, there exists a congruence subgroup $\Gamma(M) \subset 
{\rm SL}(2;\Z)_{\rm ws}$ that is contained in all those groups, under which all of 
the combinations (\ref{eq:sum-f0-L0-Hilb}) for ${}^\forall \mathfrak{K} \in {\rm Cl}_K$ 
of a string-theory realization are 
modular forms of weight 1, and $\zeta_{H_K}(s)=L(H^0_{\it et}(E),s)$ can be reproduced 
from them through the Mellin transformation, summation and multiplication.

The common subgroup is not of the form of $\Gamma_0(M)$ or $\Gamma_1(M)$ in 
${\rm SL}(2;\Z)_{ws}$, unless $C^2a_z/(f_\rho D_z) \in \Z$. It is 
isomorphic to the group $\Gamma_0(N_* {\rm LCM}(D_*, 2N_{\mathfrak{a}(\mathfrak{K})}))$ 
acting on $\tau/N_*$, where $N_*$ and $D_*$ are relatively prime integers 
satisfying $C^2a_z/(f_\rho D_z)=N_*/D_*$. It is interesting to note, from 
string-theory perspective, that there are modular transformations 
of $\Gamma_0(2N_{\mathfrak{a}(\mathfrak{K})})$ that are not captured within 
the worldsheet ${\rm SL}(2;\Z)_{ws}$ transformation. 

\subsubsection{More General Cases: $k/K$ Is a Ramified Abelian Extension 
Containing $L_{f_z}/K$}
\label{sssec:L0-nonmaximal}

Now think of an elliptic curve $[E]_\C \in Ell({\cal O}_{f_z})$ and its model 
$E/k$ over a number field $k$ that is an abelian extension of $K$ containing 
$L_{f_z}/K$. The field of definition $k$ is not necessarily the Hilbert class 
field $H_K$. We claim that $\zeta_k(s) = L(H^0_{\it et}(E),s)$ is still obtained 
through a process as in section \ref{sssec:L0-Hilbert} from the character 
functions $f_0$'s of string-theory realizations of $[E]_\C$. 
 
To see this, let us focus on string-theory realizations whose parameter 
$f_\rho$ on the complexified K\"{a}hler parameter satisfies 
\begin{align}
  {\rm LCM}( \; (\mathfrak{m}_f\mathfrak{a}(\mathfrak{K})) \; 
   {}_{\mathfrak{K} \in {\rm Cl}_K(\mathfrak{m}_f)} \; )    \supset \mathfrak{b}_z .
  \label{eq:cond-frho-L0-gen}
\end{align}
Then the projection $\pi: \Omega(\Lambda_{\rm Cardy})/\mathfrak{b}_z \rightarrow
 \Omega(\Lambda_{\rm Cardy})/\mathfrak{a}(\mathfrak{K})\mathfrak{m}_f$ is well-defined, 
and one finds that the relation 
\begin{align}
  \sum_{\pi(\alpha) = y} f_0(\tau_{ws};\alpha) =
     \vartheta_K(\tau; \mathfrak{a}(\mathfrak{K})\mathfrak{m}_f, y),
   \qquad \tau_{ws} \frac{C^2a_z}{f_\rho D_z} = \tau, 
\label{eq:sum-f0-L0-gen}
\end{align}
holds. Now (\ref{eq:Mellin-zetaDdk-theta-gen}) 
and (\ref{eq:zeta-ringL-zetaK4class}, \ref{eq:zeta-genk-zetaK4class}) 
can be used to obtain $\zeta_k(s)$ from the data $f_0$'s of the spectrum of a
string-theory realization. 

The combinations of character functions of a string realization appearing in 
the left-hand-side are weight-1 modular forms of $\Gamma(4N_{\mathfrak{a}(\mathfrak{K})})$ 
acting on $\tau$ (see Prop. \ref{props:automrph-theta0-gen}); 
note that we use $y \in [\mathfrak{a}(\mathfrak{K})]_1(\mathfrak{m}_f) \subset 
\mathfrak{a}(\mathfrak{K}) \subset {\cal O}_K$ in (\ref{eq:Mellin-zetaDdk-theta-gen}, 
\ref{eq:sum-f0-L0-gen}), which means that $y \in {\cal O}_K 
\subset f_z^{-1} {\cal O}_K^\vee \subset (\mathfrak{m}_f)^* 
\subset (\mathfrak{m}_f\mathfrak{a}(\mathfrak{K}))^*$.
The common subset of this modular transformation and ${\rm SL}(2;\Z)_{ws}$ acting on 
$\tau_{ws}$ can be worked out as in (\ref{eq:common-modular-grp}). 
It may also be possible to find an appropriate combination of $f_0(\alpha)$'s so that 
it is a modular form for a group of the form $\Gamma_0(M)$ for some $M \in \N_{>0}$ 
instead of a group $\Gamma(4N_{\mathfrak{a}(\mathfrak{K})})$; we do this only in 
the example below, however.  

\begin{exmpl}
\label{exmpl:zeta-nonmaximal}
There are two $\C$-isomorphism classes of elliptic curves with complex 
multiplication by the order ${\cal O}_{f_z = 3}$ in $K=\Q(\sqrt{-1})$; 
${\rm Cl}_K({\cal O}_{f_z = 3}) \cong \Z/(2\Z) =: \{ [0], [1]\}$.  One is for 
$z = 3i$ and the other for $z = (-1+3i)/2$. 
We choose $\mathfrak{a}([0]) = {\cal O}_K$ and $\mathfrak{a}([1]) = (1+i)_{{\cal O}_K}$ 
for the two ideal classes\footnote{Note that 
${\rm Cl}_K({\cal O}_{f_z=3}) \cong {\rm Cl}_K((f_z=3)_{{\cal O}_K})$ 
in this example.} in ${\rm Cl}_K({\cal O}_{f_z=3})$. Using the data shown in 
Table \ref{tab:non-max4Dedekind-DK4}, one can see that string-theory realizations 
with $3|f_\rho$ satisfy the condition (\ref{eq:cond-frho-L0-gen}) for 
$(f_z)_{{\cal O}_K} \mathfrak{a}([1]) = (3+3i)_{{\cal O}_K}$. 

One can work out, in such a string realization, the number of irreducible 
representations $\alpha$ of the chiral algebra contributing to $\zeta_K(s,\mathfrak{K})$, 
as follows. First, $[[\mathfrak{a}([0])]_1((f_z)_{{\cal O}_K})]_{f_z\mathfrak{a}([0])}$ 
consists of 4 elements represented by $y = 1, 2, i, 2i$ mod $f_z \mathfrak{a}([0]) = 
(3)_{{\cal O}_K}$; for each one of $y$'s, there are $\#[f_z \mathfrak{a}([0])/
\mathfrak{b}_z]$ irreducible representations $\alpha \in iReps.$ of a string-theory 
realization (see Table \ref{tab:non-max4Dedekind-DK4}) whose $f_0(\alpha)$ 
contributes to $\vartheta_K(f_z\mathfrak{a}([0]),y)$ in (\ref{eq:sum-f0-L0-gen}). 
Overall, $\zeta_K(s,[0])$ is obtained as the Mellin transform of a sum of 
$f_0(\alpha)$'s of 
$\#[{\cal O}_{f_z}^\times] \times (4/\#[{\cal O}_{f_z}^\times]) \times 
\#[f_z \mathfrak{a}([0])/\mathfrak{b}_z]$ irreducible representations of the 
chiral algebra. 
Similarly, $\zeta_K(s,[1])$ is obtained as the Mellin transform of a sum of $f_0$'s 
of $\#[{\cal O}_{f_z}^\times] \times (4/\#[{\cal O}_{f_z}^\times]) \times 
\#[f_z \mathfrak{a}([1])/\mathfrak{b}_z]$ irreducible representations. 

The subset\footnote{See the explanation below (\ref{eq:def-subset-subgroup-mod}) 
for the notation $\mathfrak{a}_\Z((f)_{{\cal O}_K})$; here, we use it for 
$\mathfrak{a} = \mathfrak{a}(\mathfrak{K})$.} 
$[\mathfrak{a}(\mathfrak{K})]_\Z((f_z)_{{\cal O}_K}) \subset {\cal O}_K$, 
over which the sum (\ref{eq:zeta-K4classes-easy-2}) runs, is not a $\Z$-lattice, 
which is why the combination (\ref{eq:sum-f0-L0-gen}) is a modular form for 
$\Gamma(4N_{f_z\mathfrak{a}(\mathfrak{K})})$, but is not guaranteed to be one for a 
group of the form $\Gamma_0(M)$ for some choice of $M \in \N_{>0}$. In this example, 
however, the set $[\mathfrak{a}([0])]_\Z((f_z)_{{\cal O}_K}) = 
[{\cal O}_K]_\Z((3)_{{\cal O}_K}) = [\mathfrak{a}([0])]_1((f_z)_{{\cal O}_K})$ can be 
decomposed into three $\Z$-lattices, ${\cal O}_{f_z}$, $i{\cal O}_{f_z}$ and 
$(f_z)_{{\cal O}_K}$, with multiplicities $+1$, $+1$ and $-2$, respectively; 
one can see this decomposition most easily by a glance at Figure~\ref{fig:L0-X}~(b). 
Similar decomposition of the set $[\mathfrak{a}([1])]_\Z((f_z)_{{\cal O}_K})$
into $\Z$-lattices will be more complicated; we did not work that out. 
$\bullet$
\end{exmpl}
\begin{table}[tbp]
\begin{center}
\begin{tabular}{cc||ccc|cc}
 $z$ & $[a,b,c]$ & $C$ & $\Omega(\Lambda_{\rm Cardy})$ & $\mathfrak{b}_z/f_\rho$ & 
 $\#[f_z\mathfrak{a}(0])/\mathfrak{b}_z]$ & $\#[f_z \mathfrak{a}([1])/\mathfrak{b}_z]$
 \\
\hline
 $3i$ & [1,0,9] & 3 & $\Z + 3w_K\Z$ & $6\Z + 2w_K\Z$ & $12(f_\rho/3)^2$ &
 $6(f_\rho/3)^2$ \\
 $\frac{(-1+3i)}{2}$ & [2,2,5] & 3 & $\Z + \frac{-1+3w_K}{2}\Z$ &
 $(3+w_K)\Z+2w_K \Z$ & $6(f_\rho/3)^2$ & $3(f_\rho/3)^2$
\end{tabular}
\caption{\label{tab:non-max4Dedekind-DK4}Two $\C$-isomorphism classes of elliptic 
curves in $Ell({\cal O}_{f_z = 3})$ where $K=\Q(\sqrt{-1})$. Their lattices 
$\Lambda_{\rm Cardy}$ and $\Lambda_{\rm winding}$ embedded in $K \subset \C$ are shown 
relatively to ${\cal O}_K = \Z + w_K \Z$, where $w_K = i$ for $K=\Q(\sqrt{-1})$ here. }
\end{center}
\end{table}

For cases with $k=L_{f_z}$ with $f_z > 1$ (and possibly for more general cases), 
there is an alternative approach, which is to use a set of proper 
${\cal O}_{f_z}$-ideals instead of ${\cal O}_K$-ideals as a set of representatives 
of the ideal class group ${\rm Cl}_K({\cal O}_{f_z})$.
Instead of $\{ \mathfrak{a}(\mathfrak{K})_{\mathfrak{K} \in {\rm Cl}_K( {\cal O}_{f_z} ) } \}$ 
introduced\footnote{cf footnote \ref{fn:ring-ray}} in section \ref{ssec:theta-zeta}, 
we can use (e.g., [Shimura \cite{Shimura-AA}, Thm 4.11], [Lang \cite{Lang-EF}, Chap. 8 \S1, Thm. 4], [Moreland \cite{Moreland}, \S5]):
\begin{align}
\left\{\mathfrak{b}(\mathfrak{K}) \right\}_{\mathfrak{K} \in {\rm Cl}_K({\cal O}_{f_z})}
  = \left\{ C^{-1} \left( \mathfrak{a}(\mathfrak{K}) \cap {\cal O}_{f_z} \right)
    \right\}_{\mathfrak{K}\in {\rm Cl}_K({\cal O}_{f_z})}.
\end{align}
Here, the factor $C^{-1}$ is not necessary for the purpose of finding a representative 
proper ${\cal O}_{f_z}$-ideal, but we include the factor $C^{-1}$, because 
$\mathfrak{b}(\mathfrak{K}) \subset C^{-1}{\cal O}_{f_z}$ still fits within 
$\Omega(\Lambda_{\rm Cardy})$. Now, instead of (\ref{eq:zeta-K4classes-easy-2}), 
we can use  
\begin{align}
 \zeta_K(s,\mathfrak{K}) 
  & \; = \frac{1}{\#[{\cal O}_{f_z}^\times]} \sum_{C \eta \in 
   [C\mathfrak{b}(\mathfrak{K})]_{\Z}((f_z)_{{\cal O}_{f_z}})} 
     \frac{\left( C^{-2}[{\cal O}_{f_z}: C\mathfrak{b}(\mathfrak{K})] \right)^s }
      {|\eta|_\C^{2s}},
  \label{eq:zeta-K4classes-easy-3}
\end{align}
where we recycle the notation $\mathfrak{a}_\Z(\mathfrak{m}_f)$ introduced 
in the discussion below (\ref{eq:def-subset-subgroup-mod}) for 
${\cal O}_{f_z}$-proper ideals $\mathfrak{a}$ and $\mathfrak{m}_f$. 

The parameter $f_\rho$ of string-theory realizations should then be chosen so that 
\begin{align}
  f_z \mathfrak{b}(\mathfrak{K})  \;  |  \;\mathfrak{b}_z
  {\rm ~for~} {}^\forall \mathfrak{K} \in {\rm Cl}_K({\cal O}_{f_z}) 
  \label{eq:cond-frho-L0-nonmax}
\end{align}
as proper ${\cal O}_{f_z}$ ideals. When this condition is satisfied, the projection 
$\pi: \Omega(\Lambda_{\rm Cardy})/\mathfrak{b}_z \rightarrow \Omega(\Lambda_{\rm Cardy})/
f_z\mathfrak{b}(\mathfrak{K})$ is well-defined, 
so we can sum over the irreducible representations $\alpha \in iReps. \cong 
\Omega(\Lambda_{\rm Cardy})/\mathfrak{b}_z$ in the fiber of this projection 
to obtain $\vartheta_K(f_z\mathfrak{b}(\mathfrak{K}),y)$. Summing them 
over $y \in [C^{-1}([C\mathfrak{b}(\mathfrak{K})]_\Z((f_z)_{{\cal O}_{f_z}}))
]_{f_z \mathfrak{b}(\mathfrak{K})}$ and carrying out the Mellin transformation, 
$\zeta_K(s,\mathfrak{K})$ can be reproduced through (\ref{eq:zeta-K4classes-easy-3}).

In the example we worked on above, the common periodicity of the lattices 
$f_z\mathfrak{b}(\mathfrak{K})$'s for $\mathfrak{K} \in {\rm Cl}_K({\cal O}_{f_z})$ 
is $(1+3i,1-3i)_{{\cal O}_{f_z}}$, so the condition (\ref{eq:cond-frho-L0-nonmax}) 
reads $1|f_\rho$ (i.e., any $f_\rho \in \N_{>0}$) for both of $[E_{z = 3i}]_\C$ and 
$[E_{z = (1+3i)/2}]_\C$ in $Ell({\cal O}_{f_z = 3})$ of $K=\Q(\sqrt{-1})$.  
That is more economical (than the requirement $3|f_\rho$ in the approach above), 
in that string realizations with a fewer number of irreducible representations 
under the chiral algebra can be used in reproducing $\zeta_k(s) = L(H^0_{\it et}(E),s)$.

\section{$L$-functions for $H^1_{et}(E)$}
\label{sec:forH1}

\subsection{Preparation from String Theory}
\label{ssec:prepare-f1}

It is known that the $L$-functions for elliptic curves with complex multiplication 
are given by the Mellin transform of some modular forms of weight 2, as we will 
review in section \ref{ssec:review-HW-L1}. Because the character functions 
$f_0(\tau_{ws};\alpha)$ of string-theory realizations are of weight 1, they are not 
useful (at least immediately) in reconstructing the $L$-functions. Here, we prepare 
weight-2 observables in string-theory realizations, which are to be used in 
section \ref{ssec:L1-combine}.

Now think of the following set of functions of $t_{ws} \in \R_{>0}$ or of 
$\tau_{ws} \in {\cal H}$, 
\begin{align}
 f_1(it_{ws};\alpha) & \; := {\rm Tr}_{V^o_\alpha} [ \Omega' q^{(L_0 - c/24)} ] 
  \times [\eta(q)]^2, \qquad q=e^{-2\pi t_{ws}}, \\
 f_1(\tau_{ws}; \alpha) & \; = {\rm Tr}_{V^-_\alpha} [ \Omega' q^{(L_0-c/24)} ] 
  \times [\eta(q)]^2, \qquad q = e^{2\pi i \tau_{ws}}, 
\end{align}
for $\alpha \in iReprs. \cong \Omega(\Lambda_{\rm Cardy})/\mathfrak{b}_z \cong 
\Gamma_+^\vee/\Gamma_+$, defined in the language of bosonic string theory. 
Newly inserted is 
\begin{align}
  \Omega' = \left\{ \begin{array}{ll} 
    \sqrt{\frac{\alpha'}{2}} k_-^\C & {\rm for~}V^-_\alpha, \\
    \sqrt{\frac{2}{\alpha'}} \frac{\Delta X^\C}{2\pi} & {\rm for~}V^o_\alpha,
        \end{array} \right.
\end{align} 
which measures a combination of the four U(1) charges of states in the 
bosonic string realizations. 

In the language of superstring theory, 
\begin{align}
f_1(\tau_{ws};\alpha) & \; = (-i) {\rm Tr}_{V_\alpha^-; {\rm R}}
    \left[\Omega' F_Le^{\pi i F_L} q^{L_0 - c/24}\right], \qquad q = e^{2\pi i \tau_{ws}}, 
   \label{eq:def-f1-closed-super} \\
f_1(it_{ws};\alpha) & \; = (-i) {\rm Tr}_{V^o_\alpha;{\rm R}}
    \left[\Omega' Fe^{\pi i F} q^{(L_0-c/24)}\right], \qquad q = e^{-2\pi t_{ws}}. 
   \label{eq:def-f1-open-super}
\end{align}
In the path-integration formulation, we define the ``partition function'' 
$Z(\tau_{ws}; u,\vec{\mu})$ obtained by modifying $\widetilde{S}$ from the one 
in (\ref{eq:modify-WSaction-1}) to 
\begin{align}
 i\widetilde{S} = iS + 2\pi i \int_\Sigma \frac{d^2\sigma}{{\rm Im}(\tau_{ws})}
  \left( u J_L + \vec{\mu} \cdot i\sqrt{\frac{2}{\alpha'}}\partial \vec{X} \right)
  + {\rm right~mover}; 
\end{align}
$\vec{\mu}$ is a doublet of complex parameters associated with the doublet 
of currents $\partial \vec{X}$, and $\vec{\bar{\mu}}$ its complex conjugates 
coupled to the currents $\bar{\partial}\vec{X}$ in the right-moving sector.
Under the worldsheet ${\rm SL}(2;\Z)_{ws}$ transformation \cite{Kachru}, 
\begin{align}
 Z\left(\frac{a\tau_{ws}+b}{c\tau_{ws}+d}, \frac{u}{c\tau_{ws}+d}, 
    \frac{\vec{\mu}}{c\tau_{ws}+d}\right) = \mathbb{E}\left[
      \frac{(1/2)c(u^2+\vec{\mu}^2)}{c \tau_{ws}+d}
    - \frac{(1/2)c(\bar{u}^2+\vec{\bar{\mu}}^2)}{c\bar{\tau}_{ws}+d} \right]
   Z(\tau_{ws},u,\vec{\mu}).
\end{align}
Taking derivatives with respect to $u$ once and $\vec{\mu}$ once, and setting 
$u=\vec{\mu} = 0$, we find 
\begin{align}
 \sum_\alpha f_1(\tau'_{ws}; \alpha) f_1(-\bar{\tau}'_{ws}; \alpha)= 
 |(c\tau_{ws}+d)|^4 \left[ \sum_\alpha f_1(\tau_{ws}; \alpha)f_1(-\bar{\tau}_{ws}; \alpha) \right],
\end{align}
a property satisfied by a vector-valued weight-2 modular form of ${\rm SL}(2;\Z)_{ws}$.

As we think of $T^2$-target diagonal rational CFT's, we know the functions 
$f_1(\tau_{ws};\alpha)$ explicitly. They are in the form of 
\begin{align}
 f_1(\tau_{ws};\alpha) = \sum_{w \in \alpha} \Omega'(w) e^{2\pi i \tau_{ws} \frac{(w,w)_{\Gamma_+}}{2}} 
  = \sum_{w \in \alpha} \Omega'(w) e^{2\pi i \tau_{ws} \frac{C^2a_z}{f_\rho D_z} \frac{\vev{w,w}_{K/\Q}}{2}}.  
\end{align}
They are a variant of congruent theta functions, and can be organized into Hecke 
theta functions; more explanations (for string theorists) follow shortly. 

\subsection{$L$-functions of Elliptic Curves with Complex Multiplication}
\label{ssec:review-HW-L1}

Here is a brief review on the Hasse--Weil $L$-function of elliptic curves 
with complex multiplication. 
We have already stated in section \ref{ssec:field-def} that any $\C$-isomorphism class 
of elliptic curves with complex multiplication, $[E]_\C \in Ell({\cal O}_{f_z})$ for 
some imaginary quadratic field $K$ and $f_z \in \N_{>0}$, has a model $E/k$ 
over some number field $k$. The Hasse--Weil $L$-function is defined for 
such an object $E/k$, not for $[E]_\C$. Section \ref{sssec:Hecke-L-Hecke-theta} 
explains how the $L$-function can be computed for a given model $E/k$, and 
we will discuss in section \ref{sssec:model} 
how to classify models $E/k$ for a given $\C$-isomorphism class $[E]_\C$.
We will restrict our attention to models where the field of definition $k$
is either i) the ring class field $k=L_{f_z}$, ii) an abelian extension $k/K$ 
containing $L_{f_z}/K$, and iii) the field $k= F_{f_z}^{[E]}$.

\subsubsection{Hecke $L$-functions and Hecke Theta Functions}
\label{sssec:Hecke-L-Hecke-theta}

The following three theorems translate computation of the $L$-function 
of an elliptic curve $E/k$ with complex multiplication into determination of 
the Hecke $L$-function associated with $E/k$.
\begin{thm}[e.g., Shimura \cite{Shimura-AA}, Props. 7.40 and 7.41]\footnote{
also [Silverman \cite{Silverman-Adv}, Thm.II.9.1. and II.9.2].} 
Let $E/k$ be an elliptic curve with complex multiplication by ${\cal O}_{f_z}$ 
where $k/K$ is an abelian extension containing $L_{f_z}/K$. 
A procedure described in the propositions of \cite{Shimura-AA} referred to above 
associates a Hecke character of the idele class group of $k$, 
$\psi_{E/k}: \mathbb{A}_k^\times/k^\times \rightarrow \C^\times$, 
with the $k$-isomorphism class of $E/k$. $\psi_{E/k}$ is of type 
$[-1/2; {\bf 1}, {\bf 0}]$; we will explain what the type of a Hecke character 
means in Definition \ref{defn:type-HeckeChara}. $\bullet$
\end{thm}

\begin{thm} [e.g., Shimura \cite{Shimura-AA}, Thm. 7.42]\footnote{
also [Silverman \cite{Silverman-Adv}, Thm. II.10.5].}
Let $E/k$ be an elliptic curve with complex multiplication by ${\cal O}_{f_z}$  
where the field of definition $k/K$ is an abelian extension containing $L_{f_z}/K$.
Then 
\begin{align}
  L(E/k,s) = L(s,\psi_{E/k}) L(s,\overline{\psi}_{E/k});
  \label{eq:HWL4H--HeckeL4H}
\end{align}
we will have more words shortly on the {\it Hecke} $L$-{\it functions} 
on the right hand side. 
Here, the reduction of $E/k$ at a prime of ${\cal O}_k$ is either 
good, or cusp that is potentially good (because $E/k$ has complex multiplication 
(Silverman \cite{Silverman-Ell}, Chap. VII \S5)), and hence all the 
fibers (good or bad) are included in (\ref{eq:HWL4H--HeckeL4H}). $\bullet$ 
\end{thm}

\begin{thm}[e.g. Shimura \cite{Shimura-ze}, Thm. 7]
Let $E/F_{f_z}^{[E]}$ be an elliptic curve with complex multiplication. 
Then its $L$-function is given by 
\begin{align}
  L(E/F_{f_z}^{[E]},s) = L(s,\psi_{E/L_{f_z}}).
  \label{eq:HWL4F--HeckeL4H}
\end{align}
Here $\psi_{E/L_{f_z}}: \mathbb{A}_{L_{f_z}}^\times / L_{f_z}^\times \rightarrow \C^\times$ 
is the Hecke character associated with the base change of $E/F_{f_z}^{[E]}$ 
with respect to ${\rm Spec}(L_{f_z}) \rightarrow {\rm Spec}(F_{f_z}^{[E]})$. $\bullet$
\end{thm}

In any of the models of elliptic curves discussed here, we need the $L$-function 
of a Hecke character $\psi_{E/k'}: \mathbb{A}_{k'}^\times / k^{'\times} 
\rightarrow \C^\times$ for an abelian extension $k'/K$ containing $L_{f_z}/K$. 
Those Hecke $L$-functions of a Hecke character of $\mathbb{A}^\times_{k'}$ can be 
written in terms of Hecke $L$-functions of Hecke characters of $\mathbb{A}_K^\times$, 
provided $E/k'$ satisfies the condition (*) stated in the following.

\begin{lemma} [Shimura \cite{Shimura-AA}, Thm. 7.44] 
\label{lemma:HeckeC-LandK}
Let $k'/K$ be an abelian extension containing $L_{f_z}/K$ and 
$\psi_{E/k'}: \mathbb{A}_{k'}^\times /k^{'\times} \rightarrow \C^\times$ 
a Hecke character associated with an elliptic curve $E/k'$ with 
complex multiplication by an order of $K$. 
Suppose that all the points of finite order of $E$ are rational over $K_{ab}$---(*).
Then there are $[k':K]$ Hecke characters of the idele class group of $K$, 
$\varphi: \mathbb{A}_K^\times /K^\times \rightarrow \C^\times$, that satisfy
\begin{align}
  \psi_{E/k'} = \varphi \cdot {\rm Nm}_{k'/K}. 
\end{align}
All those $\varphi$'s define one and the same character $[\varphi]$ on 
the image of ${\rm Nm}_{k'/K}: \mathbb{A}^\times_{k'} \rightarrow \mathbb{A}^\times_K$. 
All those $\varphi$'s are of type $[-1/2; 1,0]$. Conversely, if such $\varphi$'s exist 
for the Hecke character $\psi_{E/k'}$ of a model $E/k'$, then the model satisfies 
the condition (*).

It is known that any elliptic curve $[E_z]_\C$ with complex multiplication 
by an order of $K$ has models over some number fields that are abelian 
extensions of $K$ (incl. $K(j([E]))$) so that the condition (*) is satisfied; 
more information is found in pp.216--217 of \cite{Shimura-AA}, and also 
in section \ref{sssec:model}. $\bullet$
\end{lemma}
\begin{props}(cf. [Serre \cite{Serre-CourseA}, Prop. VI.13] and 
[Shimura \cite{Shimura-AA}, \S 7.9.A])
\label{props:L4psi-L4phi}
Here we use the same notation as in the previous Lemma, and suppose that 
a model $E/k'$ satisfies the condition (*). Then 
\begin{align}
 L(s,\psi_{E/k'}) = \prod_{\varphi \in [\varphi]} L(s,\varphi),
  \label{eq:HeckeL4H--HeckeL4K}
\end{align}
where the product runs over all the $[k':K]$ variations of $\varphi$ 
consistent with $[\varphi]$. $\bullet$ 
\end{props}

Therefore, for an elliptic curve $E/k$ defined over a number field $k$ in one of 
the class i)--iii) at the beginning of section \ref{ssec:review-HW-L1} with the 
condition (*) in Lemma \ref{lemma:HeckeC-LandK} satisfied by $E/k'$ ($k'=L_{f_z}$
for $k$ in (iii), and $k'=k$ otherwise), computation of the Hasse--Weil $L$-function 
has now been reduced to computation of the $L$-functions of type $[-1/2; 1,0]$ 
Hecke characters of the idele class group of an imaginary quadratic field $K$.
The latter---Hecke $L$-functions---is now related to the Mellin transform of 
Hecke theta functions as follows. We begin with defining the following 
functions:\footnote{
If we wish to express $L(s,\psi)$ for a Hecke character $\psi$ of $\mathbb{A}_k^\times$
of a number field $k$ with $[k:\Q] > 2$ directly as the Mellin transform of 
a modular form, rather than 
through decomposing it as in (\ref{eq:HeckeL4H--HeckeL4K}), more general form 
of theta functions (for $L=k$) needs to be introduced; 
see [Neukirch \cite{Neukirch}, VII \S8]. The general version of the theta 
functions, however, does not fit into the observation (\ref{eq:obsrv-f1}). 
For that reason, we do not exploit the general version of the theta functions 
and rely on Lemma \ref{lemma:HeckeC-LandK} and Prop. \ref{props:L4psi-L4phi} instead 
in this article.}
\begin{defn}
Let $L$ be a number field which has $r_1$ real embeddings 
$\rho_a: L \hookrightarrow \R$ ($a=1,\cdots, r_1$) and $r_2$ pairs of 
imaginary embeddings $\sigma_b: L \hookrightarrow \C$, 
$\bar{\sigma}_b= cc \circ \sigma_b$; $cc$ stands for the complex conjugation 
in $\C$. For a sublattice $\Lambda$ of the lattice 
$({\cal O}_L, \vev{-,-}_{L/\Q})$, $x \in L \otimes_\Q \R/\Lambda$ and 
${\bf p} \in (\Z/2\Z)^{r_1} \times \Z^{\oplus r_2}$, we set 
\begin{align}
  \vartheta_L^{\bf p}(\tau; \Lambda, x) & \; :=
      \sum_{w \in x} [w]_{{\bf p}} \; q^{\frac{\vev{w,w}_{L/\Q}}{2}},  \label{eq:def-theta4L-P}\\
   [w]_{{\bf p}} & \; := \prod_{a=1}^{r_1} (\rho_a(w))^{p_{\rho_a}} \; 
     \prod_{b=1}^{r_2} \left\{ \begin{array}{ll}
        (\sigma_b(w))^{p_{\sigma_b}}, & {\rm if~}p_{\sigma_b}\geq 0 \\
        (\bar{\sigma}_b(w))^{-p_{\sigma_b}}, & {\rm if~}p_{\sigma_b} < 0
     \end{array} \right\}.
\end{align}
Here, $p_{\rho_a}$'s with $a=1,\cdots, r_1$ in 
${\bf p}=(p_{\rho_1},\cdots, p_{\rho_{r_1}}, p_{\sigma_1},\cdots, p_{\sigma_{r_2}}) \in 
(\Z/2\Z)^{r_1} \times \Z^{\oplus r_2}$ are regarded as either 0 or 1, when 
they are used in defining a monomial.\footnote{
Here, monomials $[w]_{{\bf p}}$ labeled by ${\bf p} \in (\Z/2\Z)^{r_1} \times 
\Z^{\oplus r_2}$ are called spherical functions. The Poisson resummation 
(Fricke involution) formula is not messed up by insertion of such a monomial 
\cite{Iwaniec, Miyake, Neukirch}.} 
\end{defn}

Let us now complete the task of relating the Hasse--Weil $L$-function of 
an elliptic curve with complex multiplication with the congruent theta 
functions above. Let $k'/K$ be an abelian extension containing $L_{f_z}/K$, and 
$\mathfrak{m}_f$ an integral ideal of $K$ so that the ray class field 
$L_{\mathfrak{m}_f}$ contains $k'$. 

\begin{defn} [Neukirch \cite{Neukirch}, Lemma VII.7.6]
\label{defn:theta-K-1}
Let $K$ be an imaginary quadratic field, and $\varphi$ its Hecke character of type 
$[-1/2; 1, 0]$. For such a Hecke character $\varphi$ of $\mathbb{A}_K^\times/K^\times$ 
with the conductor $\mathfrak{c}_f$, one can uniquely determine a character 
$\chi_f: [{\cal O}_K/\mathfrak{c}_f]^\times \rightarrow S^1$ with respect to the 
multiplication law in $[{\cal O}_K/\mathfrak{c}_f]^\times$ and the group of complex 
phases $S^1$; we will see how $\chi_f$ is determined from $\varphi$ 
in section \ref{sssec:model}. We assume that the conductor of $\varphi$ 
satisfies $\mathfrak{m}_f | \mathfrak{c}_f$. 
Let us also choose a set of ${\cal O}_K$ integral ideals 
$\{ \mathfrak{a}(\mathfrak{K}) \}_{\mathfrak{K} \in {\rm Cl}_K(\mathfrak{m}_f)}$ 
as discussed in section \ref{ssec:theta-zeta}, with one extra condition that 
all of $\mathfrak{a}(\mathfrak{K})$ are prime to $\mathfrak{c}_f$ (not just prime 
to $\mathfrak{m}_f$). 
Using all these data, we define a {\it Hecke theta function} on $\tau \in {\cal H}$ by 
\begin{align}
  \vartheta(\tau; \varphi, \mathfrak{K}) & \; :=
     \frac{1}{\varphi(\mathfrak{a}(\mathfrak{K}))}
   \sum_{x \in [[\mathfrak{a}(\mathfrak{K})]_1(\mathfrak{m}_f)]_{\mathfrak{a}(\mathfrak{K})\mathfrak{c}_f}} 
    \chi_f(x) \; \vartheta^{1}_K\left( \frac{\tau}{\N \mathfrak{a}(\mathfrak{K})}; 
    \mathfrak{c}_f \mathfrak{a}(\mathfrak{K}), x \right). 
  \label{eq:def-Hecke-theta}
\end{align}
where $[[\mathfrak{a}(\mathfrak{K})]_1(\mathfrak{m}_f)]_{\mathfrak{a}(\mathfrak{K})\mathfrak{c}_f}$ 
is the image of $[\mathfrak{a}(\mathfrak{K})]_1(\mathfrak{m}_f)$ under the projection 
$\mathfrak{a}(\mathfrak{K}) \rightarrow 
\mathfrak{a}(\mathfrak{K})/\mathfrak{a}(\mathfrak{K})\mathfrak{c}_f$. 

For the Hecke character $\varphi: \mathbb{A}_K^\times/K^\times \rightarrow \C^\times$, 
let $\overline{\varphi}$ be the Hecke character of $\mathbb{A}_K^\times/K^\times$ 
given by $\overline{\varphi} := cc \circ \varphi$; $\overline{\varphi}$ is 
of type $[-1/2; -1,0]$, its conductor is the same as $\mathfrak{c}_f$, and 
the character of $[{\cal O}_K/\mathfrak{c}_f]^\times$ corresponding to 
$\overline{\varphi}$ is given by $\overline{\chi}_f := cc \circ \chi_f$. 
A {\it Hecke theta function for} $\overline{\varphi}$ is defined by 
\begin{align}
  \vartheta(\tau; \overline{\varphi},\mathfrak{K}) :=
      \frac{1}{\overline{\varphi}(\mathfrak{a}(\mathfrak{K}))}
   \sum_{x \in [[\mathfrak{a}(\mathfrak{K})]_1(\mathfrak{m}_f)]_{\mathfrak{a}(\mathfrak{K})\mathfrak{c}_f}}
     \overline{\chi}_f(x)\vartheta_K^{-1}
    \left( \frac{\tau}{\N \mathfrak{a}(\mathfrak{K})} ; 
       \mathfrak{c}_f\mathfrak{a}(\mathfrak{K}), x \right).
  \qquad \bullet
\end{align}
\end{defn}

Now, we are ready to write down the Hecke $L$-functions $L(s,\varphi)$ in terms 
of the Mellin transform of the Hecke theta functions. 

\begin{thm}[e.g., Koblitz \cite{Koblitz}, II \S5; Neukirch \cite{Neukirch}, VII \S7]
Let $K$ be an imaginary quadratic field, and 
$\varphi: \mathbb{A}_K^\times/K^\times \rightarrow \C^\times$ a Hecke character 
of type $[-1/2; 1,0]$. Then the Hecke $L$-function of $\varphi$ is given by 
\begin{align}
 L(s,\varphi) & \; = \sum_{\mathfrak{K} \in {\rm Cl}_K(\mathfrak{m}_f)} 
   L(s, \varphi, \mathfrak{K}),   \label{eq:sum-ClK-L1}  \\
L(s,\varphi,\mathfrak{K}) \frac{\Gamma(s)}{(2\pi)^s}  & \; =
    \frac{1}{\#[{\cal O}_K^\times]}
    \int_0^\infty \frac{dt}{t} \; t^s \vartheta(it; \varphi, \mathfrak{K}). 
   \qquad \qquad \bullet
  \label{eq:Mellin-L1}
\end{align}
\end{thm}

To summarize, the Hasse--Weil $L$-function of an elliptic curve $E/k$ with 
complex multiplication by an order of an imaginary quadratic field $K$ is  
given by combining (\ref{eq:def-theta4L-P}, \ref{eq:def-Hecke-theta}, 
\ref{eq:Mellin-L1}, \ref{eq:sum-ClK-L1}, \ref{eq:HeckeL4H--HeckeL4K}) and 
finally either (\ref{eq:HWL4H--HeckeL4H}) or (\ref{eq:HWL4F--HeckeL4H}), 
if the field of definition $k$ is either an abelian extension containing 
$L_{f_z}/K$ or $F_{f_z}^{[E]}$, and the condition (*) in Lemma \ref{lemma:HeckeC-LandK} 
is satisfied (after the base change to $L_{f_z}$, if $k=F_{f_z}^{[E]}$).

\subsubsection{Hecke Theta Functions as Modular Forms}

There is a more general version of Proposition \ref{props:automrph-theta0-gen}, 
which is stated in the form we use in this article:
\begin{thm}[Iwaniec \cite{Iwaniec}, Cor. 10.7; Miyake \cite{Miyake}, Cor. 4.9.4]
\label{thm:HeckeTheta-automorphic-indiv}
Here, we use the same notation and assumption as in Definition \ref{defn:theta-K-1}.
Then $\vartheta_K^1(\tau';\Lambda,x)$ for $x \in \Lambda^*/\Lambda$ 
[resp. $\vartheta(\tau; \varphi, \mathfrak{K})$] is a cusp form of weight 2 
for $\Gamma(4N_\Lambda) \subset {\rm SL}(2;\Z)$ [resp.\footnote{
$x \in \mathfrak{a}(\mathfrak{K})/\mathfrak{a}(\mathfrak{K})\mathfrak{c}_f$ 
implies $x \in (\mathfrak{a}(\mathfrak{K})\mathfrak{c}_f)^*/
\mathfrak{a}(\mathfrak{K})\mathfrak{c}_f$; see section \ref{sssec:L0-nonmaximal}. }  
$\Gamma(4N_{\mathfrak{a}(\mathfrak{K})\mathfrak{c}_f}) \subset {\rm SL}(2;\Z)$]. 
The group $\Gamma(4N_\Lambda)$ [resp. $\Gamma(4N_{\mathfrak{a}(\mathfrak{K})\mathfrak{c}_f})$]
acts on $\tau' \in {\cal H}$ [resp. on the combination 
$\tau/\N\mathfrak{a}(\mathfrak{K}) \in {\cal H}$] through the ordinary action 
of ${\rm SL}(2;\Z)$ on ${\cal H}$.  

When a non-trivial multiplier system is allowed, $\Gamma(4N_\Lambda)$ can be 
replaced by $\Gamma_1(4N_\Lambda)$. If $x \in {\cal O}_K/\Lambda$, then $(x,x)/2 \in \Z$, 
and the multiplier system becomes trivial. $\bullet$ 
\end{thm}

For many cases we use the theta functions $\vartheta_K^{\bf P}(\tau;\Lambda, x)$ 
in this article, $\Lambda$ is not just a general sublattice of 
$({\cal O}_K,\vev{-,-}_{K/\Q})$ (that is, not just an abelian group), but 
a subring of ${\cal O}_K$. 
In such cases, we can think of summing over $\vartheta_K^{\bf P}(\tau; \Lambda,x)$ 
over $x$ and seek for an analogue/generalization of 
Proposition \ref{props:automrph-theta0}. 
\begin{rmk}[cf. Miyake \cite{Miyake}, Thm. 4.9.3]
Let $K$ be an imaginary quadratic field, and $\mathfrak{a}$ an integral ideal 
of ${\cal O}_K$. Let $\chi_f: [{\cal O}_K/\mathfrak{c}_f]^\times \rightarrow S^1$ be a 
character with respect to the multiplication law in $[{\cal O}_K/\mathfrak{c}_f]^\times$, 
where $\mathfrak{c}_f$ is an integral ideal of ${\cal O}_K$ relatively prime 
to $\mathfrak{a}$. Then, for $\gamma = 
\left(\begin{array}{cc} a & b \\ c & d \end{array} \right)
 \in \Gamma_0(2N_{\mathfrak{a}\mathfrak{c}_f})$ and a subgroup $G_0$ of 
$[{\cal O}_K/\mathfrak{c}_f]^\times$, 
\begin{align}
&   \sum_{x \in [\mathfrak{a}_{G_0}(\mathfrak{c}_f)]_{\mathfrak{a}\mathfrak{c}_f}} 
      \chi_f(x) \vartheta_K^{\bf P}(\gamma \cdot \tau'; \mathfrak{a}\mathfrak{c}_f,x)
    \label{eq:combin-chif-HTheta}  \\
 = & \; (c\tau+d)^{1+{\rm deg}({\bf P})}
      \left( \frac{-{\rm disc}(\mathfrak{a}\mathfrak{c}_f)}{d} \right) 
    \sum_{x \in [\mathfrak{a}_{G_0}(\mathfrak{c}_f)]_{\mathfrak{a}\mathfrak{c}_f}} 
     \chi_f(x) \vartheta_K^{\bf P}(\tau'; \mathfrak{a}\mathfrak{c}_f, a \cdot x),
\end{align}
where we used the Legendre (quadratic residue) symbol $(-/-)$ in writing down 
the multiplier system explicitly; ${\rm deg}({\bf P})$ is the degree of the 
monomial $[w]_{\bf P}$.
Therefore, for the subgroup of $\Gamma_0(2N_{\mathfrak{a}\mathfrak{c}_f})$ where 
\begin{align}
(a \cdot): [\mathfrak{a}_{G_0}(\mathfrak{c}_f)]_{\mathfrak{a}\mathfrak{c}_f} \ni x 
   \longmapsto a \cdot x \in \mathfrak{a}/\mathfrak{a}\mathfrak{c}_f
\end{align}
is an isomorphism from $[\mathfrak{a}_{G_0}(\mathfrak{c}_f)]_{\mathfrak{a}\mathfrak{c}_f}$
to itself, the sum (\ref{eq:combin-chif-HTheta}) is a modular form of weight 
$1+{\rm deg}({\bf P})$, with the multiplier system given by 
$(-{\rm discr}(\mathfrak{a}\mathfrak{c}_f)/d)\chi_f(d)$. $\bullet$
\end{rmk}

When we deal with the Hecke $L$-functions for elliptic curves $E/H_K$ with 
complex multiplication by ${\cal O}_K$, we can just choose $G_0$ to be 
the entire $[{\cal O}_K/\mathfrak{c}_f]^\times$. For this case, there is 
a more definite result (cf. Prop. \ref{props:automrph-theta0}). 
\begin{thm} [Miyake \cite{Miyake}, Thm 4.8.2]
\label{thm:HeckeTheta-automorphic} 
The sum $\sum_{\mathfrak{K} \in {\rm Cl}_K} \vartheta(\varphi,\mathfrak{K})$ 
of the Hecke theta functions 
is a cusp form of weight 2 for the group $\Gamma_0(N_{\chi_f})$ acting on 
$\tau \in {\cal H}$, with a multiplier system (homomorphism) 
$\underline{\chi} : \Gamma_0(N_{\chi_f}) \rightarrow \C^\times$
(for detailed information of $\underline{\chi}$, see [Miyake \cite{Miyake}, Thm. 4.8.2]). 
The level of the group $\Gamma_0(N_{\chi_f})$ is 
$N_{\chi_f} := |D_{K/\Q}| \cdot  \N\mathfrak{c}_f = \N(\mathfrak{c}_f\mathfrak{d}_{K/\Q})$.
  $\bullet$ 
\end{thm}

In a broader context, the two Theorems above are concerned about how to extract 
single component modular forms of a subgroup $\Gamma \subset {\rm SL}(2;\Z)$ 
of some level from a (multi-component) vector-valued modular form of ${\rm SL}(2;\Z)$, 
or in the other way around. 

\subsubsection{Choice of Models}
\label{sssec:model}

We have stated that any $\C$-isomorphism class of elliptic curves with 
complex multiplication, $[E]_\C \in Ell({\cal O}_{f_z})$, has a model $E/k$
over the number field $k=L_{f_z}$, over any abelian extension $k/K$ containing 
$L_{f_z}/K$, and over $k=F_{f_z}^{[E]}$. It is not that there is just one model $E/k$ for 
a given $[E]_\C$ and $k$, however. To be more precise, one can think of classifying 
models $E/k$ defined over a number field $k$ for a given $[E]_\C$, thinking that 
two models 
$E/k$ and $E'/k$ are equivalent iff there is an isomorphism between $E$ and $E'$ 
defined over\footnote{In a colloquial language, that is whether one can find a map 
between coordinates of $E$ and $E'$ where the map in the form of polynomials 
in the coordinates have coefficients in the field $k$, not just in $\C$.} $k$. 
We will describe here how one can list up the inequivalent models over a given 
number field $k$ for a given $\C$-isomorphism class $[E]_\C$. 

We have used a notion of a type of a Hecke character of the idele class group 
$\mathbb{A}_L^\times/L^\times$ for a number field $L$ 
in section \ref{sssec:Hecke-L-Hecke-theta}, but did not explain what it is. So, here is 
\begin{defn}
\label{defn:type-HeckeChara}
Let $L$ be a number field which has $r_1$ real embeddings 
$\rho_a: L \hookrightarrow \C$ (for $a=1,\cdots, r_1$) and $r_2$ pairs of 
imaginary embeddings $\sigma_b: L \hookrightarrow \C$ and 
$\bar{\sigma}_b = cc\circ \sigma_b: L\hookrightarrow\C$ (for $b=1,\cdots, r_2$); 
$[L:\Q] = r_1+2r_2$. Any Hecke character 
$\phi: \mathbb{A}_L^\times /L^\times \rightarrow \C^\times$ can be written in the form of 
the product of continuous homomorphisms associated with all the inequivalent valuations 
of $L$: $\phi = \phi_f \cdot \phi_\infty$, and   
\begin{align}
  \phi_{\mathfrak{p}}: L_{\mathfrak{p}}^\times \rightarrow \C^\times, & \qquad 
  \left(\phi_f = \prod_{\mathfrak{p}} \phi_{\mathfrak{p}}\right) :
      \left( \prod_{\mathfrak{p}} L_{\mathfrak{p}}^\times \supset \mathbb{A}_{L,f}^\times \right)
             \rightarrow \C^\times, \\
  \phi_{v\tau}: L_{v\tau}^\times \rightarrow \C^\times & \qquad 
  \left(\phi_\infty = \prod_{v\tau} \phi_{v\tau}\right) :
       \left(\mathbb{A}^\times_{L,\infty} = \prod_{v\tau} L_{v\tau}^\times \right)
              \rightarrow \C^\times.
\end{align}
Here, $\mathfrak{p}$ runs over the set of all the non-Archimedean valuations, 
which is equivalent to the set of all the non-zero prime ideals of ${\cal O}_L$.
The index $v\tau$ runs over the set of all the inequivalent Archimedean valuations
${\rm Arch}(L)$; let $\Phi_L^{\rm real} = \{ \rho_a \; | \; a=1,\cdots r_1\}$, and 
$\Phi_L^{\rm im} \amalg \overline{\Phi}_L^{\rm im}$ be a mutually exclusive grouping of 
imaginary embeddings so that no two elements of $\Phi_L^{\rm im}$ are the complex conjugate
of the other; then the set ${\rm Arch}(L)$ is in one-to-one correspondence with 
$\Phi_L^{\rm real} \amalg \Phi_L^{\rm im}$. So, we can use 
$\tau \in \Phi_L^{\rm real}\amalg \Phi_L^{\rm im}$ as a label/index of the product. 

In this article, we say\footnote{It seems that there is no standard jargon / 
parametrization in referring to the notion we call ``type'' of a Hecke character. 
The parametrization in terms of ${\bf p}$ and ${\bf q}$ are chosen so that 
$\phi$ with $s^r=0$ reproduces the Hecke character / Gr\"{o}ssen-character 
in [Neukirch \cite{Neukirch}, VII \S6]; $\phi_\infty$ here and $\chi_\infty$ 
in \cite{Neukirch} are in the relation $\phi_\infty(a_\infty) = 
({\rm Nm}_{L/\Q}(a_\infty))^{s^r}/\chi_\infty(a_\infty)$ for 
$a_\infty \in\mathbb{A}_{L,\infty}^\times$. The parameters ${\bf p}$ and ${\bf q}$ 
of the type here are related to ${\bf m}$ and ${\bf \varphi}$ 
in [Lang \cite{Lang-ANT}, XIV] through the relation $p_\rho = m_\rho \in \Z/(2\Z)$, 
$p_\sigma = - m_\sigma \in \Z$, $q_\rho = - \varphi_\rho \in \R/(2\pi \Z)$ and 
$q_\sigma = - 2\varphi_\sigma \in \R/(2\pi \Z)$. } 
that {\it a Hecke character} $\phi: \mathbb{A}_L^\times / L^\times \rightarrow \C^\times$ 
{\it is of type} $[s^r; {\bf p}, {\bf q}]$, where 
\begin{align}
s^r \in \R, \qquad {\bf p} = (p_{\rho}, p_{\sigma}) \in [\Z/(2\Z)]^{r_1} \times \Z^{\oplus r_2},
   \qquad {\bf q} =(q_\rho, q_\sigma) \in [\R/(2\pi \Z)]^{r_1+r_2}, 
\label{eq:def-label-type4HeckeChara}
\end{align}
when $\phi_\infty = \prod_{v\tau \in {\rm Arch}(L)} \phi_{v\tau}$ is parametrized by 
\begin{align}
  \phi_{v\rho}: L_{v\rho}^\times \ni a_\rho & \mapsto
    \left(\frac{a_\rho}{|a_\rho|_\R} \right)^{-p_\rho} (|a_\rho|_\R )^{-iq_\rho}
        \times \left( |a_\rho|_\R \right)^{s^r} \in \C^\times, \\
  \phi_{v\sigma}: L_{v\sigma}^\times \ni a_\sigma & \mapsto
    \left(\frac{a_\sigma}{|a_\sigma|_\C} \right)^{-p_\sigma} (|a_\sigma|_\C )^{-iq_\rho}
        \times \left( |a_\sigma|_\C \right)^{2s^r} \in \C^\times. 
\end{align}

The type of a given Hecke character $\phi$ looks different depending 
on how the $2r_2$ imaginary embeddings are grouped into $\Phi_L^{\rm im}$ and 
$\overline{\Phi}_L^{\rm im}$. When a number field $L$ is an extension of 
an imaginary quadratic field $K$, and the pair of imaginary embeddings 
of $K$ is grouped into $\Phi_K^{\rm im} = \{ \sigma \}$ and 
$\overline{\Phi}_K^{\rm im} = \{ \bar{\sigma}\}$, we make it a rule to choose 
a canonical grouping induced from $\Phi_K^{\rm im} \amalg \overline{\Phi}_K^{\rm im}$: 
$\Phi_L^{\rm im}$ consists of imaginary embeddings of $L$ that become $\sigma$ 
upon restriction to $K$, and $\overline{\Phi}_L$ of imaginary embeddings of $L$
that become $\bar{\sigma}$ upon restriction to $K$. $\bullet$
\end{defn}

The $\phi_\infty$ part of a Hecke character is determined completely by the type 
$[s^r; {\bf p}, {\bf q}]$. For example, when $L$ is a totally imaginary field 
($r_1=0$), the type $[-1/2; {\bf 1}, {\bf 0}]$ implies that 
\begin{align}
\phi_\infty: \mathbb{A}_{L,\infty}^\times \ni (a_\sigma) \mapsto 
   \frac{1}{\prod_{\sigma \in \Phi_L^{\rm im}} a_\sigma} \in \C^\times.
  \label{eq:phi-infty-type10}
\end{align}

With this definition, we can now write down a result of classification of 
models $E/k$ modulo isomorphism over $k$, as follows:
\begin{thm} (See [Shimura \cite{Shimura-AV}, Thms. 22.1 and 19.10], 
[Lang \cite{Lang-EF}, Chap. 10 \S4] and [Shimura \cite{Shimura-ze}, Thm. 5])
\label{thm:model-E/H}
Let $K$ be an imaginary quadratic field, and $k/K$ an abelian extension 
containing a ring class field $L_{f_z}/K$ for some $f_z \in \N_{>0}$. 
For a given $[E]_\C \in Ell({\cal O}_{f_z})$, its models $E/k$ 
modulo $k$-isomorphisms are in one-to-one correspondence with 
Hecke characters $\psi: \mathbb{A}_k^\times / k^\times \rightarrow \C^\times$ 
satisfying the two conditions:
\begin{enumerate}
\item [(a)] It is of type $[-1/2; {\bf 1}, {\bf 0}]$.
\item [(b)] For any prime ideal $\mathfrak{P}$ of ${\cal O}_k$ that is prime to 
the conductor $\mathfrak{C}_f$ of $\psi$, $\psi_{\mathfrak{P}}(\pi_{\mathfrak{P}}) 
\in {\cal O}_{f_z} \subset K  \subset \C$ and 
${\rm Nm}_{k/K}(\mathfrak{P}) = (\psi_{\mathfrak{P}}(\pi_{\mathfrak{P}}))_{{\cal O}_K}$. Here, 
$\pi_{\mathfrak{P}}$ is a uniformizer of the ring of $\mathfrak{P}$-adic integers. 
%
\end{enumerate}

There is an isogeny defined over $k$ from a model $E/k$ of 
$[E]_\C \in Ell({\cal O}_{f_z})$ to a model $E'/k$ of 
$[E']_\C \in Ell({\cal O}_{f_z})$ if and only if 
$\psi_{E/k} = \psi_{E'/k}$. $\bullet$ 
\end{thm}

The $L$-function is defined for each one of $k$-isomorphism classes 
of elliptic curves over $k$ (not necessarily with complex multiplication). 
The $L$-functions $L(E/k,s)$ and $L(E'/k,s)$ of a pair of models over 
a common number field $k$ can be the same, even when there is no isomorphism 
over $k$ between $E/k$ and $E'/k$. To be more precise, 
\begin{rmk}
(See [Faltings \cite{Faltings}, p.22, Cor. 2]) 
Let $k$ be a number field and $E$ and $E'$ are elliptic curves 
defined over $k$ (not necessarily with complex multiplication). 
Then the following are equivalent:
\begin{enumerate}
\item $E$ and $E'$ are $k$-isogenous,
\item $L_v(E,s)=L_v(E',s)$ for almost all places $v$ of $k$.
\end{enumerate}
This statement holds true also for higher-dimensional abelian varieties.
As a special case of this theorem, the pair of models over $k$ that are referred to 
at the end of Thm. \ref{thm:model-E/H} have an identical $L$-function.\footnote{
That is consistent with the fact that the $L$-function is determined from 
the associated Hecke character, as in (\ref{eq:HWL4H--HeckeL4H}).}  $\bullet$
\end{rmk}

One might also be interested in the conditions for a model $E/L_{f_z}$ defined 
over the ring class field to be obtained as a base change of a model defined 
over $F_{f_z}^{[E]}$.
The answer to this question can also be phrased in terms of the Hecke character 
$\psi_{E/L_{f_z}}$:
\begin{thm} (See [Shimura \cite{Shimura-AV}, Thms. 22.2 and 20.15; \cite{Shimura-AA}, Thm. 7.46])
\label{thm:model-E/F}
Let $E/L_{f_z}$ be an elliptic curve with complex multiplication by ${\cal O}_{f_z}$, 
$\psi_{E/L_{f_z}}$ its associated Hecke character of type $[-1/2; {\bf 1}, {\bf 0}]$, 
and $\mathfrak{C}_f$ the conductor of $\psi_{E/L_{f_z}}$.
There exists a model $E/F_{f_z}^{[E]}$ whose base change to ${\rm Spec}(L_{f_z})$ 
is isomorphic to $E/L_{f_z}$ 
(here, $[E]$ is the $\C$-isomorphism class of $E/L_{f_z}$), 
if and only if the following condition\footnote{It is implicit in this condition that 
the ideal $\mathfrak{C}_f$ of ${\cal O}_{L_{f_z}}$ is invariant under $\rho_{[E]}$.}
is satisfied: 
\begin{align}
\psi_{E/L_{f_z}}(\rho_{[E]}(x)) = cc \circ \psi_{E/L_{f_z}}(x), \qquad  
{}^\forall x \in \mathbb{A}_{L_{f_z}}^\times.    
   \qquad \qquad \bullet
  \label{eq:cond-psi-real}
\end{align}
\end{thm}

\begin{rmk}
When an elliptic curve $E$ with complex multiplication by an order 
${\cal O}_{f_z}$ of $K$ is defined over a number field $k$ containing 
$L_{f_z}$, the conductor $N_{E/k}$ of the elliptic curve $E/k$ is 
given by the conductor $\mathfrak{C}_f$ of the associated Hecke character 
$\psi_{E/k}$ by [Serre--Tate \cite{ST}, Thm. 12] 
\begin{align}
 N_{E/k} = \mathfrak{C}_f^2 \in {\rm Div}({\rm Spec}({\cal O}_k)). 
\end{align}

When an elliptic curve $E$ with complex multiplication by ${\cal O}_K$ 
is defined over $F=F_{f_z=1}^{[E]}$, the conductor $N_{E/F}$ of 
the elliptic curve $E/F$ is related to the conductor 
$\mathfrak{C}_f$ of the associated Hecke character $\psi_{E/L_{f_z}}$ 
by [Gross \cite{Gross}, eq. (10.3.2)] 
\begin{align}
N_{E/F} = D_{H/F} \cdot {\rm Nm}_{H/F}(\mathfrak{C}_f) \in 
    {\rm Div}({\rm Spec}({\cal O}_F)).
  \label{eq:formula-conductor4EC-F}
\end{align}
Note that, in the case $h({\cal O}_K)=1$ (i.e., when $F=\Q$), the ideal 
$(N_{\chi_f})_\Z$ is equal to the conductor $N_{E/\Q}$. $\bullet$
\end{rmk}

Presentation so far makes it clear that the Hecke character $\psi_{E/k}$ 
for the field of definition $k$ is the crucial tool in classifying 
elliptic curves defined over number fields so far as the $L$-functions 
are concerned. In the meantime, we have written down a way to express 
the $L$-functions in terms of Hecke characters for the imaginary quadratic 
field $K$ of complex multiplication. It is therefore convenient if the 
$k$-isomorphism classification of models over $k$ (Thm. \ref{thm:model-E/H}) 
is re-stated in terms of Hecke characters $\varphi$ for $K$. 
We do so in the following, by largely following discussions 
in [Neukirch \cite{Neukirch}, VII \S6] and [Milne \cite{Milne-CFT}, Chap. V]. 
 
To start off, we study in detail the structure of the group of Hecke 
characters for a number field $L$; we intend to use the following 
discussion for $L=K$.

Let $L$ be a number field, and we follow the notation adopted in 
Definition \ref{defn:type-HeckeChara}. 
Let $\mathfrak{c}_f$ be an integral ${\cal O}_L$ ideal, and 
${\rm H.Char}[C(\mathfrak{c}_f)]$ the group of Hecke characters 
$\phi: \mathbb{A}_L^\times / L^\times \rightarrow \C^\times$ with the modulus
 $\mathfrak{c}_f$. It then follows immediately 
from [Neukirch \cite{Neukirch}, Prop. VII.6.12] that  
\begin{align}
 0 \rightarrow {\rm Char}\left[ {\rm Cl}_L(\mathfrak{c}_f) \right] 
   \rightarrow {\rm H.Char}[ C(\mathfrak{c}_f) ]
   \rightarrow {\rm Char}\left[ \mathbb{A}_{L,\infty}^\times /
        {\cal O}_{L,1}^\times(\mathfrak{c}_f) \right] \rightarrow 0
\end{align}
is exact. Here,\footnote{Notation: the group ${\cal O}_{L,1}^\times(\mathfrak{c}_f)$ 
in this article corresponds to ${\cal O}^{\mathfrak{m}}$ in \cite{Neukirch}.}
\begin{align}
  {\cal O}_{L,1}^\times(\mathfrak{c}_f) := \left\{ \epsilon \in {\cal O}_L^\times \; | 
    \; \epsilon \in 1 + \mathfrak{c}_f \right\}. 
\end{align}
The projection from ${\rm H.Char}[C(\mathfrak{c}_f)]$ to 
${\rm Char}[\mathbb{A}_{L,\infty}^\times/{\cal O}_{L,1}^\times(\mathfrak{c}_f)]$ 
is given by $\phi \mapsto \phi_\infty$. 

\begin{defn}
An integral ${\cal O}_L$-ideal $\mathfrak{c}_f$ as a modulus and 
a type $[{\bf p}, {\bf q}]$ are said to be {\it compatible}, if 
a character $\phi_\infty \in {\rm Char}[\mathbb{A}_{L,\infty}^\times]$ of type 
$[s^r; {\bf p}, {\bf q}]$ (with any $s^r \in\R$) is trivial on 
${\cal O}_{L,1}^\times(\mathfrak{c}_f)$ (the triviality condition is independent 
of $s^r$).
\end{defn}

\begin{props}[cf. Neukirch \cite{Neukirch}, VII.6.12--14]
\label{props:str-HeckeChara-givenType}
Let $\phi: \mathbb{A}_L^\times/L^\times \rightarrow \C^\times$ be a Hecke character
of type $[s^r; {\bf p}, {\bf q}]$ with a modulus $\mathfrak{c}_f$ that is compatible 
with the type $[{\bf p}, {\bf q}]$. Then at non-Archimedean valuations for 
prime ideals $\mathfrak{q}$ outside of the support of $\mathfrak{c}_f$ (i.e., 
$\mathfrak{q} \Slash{|} \mathfrak{c}_f$), 
$\phi_{\mathfrak{q}} : L_{\mathfrak{q}}^\times \rightarrow \C^\times$ are given by 
a unique unitary (i.e., $S^1$-valued) character 
$\chi: I_L(\mathfrak{c}_f) \rightarrow S^1$ for a given $\phi$:  
\begin{align}
  \phi_{\mathfrak{q}} : L_{\mathfrak{q}}^\times \ni a_{\mathfrak{q}} \mapsto 
   \chi(\mathfrak{q}^{{\rm ord}_{\mathfrak{q}}(a_{\mathfrak{q}})} ) |\pi_{\mathfrak{q}}|_{\mathfrak{q}}^{s^r}
   = \chi(\mathfrak{q}^{{\rm ord}_{\mathfrak{q}}(a_{\mathfrak{q}})} ) (\N \mathfrak{q})^{-s^r},
\end{align}
where $I_L(\mathfrak{c}_f)$ is the group of fractional ideals of ${\cal O}_L$ 
prime to $\mathfrak{c}_f$, and $\pi_{\mathfrak{q}}$ is a uniformizer of the ring of 
$\mathfrak{q}$-adic integers. 

$\phi_{\mathfrak{p}}: L_{\mathfrak{p}}^\times \rightarrow \C^\times$ for non-Archimedean 
valuations in the support of $\mathfrak{c}_f$ (i.e., $\mathfrak{p} | \mathfrak{c}_f$) 
have the following property.
For $\alpha \in {\cal O}_L$ that is prime to 
$\mathfrak{c}_f$,  define
\begin{align}
 \chi_f: \alpha \mapsto \chi_f(\alpha) & \; :=
     \left( \prod {}_{\mathfrak{q} \Slash{|} \mathfrak{c}_f} \phi_{\mathfrak{q}}(\alpha_{\mathfrak{q}})
     \right)  \cdot \phi_\infty(\alpha) 
  = \chi((\alpha)_{{\cal O}_L}) \cdot 
  \prod_{v\tau} \left(\frac{\alpha_\tau}{|\alpha_\tau|_{\C}}\right)^{-p_\tau}
         \!\!\!\!\!\! |\alpha_\tau|_\C^{-iq_\tau}  ;
\end{align}
then the triviality of $\phi$ on $L^\times \subset \mathbb{A}_L^\times$ (by def) 
implies\footnote{This property specifies $\phi_{\mathfrak{p}}$'s with 
$\mathfrak{p}|\mathfrak{c}_f$ for $a_{\mathfrak{p}} \in L_{\mathfrak{p}}^\times$ with 
${\rm ord}_{\mathfrak{p}}(a_{\mathfrak{p}})=0$, but $\phi_{\mathfrak{p}}(a_{\mathfrak{p}})$ 
remains unspecified for $a_{\mathfrak{p}}$ with ${\rm ord}_{\mathfrak{p}}(a_{\mathfrak{p}}) \neq 0$.
One should exploit the triviality of $\phi$ for $\alpha \in {\cal O}_L$ that is 
{\it not} prime to $\mathfrak{c}_f$ to determine $\phi_{\mathfrak{p}}$ completely. } 
that
\begin{align}  
  \left( \prod {}_{\mathfrak{p}|\mathfrak{c}_f} \phi_{\mathfrak{p}}(\alpha_{\mathfrak{p}}) \right) 
   = 1/\chi_f(\alpha).
\end{align}
Since $\phi$ has a modulus $\mathfrak{c}_f$, $\chi_f$ so defined is a unitary character 
of $[{\cal O}_L/\mathfrak{c}_f]^\times$. This character $\chi_f$ satisfies the condition  
\begin{align}
  \chi_f(\epsilon) 
   = \prod_{v\tau \in {\rm Arch}(L)}
      \left(\frac{\epsilon_\tau}{|\epsilon_\tau|_\C}\right)^{-p_\tau}
             |\epsilon_\tau|_\C^{-iq_\tau}, 
   \qquad {}^\forall \epsilon 
        \in {\cal O}_L^\times/{\cal O}_{L,1}^\times(\mathfrak{c}_f).
   \label{eq:cond-chif-HK-eps}
\end{align}

Let ${\rm H.Char}[C(\mathfrak{c}_f)]^{[s^r; {\bf p}, {\bf q}]}$ be the set of 
Hecke characters for $L$ with the modulus $\mathfrak{c}_f$ and type 
$[s^r; {\bf p}, {\bf q}]$. It is not empty so long as $[{\bf p}, {\bf q}]$ is 
compatible with $\mathfrak{c}_f$. For a pair of Hecke characters $\phi_1$ and $\phi_2$
in ${\rm H.Char}[C(\mathfrak{c}_f)]^{[s^r; {\bf p}, {\bf q}]}$, $\phi_1/\phi_2$ should be 
regarded as an element of ${\rm Char}[{\rm Cl}_L(\mathfrak{c}_f)]$. This is done 
by\footnote{The ratio 
$\chi_{f,1}/\chi_{f,2} : [{\cal O}_L/\mathfrak{c}_f]^\times \rightarrow S^1$ cannot 
retain more detailed information of $\phi_1/\phi_2$ than $\chi_1/\chi_2$ does, 
because $\chi_f$ is completely determined by $\chi$ (after a type is specified).} using 
$\chi_1/\chi_2: I_L(\mathfrak{c}_f) \rightarrow S^1$, which implies that $\chi_1/\chi_2$ 
factors through ${\rm Cl}_L(\mathfrak{c}_f) = I_L(\mathfrak{c}_f)/P_{L,1}(\mathfrak{c}_f)$. 
$\bullet$ 
\end{props}

Using an injective homomorphism $c: I_L(\mathfrak{c}_f) \hookrightarrow
 \mathbb{A}_L^\times/L^\times$ given by sending a prime ideal $\mathfrak{q} \in I_L(\mathfrak{c}_f)$ 
to $(1_{\mathfrak{p}},\pi_{\mathfrak{q}},1_\tau) \in (\prod_{\mathfrak{p} \neq \mathfrak{q}} L^\times_{\mathfrak{p}}) 
\times L^\times_{\mathfrak{q}} \times \mathbb{A}_{L,\infty}^\times$, 
a character $\phi \circ c: I_L(\mathfrak{c}_f) \rightarrow \C^\times$ is induced; 
we will abuse the notation and use $\phi$ also for the character $\phi \circ c$ of 
$I_L(\mathfrak{c}_f)$ in this article. 

\begin{props}[Neukirch \cite{Neukirch}, VII \S6]
Let $\phi: I_L(\mathfrak{c}_f) \rightarrow \C^\times$ be a character 
corresponding to a Hecke character $\phi: \mathbb{A}_L^\times/L^\times \rightarrow \C^\times$
with a modulus $\mathfrak{c}_f$. Then a principal ideal $(\xi)_{{\cal O}_L}$ prime to 
$\mathfrak{c}_f$ has the following value: 
\begin{align}
 \phi((\xi)_{{\cal O}_L}) = \chi((\xi)_{{\cal O}_L}) 
 \prod {}_{\mathfrak{q} \Slash{|}\mathfrak{c}_f}  |\xi_{\mathfrak{q}}|_{\mathfrak{q}}^{s^r}
 = \frac{\chi_f(\xi)}{\phi_\infty(\xi)}. 
\end{align}
This result, combined with (\ref{eq:phi-infty-type10}), determines 
the expression on the right-hand-side of (\ref{eq:def-Hecke-theta}). $\bullet$
\end{props}

We wish to use Prop. \ref{props:str-HeckeChara-givenType} in classifying Hecke 
characters $\varphi$ for the imaginary quadratic field $K$ of complex multiplication. 
Remember that a Hecke character $\psi_{E/k'}$ of type $[-1/2; {\bf 1}, {\bf 0}]$ 
of an abelian extension $k'$ over $K$ corresponds to a set of $[k':K]$ 
different Hecke characters of type $[-1/2; 1, 0]$ for $K$, provided that 
$E/k'$ satisfies the condition in Lemma \ref{lemma:HeckeC-LandK}.  
Therefore, we wish to introduce an equivalence relation where 
$\chi \sim \chi'$ if they are different only by ${\rm Char}[{\rm Gal}(k'/K)]$, so that 
we can classify Hecke characters for $k'$ by dealing with characters associated 
with the imaginary quadratic field $K$.

Let us first develop a result that is useful in dealing with the cases 
of $k'=H_K$ (so $f_z = 1$). Now, remember  
\begin{lemma}[Milne \cite{Milne-CFT}, Thm. V.1.5]
The following sequence is exact:
\begin{align}
 0 \rightarrow {\cal O}_L^\times/{\cal O}_{L,1}^\times(\mathfrak{c}_f) \rightarrow 
   [{\cal O}_L/\mathfrak{c}_f]^\times \rightarrow {\rm Cl}_L(\mathfrak{c}_f)
  \rightarrow {\rm Cl}_L \rightarrow 0. \qquad \qquad \bullet
\end{align}
\end{lemma}
Using this fact, we arrive at 
\begin{thm}
\label{thm:model-classify-HK}
Let $\mathfrak{c}_f$ be an integral ${\cal O}_L$ ideal to be used as a modulus 
of Hecke character for a number field $L$, and $[{\bf p}, {\bf q}]$ a type 
compatible with $\mathfrak{c}_f$. 
Then the set ${\rm H.Char}[C(\mathfrak{c}_f)]^{[s^r;{\bf p},{\bf q}]}$ modulo difference 
by ${\rm Char}[{\rm Cl}_L]$ is in one-to-one with the set of unitary characters 
$\chi_f: [{\cal O}_L/\mathfrak{c}_f]^\times \rightarrow S^1$ that satisfy 
the condition (\ref{eq:cond-chif-HK-eps}).
\end{thm}

We are also interested in cases where the field of definition is $k' = L_{f_z}$
with $f_z > 1$, and also in cases where $k'/K$ is an abelian extension 
containing $L_{f_z}/K$.  

\begin{lemma}
\label{lemma:Milne-exactseq-modified}
Let $k$ be an abelian extension of $L$ that is contained in the ray class 
field $L_{\mathfrak{c}_f}$ of $L$. Then appropriate subgroups of 
$[{\cal O}_L/\mathfrak{c}_f]^\times$ and 
${\cal O}_L^\times/{\cal O}_{L,1}^\times(\mathfrak{c}_f)$ are determined so that 
\begin{align}
  0 \rightarrow [{\cal O}_L^\times/{\cal O}_{L,1}^\times(\mathfrak{c}_f)]_k
  \rightarrow [{\cal O}_L/\mathfrak{c}_f]^\times_k \rightarrow 
  {\rm Cl}_L(\mathfrak{c}_f) \rightarrow {\rm Gal}(k/L) \rightarrow 0
\end{align}
is exact. 
\end{lemma}
\begin{thm}
\label{thm:model-classify-gen}
The notation and assumption being the same as in Thm. \ref{thm:model-classify-HK} 
and Lemma \ref{lemma:Milne-exactseq-modified}, the set 
${\rm H.Char}[C(\mathfrak{c}_f)]^{[s^r; {\bf p},{\bf q}]}$ modulo 
difference by ${\rm Char}[{\rm Gal}(k/L)]$ is in one-to-one with the set of 
unitary characters $\chi_f: [{\cal O}_L/\mathfrak{c}_f]^\times \rightarrow S^1$ 
satisfying (\ref{eq:cond-chif-HK-eps}) evaluated on 
$[{\cal O}_L/\mathfrak{c}_f]^\times_k \subset [{\cal O}_L/\mathfrak{c}_f]^\times$. 
\end{thm}

The condition (a) in Thm. \ref{thm:model-E/H} can be implemented in the 
language of Hecke characters $\varphi$ of $\mathbb{A}_K^\times/K^\times$ through 
Thms. \ref{thm:model-classify-HK} and \ref{thm:model-classify-gen} for $L=K$. 
The condition (b) is implemented as follows:
\begin{thm}
\label{thm:model-classify-condB}
Let $K$ be an imaginary quadratic field, and $k/K$ an abelian extension 
containing $L_{f_z}/K$ for some $f_z \in \N_{>0}$. Suppose that 
an elliptic curve $E/k$ has complex multiplication by ${\cal O}_{f_z}$, 
and satisfies the condition (*) in Lemma \ref{lemma:HeckeC-LandK}. 
If the conductor $\mathfrak{C}_f$ of the Hecke character $\psi_{E/k}$
of $\mathbb{A}_k^\times/k^\times$ has the same support as $\pi^*(\mathfrak{c}_f)$ 
for an integral ${\cal O}_K$ ideal $\mathfrak{c}_f$ satisfying 
$L_{\mathfrak{c}_f} \supset k$, where 
$\pi: {\rm Spec}({\cal O}_k) \rightarrow {\rm Spec}({\cal O}_K)$, 
then the $[k:K]$ Hecke characters $\varphi$'s of $\mathbb{A}_K^\times/K^\times$
in Lemma \ref{lemma:HeckeC-LandK} admit $\mathfrak{c}_f$ as a modulus, 
\begin{align}
  {\cal O}_K^\times \cdot 
   \left( [{\cal O}_{f_z}] \cap [{\cal O}_K/\mathfrak{c}_f]^\times_k \right)
  = [{\cal O}_K/\mathfrak{c}_f]^\times_k, 
  \label{eq:cond-cf-nonmax}
\end{align}
is of type $[-1/2; 1,0]$, and satisfy one more condition 
\begin{align}
  \chi_f([\alpha]) \in {\cal O}_{f_z}^\times \qquad 
  {}^\forall [\alpha] \in [{\cal O}_{f_z}] \cap 
                        [{\cal O}_K/\mathfrak{c}_f]^\times_k
   \subset [{\cal O}_K/\mathfrak{c}_f]^\times.  
  \label{eq:cond-chif-OfX-valued}
\end{align}

Conversely, suppose that $\varphi$ is a Hecke character of 
$\mathbb{A}_K^\times/K^\times$ that admits $\mathfrak{c}_f$ as 
a modulus, which satisfies $k \subset L_{\mathfrak{c}_f}$ and 
the condition (\ref{eq:cond-cf-nonmax}). Suppose further that $\varphi$ 
is of type $[-1/2;1,0]$ and also satisfies (\ref{eq:cond-chif-OfX-valued}). 
Then $\psi := \varphi \circ {\rm Nm}_{k/K}$ is a Hecke character 
of $\mathbb{A}_k^\times/k^\times$ that admits $\pi^*(\mathfrak{c}_f)$ 
as a modulus, and is of type $[-1/2; {\bf 1}, {\bf 0}]$. The 
condition (b) of Thm. \ref{thm:model-E/H} is satisfied for 
prime ideals $\mathfrak{P}$ prime to $\pi^*(\mathfrak{c}_f)$. 
So, if the Hecke character $\psi$ constructed in that way 
has a conductor $\mathfrak{C}_f$ whose support is the same as 
that of $\pi^*(\mathfrak{c}_f)$, then there is a model $E/k$ whose 
associated Hecke character is this $\psi$.  $\bullet$
\end{thm}

The condition (\ref{eq:cond-psi-real}) is also translated as follows.
\begin{thm}
\label{thm:model-classify-real}
Let $E/L_{f_z}$ be a model of $[E] \in Ell({\cal O}_{f_z})$ and $\psi_{E/L_{f_z}}$ its 
Hecke character. Suppose that this model has the property (*) in Lemma \ref{lemma:HeckeC-LandK}. 
Then the condition (\ref{eq:cond-psi-real}) is replaced by 
\begin{align}
  \chi_f(\rho_{[E]}(\alpha)) = cc \circ \chi_f(\alpha), \qquad 
   {}^\forall [\alpha] \in 
    [{\cal O}_{f_z}] \cap [{\cal O}_K/\mathfrak{c}_f]^\times_{L_{f_z}}
    \subset [{\cal O}_K/\mathfrak{c}_f]^\times, 
  \label{eq:cond-chif-real}
\end{align}
where $\mathfrak{c}_f$ is the conductor of the Hecke character $\varphi$ 
of $\mathbb{A}_K^\times/K^\times$. 
Here, $\rho_{[E]}$ is regarded as an element of 
${\rm Gal}(L_{f_z}/F_{f_z}^{[E]}) \subset {\rm Gal}(L_{f_z}/\Q)$, which maps 
$K \subset L_{f_z}$ to itself. It is implicit in the condition here 
that the integral ${\cal O}_K$ ideal $\mathfrak{c}_f$ is invariant under 
$\rho_{[E]}$.
\end{thm}

Therefore, the conditions (a, b) in Thm. \ref{thm:model-E/H} and (\ref{eq:cond-psi-real}) 
in Thm. \ref{thm:model-E/F} can be implemented by 
Thms. \ref{thm:model-classify-HK}, \ref{thm:model-classify-gen}, \ref{thm:model-classify-condB} 
and \ref{thm:model-classify-real}, respectively, purely in the language of 
the imaginary quadratic field $K$ [Koblitz \cite{Koblitz}, II \S5],\footnote{
memo: $\chi'$ in \cite{Koblitz} corresponds to $\chi_f$ in this article, 
and $\tilde{\chi}$ in \cite{Koblitz} to $\varphi$ or $\psi$ in this article.}
at the cost of introducing a restriction (*) on models.  

\subsection{Combining Them Together}
\label{ssec:L1-combine}

\subsubsection{Models Defined over $H_K$ or $F_K^{[E]}$}
\label{sssec:L1-Hilbert}

Think of a $\C$-isomorphism class of elliptic curves $[E_z]_\C \in Ell({\cal O}_K)$ 
for some imaginary quadratic field $K$. It is realized in superstring theory 
in the form of rational diagonal $T^2$-target ${\cal N}=(2,2)$ SCFT, parametrized 
by $f_\rho \in \N_{>0}$, which controls the choice of complexified 
K\"{a}hler form on $[E_z]_\C$. Now, we observe that the functions 
$f_1(\tau_{ws};\alpha)$ and $\vartheta^1_K(\tau;\Lambda,x)$ are almost the same. In fact, 
\begin{align}
  f_1(\tau_{ws};\alpha) = \frac{C}{i}\sqrt{\frac{2a_z}{f_\rho D_z}}
    \vartheta^1_K\left(\frac{C^2a_z}{f_\rho D_z} \tau_{ws} \; ;  \; 
       \Omega(\Lambda_{\rm winding}), \alpha \right).
\label{eq:obsrv-f1}
\end{align}

Now, think of a model $E/H_K$ or possibly $E/F_K^{[E]}$ of $[E_z]_\C$ 
that corresponds to a multiplicative character 
$\chi_f: [{\cal O}_K/\mathfrak{c}_f]^\times \rightarrow S^1$, where $\mathfrak{c}_f$ 
is an integral ${\cal O}_K$ ideal (Thm. \ref{thm:model-classify-HK}, with $L=K$).  
Focus on string-theory realizations with any $f_\rho \in \N_{>0}$ chosen so that 
\begin{align}
 \mathfrak{c}_f \cdot {\rm LCM}\left(  \mathfrak{a}(\mathfrak{K})_{\mathfrak{K} \in {\rm Cl}_K} 
         \right) \; | \; \mathfrak{b}_z .
   \label{eq:choice-frho-L1-OK}
\end{align}
This condition is a little more restrictive than (\ref{eq:cond-frho-L0-Hilb})
for re-construction of $L(H^0_{et}(E),s) = \zeta_k(s)$ from $f_0(\tau_{ws};\alpha)$'s 
available in string-theory realizations.

For string-theory realizations of $[E_z]_\C$ satisfying the 
condition (\ref{eq:choice-frho-L1-OK}), 
\begin{align}
  f_1(\tau_{ws}; \mathfrak{a}(\mathfrak{K})\mathfrak{c}_f, x) & \; := 
   \sum_{\pi(\alpha) = x} f_1(\tau_{ws}; \alpha), \label{eq:sum-4f1} \\
  \vartheta^1_K(\tau; \mathfrak{a}(\mathfrak{K})\mathfrak{c}_f, x)
   & \; = \sum_{\pi(\alpha) = x} \vartheta^1_K(\tau; \mathfrak{b}_z, \alpha), 
    \label{eq:sum-4theta1}
\end{align}
where $\pi: \mathfrak{a}(\mathfrak{K})/\mathfrak{b}_z \rightarrow 
\mathfrak{a}(\mathfrak{K})/\mathfrak{a}(\mathfrak{K})\mathfrak{c}_f$ is the projection. 
Therefore, the observation (\ref{eq:obsrv-f1}) implies the main result of this article, 
\begin{thm}
\label{thm:main-1}
For a $\C$-isomorphism class $[E_z]_\C \in Ell({\cal O}_K)$, think of a 
model $E/H_K$ that satisfies the condition (*) in Lemma \ref{lemma:HeckeC-LandK}.
Then the Hasse--Weil $L$-function of $E/H_K$ is obtained from the Boltzmann-weighted 
sum of U(1)-charges $f_1(\tau_{ws};\alpha)$ 
in (\ref{eq:def-f1-closed-super}, \ref{eq:def-f1-open-super})
in any one of string realizations of $[E_z]_\C$ so long as the parameter $f_\rho$ 
satisfies the condition (\ref{eq:choice-frho-L1-OK}). The procedure is to use 
(\ref{eq:obsrv-f1}), (\ref{eq:sum-4f1}, \ref{eq:sum-4theta1}), (\ref{eq:def-Hecke-theta}), 
the Mellin transformation (\ref{eq:Mellin-L1}), (\ref{eq:sum-ClK-L1}), 
(\ref{eq:HeckeL4H--HeckeL4K}), and finally (\ref{eq:HWL4H--HeckeL4H}). 

If $\chi_f$ satisfies one more condition, (\ref{eq:cond-chif-real}), then the model 
$E/H_K$ is obtained as a base change of a model $E/F_K^{[E]}$ whose $L$-function 
is obtained by using (\ref{eq:HWL4F--HeckeL4H}) instead of (\ref{eq:HWL4H--HeckeL4H}). 

The numbers of vertex operators with given sets of conformal weight and U(1) charge 
are related to the numbers of solutions of arithmetic models reduced over 
prime ideals through (\ref{eq:obsrv-f1}). $\bullet$ 
\end{thm}

From a slightly different perspective, the same result can be stated also as follows:
\begin{thm}
\label{thm:main-2}
Think of a diagonal rational $T^2$-target CFT (or ${\cal N}=(2,2)$ SCFT) corresponding 
to a $\C$-isomorphism class of elliptic curves $[E_z]_\C \in Ell({\cal O}_K)$ and 
a complexified K\"{a}hler parameter $\rho = f_\rho a_z z$. Then linear combinations 
of the Boltzmann-weighted sum of U(1) charges of irreducible representations (i.e., 
$f_1(\tau_{ws}; \alpha)$'s) in that CFT yield the Hasse--Weil $L$-function in the 
procedure outlined above for models $E/H_K$ of $[E_z]_\C$ with the property (*),  
as long as the conductor 
$\mathfrak{c}_f$ of Hecke characters $\varphi$ of $\mathbb{A}_K^\times/K^\times$ 
in Lemma \ref{lemma:HeckeC-LandK} satisfies the condition (\ref{eq:choice-frho-L1-OK}). 
The number of such models of $[E_z]_\C$ over $H_K$ is finite, once the complex 
structure (i.e., $[E_z]_\C \in Ell({\cal O}_K)$) and the K\"{a}hler parameter 
(i.e., $f_\rho \in \N_{>0}$) are fixed. 

The same procedure also yields the $L$-functions of a finite number of models 
over $F_K^{[E]}$, if there is any such model within the restriction on $\mathfrak{c}_f$ 
set by $f_\rho$ through (\ref{eq:choice-frho-L1-OK}).  $\bullet$
\end{thm}

We have posed a question in Introduction how a $T^2$-target CFT, which depends 
only on the $\C$-isomorphism class $[E]_\C$ (and a K\"{a}hler parameter $\approx f_\rho$), 
can contain 
information of the $L$-functions of multiple different models of $[E]_\C$ defined 
over some number fields. When $[E]_\C$ has complex multiplication, the answer is now clear. 
Such a CFT has finitely many irreducible representations of the chiral algebra labeled 
by $\alpha \in iReps. \cong \Gamma_+^\vee/\Gamma_+ \cong \Lambda_{\rm Cardy}/\Lambda_{\rm winding}$.
Multiple different linear combinations of 
$\{ f_1(\tau_{ws};\alpha) \; | \; \alpha \in iReps.\}$ yield the $L$-function of multiple 
different models of $[E]_\C$. 

The other question posed in Introduction was how the $L$-function of a model $E/k$ 
become independent of the choice of a K\"{a}hler parameter ($f_\rho$) 
in a string-theory realization. The procedure 
summarized in Thm. \ref{thm:main-1} allows us to reproduce the $L$-function 
of $E/k$ from the functions $\{ f_1(\tau_{ws};\alpha) \; | \; \alpha \in iReps.\}$ 
of a string realization with the parameter $f_\rho$ that 
can be chosen in infinitely many different ways. Although the set $iReps.$ and 
the set of functions $\{f_1(\alpha) | \alpha \in iReps.\}$ depend on the choice 
of $f_\rho$, the $L$-function so computed should remain independent of $f_\rho$. 
The procedure achieves $f_\rho$-independence technically, although it is not 
clear how the independence comes about (without exploiting relations 
among theta functions). 

Note also that the study in this article restricted string realizations 
of an elliptic curve $[E]_\C$ with complex multiplication to be within 
the class of {\it diagonal} rational CFT's; the extra constraint 
of being diagonal made our problem easier in that $f_0$'s and $f_1$'s are 
the same regardless of whether we use the left-moving, right-moving 
or the open string sector. The degeneracy among $f_0$'s and $f_1$'s in 
those three sectors also obscured, at the same time, 
how the metric independence has been achieved. 

As a side remark, Thms. \ref{thm:HeckeTheta-automorphic-indiv} 
and \ref{thm:HeckeTheta-automorphic} can be used to see the following.
\begin{thm}
\label{thm:main-3}
Think of any diagonal rational $T^2$-target ${\cal N}=(2,2)$ SCFT where the parameter 
$f_\rho$ satisfies the condition (\ref{eq:choice-frho-L1-OK}) for some integral 
${\cal O}_K$-ideal $\mathfrak{c}_f$ compatible with type $[1,0]$.
Suppose that $\mathfrak{c}_f$ admits a non-trivial character 
$\chi_f:[{\cal O}_K/\mathfrak{c}_f]^\times \rightarrow S^1$ satisfying the 
conditions (\ref{eq:cond-chif-HK-eps}) and (\ref{eq:cond-chif-OfX-valued}). 
Then individual $f_1(\tau_{ws};x)$'s with 
$x \in \mathfrak{a}(\mathfrak{K})/\mathfrak{a}(\mathfrak{K})\mathfrak{c}_f$ are modular 
forms of weight 2 for $\Gamma(4N_{\mathfrak{a}(\mathfrak{K})\mathfrak{c}_f})$. 
Because the group $\Gamma(4N_{\mathfrak{a}(\mathfrak{K})\mathfrak{c}_f}) \subset 
{\rm SL}(2;\Z)$ acts ordinarily on the combination $\tau = 
\tau_{ws} \frac{C^2 a_z}{f_\rho D_z}$, the common subgroup 
$\Gamma(4N_{\mathfrak{a}(\mathfrak{K})\mathfrak{c}_f}) \cap {\rm SL}(2;\Z)_{ws}$ is 
\begin{align}
  {\rm SL}(2;\Z)_{ws} \cap \left[ 
   \diag \left( \frac{f_\rho D_z}{C^2 a_z}, 1\right) \cdot 
     \Gamma(4N_{\mathfrak{a}(\mathfrak{K})\mathfrak{c}_f}) \cdot
    \diag \left( \frac{C^2a_z}{f_\rho D_z}, 1\right) \right].
\end{align}
See also the comment at the end of section \ref{sssec:L0-Hilbert}. 

So long as we deal only with models defined over $F_K$ or $H_K$, and 
$[E] \in Ell({\cal O}_K)$, the following combination,
\begin{align}
  \sum_{\mathfrak{K} \in {\rm Cl}_K} \sum_{\alpha \in \mathfrak{a}(\mathfrak{K})/\mathfrak{b}_z}  
    \chi_f(\pi(\alpha)) f_1(\tau_{ws}; \alpha), 
 \label{eq:sum-w-chif}
\end{align}
where $\pi: \mathfrak{a}(\mathfrak{K})/\mathfrak{b}_z \rightarrow 
\mathfrak{a}(\mathfrak{K})/\mathfrak{a}(\mathfrak{K})\mathfrak{c}_f$ is the 
projection, is a cusp form of weight 2 for $\Gamma_0(N_{\chi_f})$ for some 
multiplier system (homomorphism) 
$\underline{\chi}: \Gamma_0(N_{\chi_f}) \rightarrow \C^\times$. $\bullet$
\end{thm}

What is interesting is that, in string theory, there is no theoretical 
motivation to think of a multiplication law among the irreducible 
representations of the chiral algebra ${\cal A}_- \times {\cal A}_+$ 
or ${\cal A}_+$. 
The addition law among ${\cal O}_K/\Omega(\Lambda_{\rm winding})$, not the 
multiplication law, corresponds to the fusion algebra\footnote{Generally in a 
rational CFT, the fusion algebra introduces a structure of algebra into $\Z[iReps.]$. 
In the case of a $T^2$-target rational CFT, however, the multiplication law on 
$\Z[iReps.]$
can be induced from an abelian group law on $iReps.$. The addition law referred to 
in the main text corresponds to this abelian group law on $iReps.$.} 
on $iReps. = \Lambda_{\rm Cardy}/\Lambda_{\rm winding}$. 
Here, however, it is crucial to use the character 
of the group $[{\cal O}_K/\mathfrak{b}_z]^\times$ in multiplication, not 
in the addition of ${\cal O}_K/\mathfrak{b}_z$, in order to construct 
the modular forms (\ref{eq:sum-w-chif}) of weight 2 for $\Gamma_0(N_{\chi_f})$.
As a reminder, though, individual $f_1(\tau_{ws};x)$'s are modular forms for 
a group of the form $\Gamma(4N_{\mathfrak{a}(\mathfrak{K})\mathfrak{c}_f})$ without a sum 
with the character $\chi_f$, if not\footnote{
They {\it are} for a group $\Gamma_0(M)$ for an $M$ that scales as 
$N_{\mathfrak{a}(\mathfrak{K})\mathfrak{c}_f}^2$.} for a group of the form of $\Gamma_0(M)$
for some $M \in \N$ that scales as $N_{\mathfrak{a}(\mathfrak{K})\mathfrak{c}_f}$. 

Here, we illustrate by examples how Theorems \ref{thm:main-1} and \ref{thm:main-2} 
work in practice. 
\begin{exmpl}
\label{exmpl:L1-1}
For an imaginary quadratic field $K=\Q(\sqrt{-1})$, $[E_z]$ with 
$z = \sqrt{-1}=i$ is the only $\C$-isomorphism class of elliptic curves 
with complex multiplication by ${\cal O}_K$. 
It has multiple models defined over the Hilbert class field $H_K=K$, 
and those over $F_K=\Q$. 
A modulus $\mathfrak{c}_f$ of a Hecke character $\psi_{E/H_K}=\varphi$
for $H_K=K$ is compatible with the type $[1,0]$ iff 
$\mathfrak{c}_f \Slash{\, | } (2)_{{\cal O}_K}$ in this case. 

For example, two choices $\mathfrak{c}_f = (2+2i)_{{\cal O}_K}$ and 
$\mathfrak{c}_f = (4)_{{\cal O}_K}$ have their own unique primitive multiplicative 
character $\chi_f$ satisfying (\ref{eq:cond-chif-HK-eps}) 
and (\ref{eq:cond-chif-OfX-valued}) for $k=H_K=K$. Those two characters $\chi_f$ satisfy 
(\ref{eq:cond-chif-real}), so the corresponding two models can be defined over $\Q$.
It is known that the two models correspond to defining equations 
$y^2 = x^3 -n^2 x$ with $n=1,2$, respectively [Koblitz \cite{Koblitz}, Chap. II].
On the other hand, the choice $\mathfrak{c}_f = (2+i)_{{\cal O}_K}$ has a unique 
multiplicative character satisfying the two conditions (\ref{eq:cond-chif-HK-eps}, 
\ref{eq:cond-chif-OfX-valued}) for $k=H_K=K$, so there exists a model over $K$, but
it is not obtained as a base change from a model over $\Q$, 
because (\ref{eq:cond-chif-real}) is not satisfied.  
When we choose $\mathfrak{c}_f = (3)_{{\cal O}_K}$, there are two inequivalent 
characters $\chi_f: [{\cal O}_K/\mathfrak{c}_f]^\times \rightarrow 
S^1$ satisfying (\ref{eq:cond-chif-HK-eps}). The two characters do not satisfy 
(\ref{eq:cond-chif-OfX-valued}), however. So, there is no model of $[E_{z = i}]_\C$ 
over $k=H_K=K$ with the conductor $\mathfrak{C}_f = (3)_{{\cal O}_K}$. 

For both of the two models corresponding to $\mathfrak{C}_f = (2+2i)_{{\cal O}_K}$ and 
$(4)_{{\cal O}_K}$, the condition (\ref{eq:choice-frho-L1-OK}) on the parameter 
$f_\rho$ of string realizations is $2|f_\rho$; remember that 
$\Omega(\Lambda_{\rm winding})/f_\rho = (2)_{{\cal O}_K}$ for the case of $[E_z]$ here
(see Table \ref{tab:123}).
In the minimal realization\footnote{The $L$-function written down 
in \cite{Schimmrigk-05} for this $\C$-isomorphism class of elliptic curves, 
$[E_z]_\C$ with $z = i$, is the one for $\mathfrak{C}_f = (4)_{{\cal O}_K}$. 
The conductor of the elliptic curve over $\Q$ is $N_{E/\Q} = N_{\chi_f} = 
4 \times {\rm Nm}_{K/\Q}(\mathfrak{c}_f) = (64)_\Z$, as in \cite{Schimmrigk-05}. 
The modular forms (\ref{eq:exmpl-1})---we know that they are (due 
to Theorem \ref{thm:main-3})---are obtained from string realizations with $f_\rho = 2$ 
or any $2|f_\rho$, not from the one with $f_\rho = 1$.  \\
It should be noted, on the other hand, that the two Gepner constructions 
$0^{\otimes 1}2^{\otimes 2}/(\Z/4)$ and $0^{\otimes 1}2^{\otimes 2}/((\Z/4) \times (\Z/4\Z))$
yield the same ${\cal N} = (2,2)$ SCFT, as one can see by computing the spectrum. 
This string realization corresponds to $z = \rho = i$, so $f_\rho = 1$. See also 
footnote \ref{fn:not-minimal-model}.} using $f_\rho = 2$, for example, there are 16 irreducible 
representations of the chiral algebra ${\cal A}_- \cong {\cal A}_+$. 
Out of the Boltzmann-weighted sum of U(1)-charges $f_1(\tau_{ws};\alpha)$ of 
those 16 representations, the following linear combinations yield the $L$-function
through the Mellin transformation: 
\begin{align}
& \frac{[f_1(1)-if_1(i)-f_1(-1)+if_1(-i)]}{4}
\pm i \frac{[f_1(2+i)-if_1(-1+2i)-f_1(-2-i)+if_1(1-2i)]}{4}, \nonumber \\
&\; = f_1(1) \pm i f_1(2+i);
  \label{eq:exmpl-1}
\end{align}
here, only the second argument $\alpha \in iReps.$ is retained in the expression 
above, and a complex number in a unit cell of $\C/\mathfrak{b}_z$ is used to refer 
to an irreducible representation $\alpha$ by exploiting the embedding 
$\Omega: iReps. \cong \Omega(\Lambda_{\rm Cardy})/\mathfrak{b}_z
 \subset K/\mathfrak{b}_z \subset \C/\mathfrak{b}_z$. 

For the model over $K$ corresponding to the multiplicative character with $\mathfrak{C}_f 
= (2+i)_{{\cal O}_K}$, we find that string realizations with $5|f_\rho$ need to be used 
to obtain the modular forms to be Mellin-transformed.  $\bullet$
\end{exmpl}

\begin{exmpl}
For an imaginary quadratic field $K=\Q(\sqrt{-3})$, $Ell({\cal O}_K)$ consists of 
just one $\C$-isomorphism class, $[E_z]_\C$ with $z = \zeta_3$, where 
$\zeta_N:= e^{2\pi i/N}$. The three Gepner constructions,\footnote{
\label{fn:not-minimal-model}
An orbifold construction is a procedure of using a modular invariant SCFT to build 
a modular invariant SCFT that is different from the original one, as well as 
the SCFT so constructed. Gepner construction is a special version of that.  
In this article, we study the relation between the $L$-functions of a variety $X$ 
and the spectrum (i.e., $f_0$'s and $f_1$'s) of the $X$-target SCFT, for natural reason, 
instead of the spectrum of the SCFT (such as $N=2$ minimal models) which one may use to 
construct the $X$-target SCFT.
The fact that multiple different Gepner constructions may give rise to the same 
geometry-target SCFT motivates the way we set the problem in this article.} 
$1^{\otimes 3}/(\Z/3)$, $1^{\otimes 3}/((\Z/3)\times (\Z/3))$ and 
$0^{\otimes 1}1^{\otimes 1}4^{\otimes 1}/(\Z/6)$, all give rise to a single ${\cal N}=(2,2)$
SCFT, as one can see by computing and comparing the spectrum for those Gepner 
constructions; the common spectrum reveals that it is a string-theory realization 
of this $[E_z]_\C$ with $z = \zeta_3$, with $f_\rho = 1$ (i.e., $\rho = \zeta_3$). 

The compatibility condition with the type [1, 0] rules out any choice with 
$\mathfrak{c}_f | (1+\zeta_6)_{{\cal O}_K}$ or $\mathfrak{c}_f | (2)_{{\cal O}_K}$.

Ref. \cite{Schimmrigk-05} picks up two models over $\Q$ for this $\C$-isomorphism class 
$[E_z]_\C$. Here is how we understand the two models. First, let us choose the modulus 
$\mathfrak{c}_f=(3)_{{\cal O}_K}$; there is just one unitary character of 
$[{\cal O}_K/\mathfrak{c}_f]^\times$ satisfying the conditions (\ref{eq:cond-chif-HK-eps}, 
\ref{eq:cond-chif-OfX-valued}) then. The corresponding Hecke character $\varphi = \psi: 
\mathbb{A}_K^\times / K^\times \rightarrow \C^\times$ must be primitive, since any 
${\cal O}_K$-ideals dividing $(3)_{{\cal O}_K}$ cannot be compatible with the type [1, 0]. 
This Hecke character $\psi_{E/K}$ therefore has a conductor 
$\mathfrak{C}_f = \mathfrak{c}_f = (3)_{{\cal O}_K}$. The condition (\ref{eq:cond-chif-real})
is satisfied, and hence this model $E/K$ is obtained as a base change from a model $E/\Q$. 
The $L$-function of this model $E/\Q$ is $[1/1^s-2/4^s-1/7^s + \cdots]$ and the conductor 
of the elliptic curve over $\Q$ is $N_{E/\Q} = (27)_\Z$, where we 
used (\ref{eq:formula-conductor4EC-F}). This reproduces the $L$-function and $N_{E/Q}$ 
of one of the two models over $\Q$ discussed in \cite{Schimmrigk-05}.

The other model over $\Q$ is found by setting $\mathfrak{c}_f = (4(1+\zeta_6))_{{\cal O}_K}$. 
There are four unitary characters of $[{\cal O}_K/\mathfrak{c}_f]^\times$ in this case, 
but one of them is induced from a unitary character for 
$\mathfrak{c}_f = (2(1+\zeta_6))_{{\cal O}_K}$ and two others from two unitary characters 
for $\mathfrak{c}_f = (4)_{{\cal O}_K}$. There is just one unitary character where 
the choice of modulus $\mathfrak{c}_f = (4(1+\zeta_6))_{{\cal O}_K}$ is primitive. 
So, there is a unique Hecke character for $K = \Q(\sqrt{-3})$ where the conductor 
is $\mathfrak{C}_f = (4(1+\zeta_6))_{{\cal O}_K}$. It satisfies (\ref{eq:cond-chif-real}), 
so the corresponding model is obtained as a base change of a model $E/\Q$. 
The $L$-function and the conductor $N_{E/\Q}$ of such a model $E/\Q$ are computed to be 
$[1+4/7^s+2/13^s-8/19^s-5/25^s + \cdots]$, and $N_{E/\Q} = (144)_\Z$, respectively. 
So, the other model of $[E_{z = \zeta_3}]_\C$ in \cite{Schimmrigk-05} is reproduced. 

In order to obtain the modular forms in the procedure in Theorem \ref{thm:main-1}, 
we need a string realization with $3|f_\rho$, and $4|f_\rho$ for the models 
corresponding to $\mathfrak{C}_f = (3)_{{\cal O}_K}$ and $(4(1+\zeta_6))_{{\cal O}_K}$, 
respectively. $\bullet$
\end{exmpl}

There are nine imaginary quadratic fields $K=\Q(\sqrt{-d_0})$ where $h({\cal O}_K) = 1$
 ($d_0 = 1,2,3,7,11,19,43,67,163$). That is when 
just one linear combination of $f_1(\tau_{ws};\alpha)$'s (for 
$\vartheta(\tau; \varphi, \mathfrak{K})$ with $\{\mathfrak{K} \} = {\rm Cl}_K$) 
is enough in constructing the $L$-functions of models over $\Q$ or over $K$. 
The two examples above, where $K=\Q(\sqrt{-1})$ and $\Q(\sqrt{-3})$, 
are both in this category. 
Let us also take a look at an example of imaginary quadratic fields other than these 
nine special ones. 

\begin{exmpl}
For an imaginary quadratic field $K=\Q(\sqrt{-5})$, a modulus $\mathfrak{c}_f$ of 
a Hecke character of $\mathbb{A}_K^\times/K^\times$ is compatible with the type $[1,0]$ 
iff $\mathfrak{c}_f \Slash{|} (2)_{{\cal O}_K}$. 

If we choose $\mathfrak{c}_f = (3, + 1+\sqrt{5}i)_{{\cal O}_K}$, or 
$\mathfrak{c}_f = (3,-1+\sqrt{5}i)_{{\cal O}_K}$, there is one unique character 
$\chi_f$ satisfying (\ref{eq:cond-chif-HK-eps}, \ref{eq:cond-chif-OfX-valued}) 
for each case. $\pi^*(\mathfrak{c}_f)$ is a prime ideal of ${\cal O}_{H_K}$ for 
any one of the two choices of $\mathfrak{c}_f$ above, and hence 
$\psi = \varphi \circ {\rm Nm}_{H_K/K}$ is primitive; 
$\mathfrak{C}_f = \pi^*(\mathfrak{c}_f)$.
Thus, there is one model over $H_K$ for each choice of $\mathfrak{c}_f$ 
and for each one of $Ell({\cal O}_K)$.
The condition (\ref{eq:choice-frho-L1-OK}) implies that we need to employ string 
realizations with $3|f_\rho$ in order to write down the inverse Mellin transform 
of the Hecke $L$-functions in terms of $f_1$'s. The four models over $H_K$ here, 
corresponding to two possible $\mathfrak{c}_f$'s and two in $Ell({\cal O}_K)$, 
cannot be obtained by the base change of a model over $F_K=\Q(\sqrt{5})$, because 
the two $\chi_f$'s do not satisfy (\ref{eq:cond-chif-real}). $\bullet$
\end{exmpl}

\subsubsection{Models Defined over a Ramified Extension of $K$}
 \label{sssec:L1-general}

Even for arithmetic models of elliptic curves with complex multiplication 
by a non-maximal order in general, or for models of elliptic curves with 
complex multiplication whose field of definition $k$ is an abelian extension 
over $K$ containing the ring class field as a proper subfield, we can use 
Thms. \ref{thm:model-classify-gen} and \ref{thm:model-classify-condB}
to list up inequivalent models over a number field $k$ in the class we consider 
in this article. For those models, general algorithm for the computation of 
the $L$-functions reviewed in section \ref{sssec:Hecke-L-Hecke-theta} still works 
(as stated there); 
the relation (\ref{eq:obsrv-f1}) also holds for these models; therefore, 
the functions $f_1$'s obtained from string realizations of these arithmetic 
models can be organized by using the character $\chi_f$ as linear combination 
coefficients so that the Mellin transform of the linear combinations 
become Hecke $L$-functions to be used in rebuilding the Hasse--Weil $L$-function. 

The only one change we have to make is to replace the 
condition (\ref{eq:choice-frho-L1-OK}) on the parameter $f_\rho$ of 
string realizations by 
\begin{align}
  \mathfrak{c}_f \cdot {\rm LCM}\left(
     \mathfrak{a}(\mathfrak{K})_{\mathfrak{K} \in {\rm Cl}_K(\mathfrak{m}_f)} \right) 
   \supset \mathfrak{b}_z,
\end{align}
because $\mathfrak{b}_z$ is a proper ${\cal O}_{f_z}$-ideal for elliptic 
curves in $Ell({\cal O}_{f_z})$ and is not necessarily an ${\cal O}_K$ ideal, 
when $f_z > 1$. We do not reiterate Thms. \ref{thm:main-1}, \ref{thm:main-2} 
and the first half of \ref{thm:main-3} modified for these general cases, 
because the necessary modification is obvious. 

\begin{exmpl}
For an imaginary quadratic field $K=\Q(\sqrt{-1})$, $k=K(\sqrt{3})$ is 
an abelian extension of $K=L_{f_z=1} = L_{f_z=2}$ as well as $L_{f_z=3}=K(\sqrt{3})$ 
(see Table \ref{tab:cft-qif}).
It is also isomorphic to the ray class field $L_{\mathfrak{m}_f}$ for 
$\mathfrak{m}_f = (3)_{{\cal O}_K}$. 

Here, we look at models over $k=K(\sqrt{3})$ of elliptic curves in 
$Ell({\cal O}_K) = \{ [E_{z = i}] \}$ and $Ell({\cal O}_{f_z =3}) = \{ [E_{z = 3w_K}], 
[E_{z = (1+3w_K)/2}] \}$ where the Hecke characters $\varphi$ of 
$\mathbb{A}_K^\times/K^\times$ in Lemma \ref{lemma:HeckeC-LandK} have a conductor $\mathfrak{c}_f = (3)_{{\cal O}_K}$. 
We have already seen in Example \ref{exmpl:L1-1}
that $\mathfrak{c}_f = (3)_{{\cal O}_K}$ as a modulus is compatible with the type 
$[1,0]$. Moreover, the condition (\ref{eq:cond-cf-nonmax}) is satisfied for 
$f_z = 3$; trivial for $f_z = 1$. 

There are two inequivalent characters $\chi_f:[{\cal O}_K/\mathfrak{c}_f]^\times 
\rightarrow S^1$ satisfying (\ref{eq:cond-chif-HK-eps}):
\begin{align}
 \chi_f(1) = 1, \quad \chi_f(i)=-i, \quad \chi_f(2) = -1, \quad \chi_f(2i)=i, \\
 \chi_f(1+i)= a, \quad \chi_f(2+i)=-ia, \quad \chi_f(2+2i)=-a, \quad \chi_f(1+2i)=ia, 
\end{align}
where $a=e^{\pi i /4}$ or $-e^{\pi i/4}$. The two $\chi_f$'s become the same,
when they are restricted to $[{\cal O}_K/\mathfrak{c}_f]^\times_k = 
\{ [1], [2], [i], [2i]\}$, however, and the value of $\chi_f$ remains within 
${\cal O}_K^\times$ (the condition (\ref{eq:cond-chif-OfX-valued}) for $k$); 
because $\pi^*((3)_{{\cal O}_K})= \mathfrak{P}_3^2$ for a norm-3 prime ideal 
$\mathfrak{P}_3$ of ${\cal O}_k$, the support of the conductor of 
$\psi = \varphi \circ {\rm Nm}_{k/K}$ cannot be different from that of 
$\pi^*(\mathfrak{c}_f)$; 
this means that there is just one model over $k$ for $[E_{z = i}]_\C$ 
(Thms. \ref{thm:model-classify-gen} and \ref{thm:model-classify-condB}).  
There is also just one model over $k$ for each one of $[E_{z = 3w_K}]_\C$ 
and $[E_{(1+3w_K)/2}]_\C$, since the value of $\chi_f$ remains to be within 
${\cal O}_{f_z}^\times = \{ \pm 1\}$ for $[{\cal O}_{f_z}] \cap 
[{\cal O}_K/\mathfrak{c}_f]^\times = \{ [1], [2] \}$. 
The Hecke $L$-function of $\psi_{E/k}$ of $\mathbb{A}_k^\times/k^\times$ is given by 
$L(s,\psi_{E/k}) = \prod_{a \in \{ \pm e^{\pi i/4} \} } L(s,\varphi_a)$ for all the 
three models over $k$; the two $\varphi_a$'s are the Hecke characters of 
$\mathbb{A}_K^\times/K^\times$ corresponding to the two $\chi_f$'s. 
The Hecke $L$-function $L(s,\varphi_a)$ is given by the Mellin transform of 
appropriate linear combinations of $f_1$'s of string realizations of the 
three models, if $3|f_\rho$, $3|f_\rho$ and $3|f_\rho$, respectively. $\bullet$
\end{exmpl}

\section{The $L$-function for $H^2_{\it et}(E)$}

The $L$-function is defined for an elliptic curve $E/k$ defined over 
a number field $k$ also in association with $H^2_{\it et}(E)$. Because 
$L(H^2_{\it et}(E),s) = \zeta_k(s-1)$, this $L$-function does not carry 
any information not contained in $L(H^0_{\it et}(E),s) = \zeta_k(s)$. 

If one is interested in reconstructing $L(H^2_{\it et}(E),s)$ directly 
from some data available in string-theory realizations of $E/k$, than 
shifting the argument of $L(H^0_{\it et}(E),s)$, then we can use 
\begin{align}
f_2(it_{ws};\alpha) := (-i) 
     {\rm Tr}_{V^o_\alpha;R} \left[ F e^{\pi i F} \Omega' \overline{\Omega}' 
    q^{L_0-c/24} \right],  \quad q=e^{-2\pi t_{ws}}
\end{align}
for $\alpha \in iReps.$, instead of $f_0$'s. By replacing $f_0$'s 
that appear in section \ref{sec:forH0} by $f_2$'s, we simply\footnote{
The corresponding theta functions, however, do not fall into the category of 
congruent theta functions introduced in (\ref{eq:def-theta4L-P}); the Poisson 
resummation formula is messed up by insertion of a monomial $|w|_\C^2$. }
obtain $\zeta_k(s-1)$.

\subsection*{Acknowledgments}

We are grateful to T. Abe and K. Hori for discussion and useful comments. 
This work is supported by WPI Initiative, MEXT, Japan (SK, TW). 


%

\end{document}